%% file: paper_combined.tex
\documentclass[12pt]{article}
\usepackage{lscape}
\usepackage{geometry}
\usepackage[onehalfspacing]{setspace}
\usepackage{amssymb}
\usepackage{amsfonts}
\usepackage{graphicx}
\usepackage{amsmath}
\usepackage{bbm}
\usepackage{epstopdf}
\usepackage{natbib}
\usepackage{bibunits}
\usepackage{morefloats}
\usepackage{float}
\usepackage{placeins}  % For \FloatBarrier
\usepackage{comment}
\usepackage{pdflscape}
\usepackage{adjustbox}
\usepackage{subcaption}
\usepackage{longtable}
\usepackage{apalike}
\usepackage{xfrac}
\usepackage[dvipsnames,table]{xcolor}

\usepackage{etoolbox,siunitx}
\robustify\bfseries

\usepackage{hhline,multirow}
\usepackage{yfonts}
\usepackage[ruled,vlined]{algorithm2e}
\SetAlFnt{\fontfamily{phv}\selectfont\small}
\SetAlCapFnt{\fontfamily{phv}\selectfont\normalsize\bfseries}
\SetAlCapNameFnt{\fontfamily{phv}\selectfont\normalsize\bfseries}
\SetKwSty{textbf}
\SetArgSty{textnormal}
\SetAlgoSkip{bigskip}
\SetAlCapSkip{1em}
\definecolor{algoblue}{RGB}{0,51,102}
\SetKwComment{tcp}{\fontfamily{phv}\selectfont\color{dpd}$\vartriangleright$~}{}
\newcommand{\algostep}[1]{\textcolor{algoblue}{\fontfamily{phv}\selectfont\textbf{#1}}}

\usepackage{longtable}
\usepackage{threeparttablex}

\newcolumntype{d}[1]{D{.}{.}{#1}}

\usepackage{qtree}
\usepackage{multicol}
\usepackage{booktabs}
\usepackage{threeparttable}
\usepackage{tabularx}
\usepackage{dcolumn}
\newcolumntype{d}[1]{D..{#1}} % for alignment of numbers on decimal marker

\usepackage{siunitx}
\usepackage{mathpazo} % math & rm
\usepackage[scaled]{helvet} % ss
\usepackage{courier} % tt
\normalfont
\usepackage[T1]{fontenc}
\usepackage{listings}
\usepackage{epigraph}
\def\sym#1{\ifmmode^{#1}\else\(^{#1}\)\fi}

\usepackage[pdftex,bookmarks,colorlinks]{hyperref}
\hypersetup{
  colorlinks,
  citecolor=darkpowderblue,
  linkcolor=red,
  urlcolor=darkpowderblue}
\usepackage{cleveref}

\usepackage{etoolbox}

\definecolor{dukeblue}{rgb}{0.0, 0.0, 0.61}
\definecolor{darkred}{rgb}{0.8,0,0}

\makeatletter
\patchcmd{\epigraph}{\@epitext{#1}}{\itshape\@epitext{#1}}{}{}
\makeatother

\makeatletter
\def\munderbar#1{\underline{\sbox\tw@{$#1$}\dp\tw@\z@\box\tw@}}
\makeatother

\usepackage{graphicx}
\usepackage{wrapfig}

\usepackage{xspace}
\usepackage{microtype}

\definecolor{bred}{RGB}{122, 0, 0}
\definecolor{darkpowderblue}{rgb}{0.0, 0.05, 0.5}

\definecolor{dpd}{rgb}{0.0, 0.05, 0.5}
{}
 \definecolor{dpd2}{rgb}{0.0, 0.043, 0.43}

\makeatletter
\newcommand\halftiny{\@setfontsize\halftiny\@vipt\@viipt}
\newcommand\notsotiny{\@setfontsize\notsotiny{6.99}{9.2828}}
\newcommand\notsolarge{\@setfontsize\notsolarge{12}{14}}
\renewcommand\paragraph{\@startsection{paragraph}{4}{\z@}%
  {1.5ex \@plus .5ex \@minus .2ex}%
  {-1em}%
  {\normalfont\normalsize\bfseries}}
\makeatother

\usepackage{soul}

\renewenvironment{abstract}
 {\small
  \begin{center}
  \bfseries \abstractname\vspace{-.5em}\vspace{0pt}
  \end{center}
  \list{}{
    \setlength{\leftmargin}{1.8cm}    \setlength{\rightmargin}{\leftmargin}  }  \item\relax}
 {\endlist}
\usepackage{graphicx}
\usepackage{wrapfig}
 
\allowdisplaybreaks
 
\usepackage{enumitem}

\usepackage{tikz}
\usetikzlibrary{positioning}

\definecolor{ForestGreen}{RGB}{34,139,34}
\definecolor{dpd2}{rgb}{0.0, 0.035, 0.38}
\definecolor{RowAlt}{gray}{0.975}

\newcommand{\mn}[1]{{\fontfamily{phv}\selectfont #1}} % FRED-QD mnemonic formatting

\begin{document}
\sloppy

\title{\vspace*{-0.5cm} \huge  \textbf{\color{dpd} LGB+: \\ A Macroeconomic Forecasting Road Test \\ \phantom{.}} \vspace*{0.25cm}}  

\author{\hspace*{-0.3cm} Philippe Goulet Coulombe\thanks{%
Département des Sciences Économiques, \href{mailto:p.gouletcoulombe@gmail.com}{{\fontfamily{phv}\selectfont p.gouletcoulombe@gmail.com}}.
For helpful comments, I thank Karin Klieber, Maximilian Göbel, and Anne Valder.
R and Python implementations are available on \href{https://github.com/philgoucou/lgbplus}{{\fontfamily{phv}\selectfont  GitHub}}.
This draft: \today.}\\[-0.2cm] 
\hspace*{-0.3cm} \textbf{\texttt{\fontfamily{phv}\selectfont \notsolarge Université du Québec à Montréal}} 
}

\date{\vspace{0.2cm}
\small
\small
\vspace{0.0025cm}
\large
\vspace{0.15cm}
%{\normalsize\color{red}\textbf{Preliminary and incomplete. Please do not share without permission.}}
}
\maketitle

\date{\vspace{0.002cm}
\small
\small
\vspace{-0.15cm}
}
%\maketitle
\begin{abstract}

\noindent Needless to say, linear dynamics are pervasive in economic time series, particularly autoregressive ones. While gradient boosting with trees is naturally geared toward nonlinearities, it is inefficient in small samples when much of the predictive content is linear, contorting splits to approximate relationships better captured by simple linear terms. This paper proposes LGB+, a boosting procedure operating on a more inclusive set of basis functions. The idea comes in two flavors. LGB+ evaluates a tree \textit{and} a linear candidate at each step against out-of-bag data; only the winner advances. The simpler variant, LGB$^{\texttt{A}}$+, alternates on a fixed schedule: a block of tree updates, then a greedy linear correction, repeat. Both designs avoid ex ante commitments to any particular functional form or predictor selection. Because the prediction is the sum of a linear and a tree component, forecasts decompose natively into linear and nonlinear contributions, and so do permutation-based variable importance and historical proximity weights. In a quarterly U.S.\ macroeconomic forecasting exercise, LGB+ delivers strong gains for targets with pronounced autoregressive dynamics or mixed linear--nonlinear signals. Variables dominating the linear channel are those operating through autoregressive persistence or near-accounting relationships to the target (e.g., initial claims for unemployment and building permits for housing starts).

\end{abstract}

\thispagestyle{empty}
%\noindent \textit{JEL Classification: C53, C55, E37, C32} 

%\noindent \textit{Keywords: Random Forest, Trees, Regime Switching, Structural Breaks, Time-varying Parameters}

%EndExpansion

%\noindent \textit{JEL codes: C12, C14, G11, G12, G17}

%\noindent \textit{Keywords: Portfolio optimization,  Machine Learning,  Statistical Arbitrage,  Nonlinear time series}

\clearpage

%\thispagestyle{empty}
%\tableofcontents
%\thispagestyle{empty}

\clearpage 
\setcounter{page}{1}

\newgeometry{left=2 cm, right= 2 cm, top=2.3 cm, bottom=2.3 cm}

\section{Introduction}

Linear dynamics---autoregressive structures especially---are the dominant feature of most macroeconomic forecasting problems, and tree-based methods are not built for this. By construction, random forests and gradient boosting machines devote a substantial fraction of their capacity approximating effects that a handful of linear terms would capture far more cheaply. A familiar scene in applied macro forecasting: one assembles a large predictor panel, feeds it into, say, a random forest, and is thoroughly defeated by a four-lag autoregression. The model is simply asked to do too much with too little. In macroeconomic forecasting, where small samples are the rule rather than the exception, approximating something simple with something complicated is a luxury we cannot afford.

This, of course, has been observed many times in the past. MacroRF \citep{gouletcoulombe2020macroeconomic} addressed it by modeling nonlinearities as time variation in otherwise linear relationships. Partially linear and local linear forest methods take a related approach through locally linear adjustments \citep{friedberg2021local,athey2019generalized}. In each case, the linear specification must be committed to before the trees see the data which, inevitably, is a specification gamble.

%This issue has been noted in previous work, like \cite{gouletcoulombe2020macroeconomic}, and others. 

%including my own analysis of random forests for macroeconomic forecasting, which proposed hybrid approaches designed to accommodate linear structure more explicitly. While such methods can be effective, they introduce two limitations when the primary objective is forecasting accuracy rather than structural interpretation. First, they typically require the researcher to specify the linear component ex ante—an easy task in some settings, but ambiguous in others. Second, existing implementations can be computationally intensive in large datasets, limiting their practicality as routine benchmarks.

\paragraph{LGB+.} The fix is surprisingly simple, and builds directly on LightGBM \citep{ke2017lightgbm}---hence the name. Within a single unified boosting loop, there is no need to pre-commit to a linear specification: adding a greedy univariate linear update as a candidate at each step---alongside the usual tree update---lets the algorithm discover on its own what should enter linearly and what should not. The idea comes in two flavors that form a hybrid boosting family. LGB+ formalizes this as a per-step competition: a tree candidate and a linear candidate are each evaluated on out-of-bag data; only the winner advances. LGB$^{\texttt{A}}$+ takes a simpler route, alternating on a fixed schedule: a block of tree updates, then one linear correction, repeat. Both share the same core mechanism---interleaving linear and tree updates within the boosting loop---and perform comparably across a wide range of settings.
Crucially, the linear component is never pre-specified: it is built incrementally during training, one variable at a time, as the algorithm identifies which predictors help most when entered linearly. Unlike residualization or standard additive models, nothing needs to be committed ex ante. Despite this simplicity, the strategy is largely absent from macroeconomic forecasting benchmarks---yet the evidence will show it has a pretty desirable ROI.

Beyond accuracy, LGB+'s additive structure opens the door to something infrequent in machine learning: a direct decomposition of what is linear and what is not. First, because the prediction sums a linear and a tree component, each forecast decomposes directly into a linear contribution and a tree-based one. Second, permutation-based variable importance, usually computed for the full prediction, can be decomposed into three terms---linear, trees, and an interaction cross-term---since importance is MSE-based and squaring the additive sum produces the cross-term. This reveals which variables matter and through which channel they operate. Third, following \citet{gouletcoulombe2024dual}, the prediction can be written as a weighted combination of training outcomes, with weights interpretable as proximity between current conditions and historical episodes; these observation weights themselves decompose into linear versus nonlinear proximity. Together, these tools help diagnose both successes and failures, and whether linearity or nonlinearity has to do with it.

%; here, the partition of explanatory power between linear and nonlinear channels is itself learned from the data.}

\paragraph{Results.} I evaluate the method in two settings. Monte Carlo simulations spanning eight data-generating processes, from purely linear to highly nonlinear, demonstrate that {both variants (LGB+ and LGB$^{\texttt{A}}$+)} perform robustly throughout, trailing OLS only modestly in linear environments while matching or exceeding pure tree methods in nonlinear settings. Importantly, the linear component does not hurt even when it is useless: if residuals have no linear structure, the greedy updates shrink toward zero and, in the competition variant, simply lose to the tree candidate at each step.

Empirically, I apply LGB+ to forecasting six U.S.\ macroeconomic aggregates (GDP, unemployment, inflation, industrial production, housing starts, and the term spread) at the quarterly frequency using the standard FRED-QD database as input. Four main findings emerge from this exercise. First, pre-COVID (2007--2019), LGB+ delivers its strongest gains where autoregressive dynamics matter most---unemployment at $h=1$ (RMSE 33\% below the AR benchmark) and industrial production at $h=1$ (12\% below)---with these improvements driven primarily through the linear channel. The linear channel excels at short horizons because it captures two types of relationships that trees approximate only inefficiently: autoregressive persistence (time smoothing) and approximate accounting identities. For instance, initial unemployment claims (\mn{CLAIMSx}) predict unemployment almost mechanically, and building permits predict housing starts by near-definitional precedence---both enter overwhelmingly through the linear channel.

Second, the decomposition reveals meaningful heterogeneity beyond these accounting-like relationships: for unemployment, lagged values contribute both linearly and through trees (consistent with asymmetric business-cycle dynamics), whereas for industrial production the autoregressive structure is almost entirely linear. Third, post-COVID (2021/22--2025), the linear channel becomes actively harmful for several real-activity targets: variables like nondurable materials production signal a recession following the 2022 rate hikes that did not materialize. Housing starts is the notable exception, where the rate-driven slowdown did materialize and mixed linear and nonlinear channels (inventory ratios, corporate bond rates) correctly captured it. Fourth, at longer horizons, variable importance shifts from autoregressive toward more forward-looking financial indicators (S\&P~500, Case-Shiller, credit conditions), with the tree component gaining relative importance.

\paragraph{Related literature.} This paper connects to four strands of literature.

\vspace{0.3em}
\noindent\textit{Macroeconomic forecasting and econometrics.} A long-standing empirical regularity in macroeconomic forecasting is the dominance of low-dimensional linear dynamics, particularly autoregressive behavior at short horizons. This regularity helps explain the persistent competitiveness of linear benchmarks, even in data-rich environments \citep{stockwatson2002forecasting}. More recent work nonetheless shows that machine learning methods can improve forecasting performance for key macroeconomic variables when applied carefully to large predictor sets \citep{goulet2022machine,medeiros2021forecasting}, among many others.

\vspace{0.3em}
\noindent\textit{Tree-based methods and linear structure.} Tree-based ensemble methods play a central role in this literature because they accommodate nonlinearities, interactions, and state dependence without requiring explicit pre-specification. However, several contributions note that when predictive content is largely linear, such methods may be statistically inefficient in small samples, expending model capacity approximating dynamics that could be captured directly. In fact, this is precisely the problem MacroRF was designed to address: rather than abandoning linearity altogether, it models nonlinearities as time variation in the coefficients of an otherwise linear and interpretable structure, yielding gains in both interpretability and forecasting performance \citep{gouletcoulombe2020macroeconomic}. Closely related approaches include partially linear and local linear forest methods, which correct forests for smooth or linear structure through local linear adjustments \citep{friedberg2021local,athey2019generalized}. These methods are primarily motivated by statistical efficiency and inference, and rely on local approximations induced by forest weights. Related ideas have since appeared in the econometric literature \citep{masini2025balancing,hauzenberger2023bayesian}. While similar in spirit, these approaches differ from the present contribution in both objective and implementation: LGB+ introduces global, sequential linear updates directly into the boosting loop, with a sole focus on forecasting performance. Its treatment of the linear component is deliberately agnostic—come what may.

A closely related contribution is \citet{chinn2023nowcasting}, who clearly identify the challenge of determining what should enter linearly in a machine learning forecast---and propose one compelling way forward: a data-driven pre-selection step that feeds into a MacroRF-type framework for world trade nowcasting. This is a meaningful step, though the two-stage design can be somewhat involved and the linear and tree components are not jointly optimized, leaving some potential gains on the table. LGB+ takes a different route, integrating variable selection and model fitting into a single unified loop.

\vspace{0.3em}
\noindent\textit{Structured boosting in macroeconomics.} Recent work also emphasizes structured and interpretable boosting schemes in macroeconomic applications. For example, blockwise boosting approaches alternate updates across economically defined predictor blocks to enhance transparency in inflation forecasting \citep{buckmann2025blockwise}. While these methods share the idea of staged updates, they remain entirely tree-based and focus on economic decomposition rather than on separating linear from nonlinear predictive content. 

\vspace{0.3em}
\noindent\textit{Boosting and machine learning.} From a machine learning perspective, boosting is naturally interpreted as functional gradient descent, where predictive functions are constructed through sequential updates using simple base learners \citep{friedman2001greedy,friedman2002stochastic}. Within this framework, linear base learners and coordinate-wise updates are well-understood building blocks---often selected based on correlation with current residuals. Their close connection to regularization through shrinkage and early stopping has been studied extensively \citep{buhlmann2003boosting,buhlmann2007boosting}. In that broader sense, the choice of base learner does matter. It is simply that the combination of boosting with trees has proven so potent on many tabular datasets—often outperforming more general approaches such as MARS \citep{friedman1991mars}—that the landscape has become largely dominated by tree-based methods. Whether this dominance carries over to the peculiar case of macroeconomic forecasting, however, is less clear.

The statistics literature has explored model-based boosting approaches that allow practitioners to mix linear, smooth, and tree-based components within a single formula \citep{hothorn2010mboost}. While conceptually related, LGB+ differs in two respects. First, it restricts updates to just two types---a shallow tree and a univariate linear update---rather than selecting from an open-ended menu of smooth and parametric components. This is a deliberate design choice: trees and linear terms represent two polar extremes of functional form, and mixing just these two keeps the model parsimonious and preserves a clean decomposition of what is linear and what is not. Adding intermediate basis functions would require either a more complex alternating schedule or evaluating additional candidates in the competition variant, increasing computational cost and overfitting risk with unclear forecasting payoff. Second, in the competition variant, out-of-bag evaluation adjudicates between candidates, whereas mboost selects the component that best fits residuals on the training sample. In both designs, linear structure is explicitly and regularly revisited rather than potentially being crowded out by trees.

As a consequence of tree dominance, modern gradient boosting implementations such as XGBoost and LightGBM typically require the user to choose \emph{ex ante} between tree-based and linear boosters, rather than allowing the two to be interleaved within a single training loop \citep{chen2016xgboost,ke2017lightgbm}. Applied workflows therefore often rely on two-stage procedures, such as residualizing outcomes using linear models before applying machine learning. Such approaches can be fragile, require a consequential ordering choice, and may leave little structure for trees to learn. The approach proposed here instead allows linear and nonlinear components to compete gradually during optimization, without being wed to any particular structure.

\paragraph{Outline.} Section~\ref{sec:algorithm} introduces the LGB+ algorithm and provides a simulation cross-check demonstrating its robustness across a range of data-generating processes. Section~\ref{sec:empirical} presents the main empirical forecasting results on U.S.\ macroeconomic aggregates, followed by a linear--nonlinear decomposition of variable importance (Section~\ref{sec:interpretability}) and a dual interpretation of selected forecasts (Section~\ref{sec:dual}). Detailed simulation results are relegated to Appendix~\ref{app:simulation}.

\section{The LGB+ Algorithm}\label{sec:algorithm}

This section introduces the LGB+ estimation procedure. Two design philosophies are available: \textit{competition}, where tree and linear candidates are evaluated at each step and only the winner is applied (LGB+), and \textit{alternating}, where blocks of tree updates are deterministically interleaved with linear corrections (LGB$^{\texttt{A}}$+). Both combine tree boosting with greedy linear updates; they differ in how the two types of updates are scheduled. I then provide a simulation cross-check to verify that both variants behave as expected across a range of data-generating processes, signal-to-noise ratios, and sample sizes.

\begin{algorithm}[t]
\SetAlgoLined
\DontPrintSemicolon
      \vspace{0.35em}
\SetKwInput{KwData}{Input}
\SetKwInput{KwResult}{Output}
\KwData{Data $\{(X_i, y_i)\}_{i=1}^N$, steps $M$, rate $\eta$, subsample $q$, candidate fraction $\rho$, selection method}
\KwResult{Prediction function $F_M(\cdot)$}
% \\[1.8em]  % was silently dropped in original SSRN compile (TeX "no line here to end")
      \vspace{0.65em}
\algostep{\color{dpd}Initialize:} $F_0(\cdot) \gets \bar{y}$\;
% \\[0.5em]  % was silently dropped in original SSRN compile (TeX "no line here to end")
\For{$t = 1, \ldots, M$}{
  % \\[2.89em]  % was silently dropped in original SSRN compile (TeX "no line here to end")
      \vspace{0.15em}
  \tcp{\algostep{\color{dpd}Sample}}
    \vspace{0.065em}
  \begin{itemize} \itemsep 0.03em
      \item[] Draw subsample $S_t$ of size $\lfloor qN\rfloor$
      \item[] Compute residuals: $r_i \gets y_i - F_{t-1}(X_i)$ for $i \in S_t$
  \end{itemize}
  \vspace{-0.25em}
  \tcp{\algostep{\color{dpd}Candidate Tree Update}}
  \vspace{-0.85em}
  \begin{itemize} \itemsep 0.03em
      \item[] Fit shallow tree $h_t(\cdot)$ to $\{(X_i, r_i)\}_{i\in S_t}$
      \item[] Candidate: $F^{\text{tree}}(\cdot) \gets F_{t-1}(\cdot) + \eta\, h_t(\cdot)$
  \end{itemize}
  \vspace{-0.25em}
  \tcp{\algostep{\color{dpd}Candidate Linear Update}}
  \vspace{-0.85em}
  \begin{itemize} \itemsep 0.03em
      \item[] Sample feature subset $\mathcal{J}_t$ with $|\mathcal{J}_t|=\lfloor \rho P\rfloor$
      \item[] Select feature: $k^* \gets \arg\max_{k\in\mathcal{J}_t} |\mathrm{Corr}(X_{\cdot,k},\boldsymbol{r})|$
      \item[] Fit $r = \hat{\alpha} + \hat{\beta}\, X_{k^*}$ on $S_t$ (univariate OLS)
      \item[] Candidate: $F^{\text{lin}}(\cdot) \gets F_{t-1}(\cdot) + \eta\, (\hat{\alpha} + \hat{\beta}\, X_{\cdot,k^*})$
  \end{itemize}
  \vspace{-0.25em}
  \tcp{\algostep{\color{dpd}Selection (Competition)}}
  \vspace{-0.85em}
  \begin{itemize} \itemsep 0.03em
      \item[] Evaluate losses $L_{\text{tree}}$ and $L_{\text{lin}}$ on judge set (OOB/validation/training)
      \item[] If $L_{\text{tree}} \le L_{\text{lin}}$, set $F_t \gets F^{\text{tree}}$; else set $F_t \gets F^{\text{lin}}$
  \end{itemize}
      \vspace{-0.65em}
}
\Return{$F_M(\cdot)$}
      \vspace{0.35em}
\caption{{\fontfamily{phv}\selectfont\textbf{LGB+}}}
\label{alg:lgbplus}
\end{algorithm}

\subsection{Estimation Procedure}

Let $\boldsymbol{X} \in \mathbb{R}^{N \times P}$ denote the predictor matrix and $\boldsymbol{y} \in \mathbb{R}^{N}$ the response vector, where observations are indexed by $i = 1, \dots, N$. LGB+ constructs the prediction function $F(\cdot)$ by sequentially applying one of two candidate updates at each boosting step: a tree update $h_t(\cdot)$ and a greedy univariate linear update on a selected predictor.

 At step $t$, a subsample $S_t$ is drawn (row subsampling) and residuals $r_i = y_i - F_{t-1}(X_i)$ are computed on that subsample. A shallow regression tree is fit to $\{(X_i, r_i)\}_{i\in S_t}$, yielding a candidate tree update $F^{\text{tree}}(\cdot) = F_{t-1}(\cdot) + \eta\, h_t(\cdot)$. In parallel, a candidate linear update is constructed by first randomly sampling a subset of predictors $\mathcal{J}_t$ of size $\lfloor \rho P\rfloor$ (with $\rho \in (0,1]$), selecting the feature $k^* \in \mathcal{J}_t$ most correlated with the current residuals, and fitting the univariate OLS regression $r = \alpha + \beta X_{k^*}$ on $S_t$. This yields the candidate linear update $F^{\text{lin}}(\cdot) = F_{t-1}(\cdot) + \eta\, (\hat\alpha + \hat\beta\, X_{\cdot,k^*})$.

Crucially, LGB+ does not apply both candidates. Instead, it evaluates which candidate reduces loss the most and applies only the winner. In macroeconomic forecasting I recommend an out-of-bag (OOB) selection rule: evaluate each candidate on the observations not used in $S_t$ (i.e., the OOB set) and choose the update with lower OOB mean squared error. This step-by-step external evaluation mitigates overfitting and prevents the more complex tree candidate from winning on in-sample data simply by virtue of having more parameters to fit.

\paragraph{Ensembling.} Boosting remains inherently greedy: at each step the algorithm fits a single tree and a single linear candidate and selects the better of the two. This introduces variance across training runs because subsamples $S_t$ and candidate feature sets $\mathcal{J}_t$ are random. As in standard boosting, this randomness can be turned into a feature: averaging predictions across a small number of independent runs further stabilizes the learned tree/linear mix and yields a smoother, less brittle forecast rule.

A side effect of stacking multiple regularization devices---subsampling, shrinkage, ensembling---is that forecasts can become systematically shrunk toward the training mean. When this occurs, a simple post-hoc calibration---regressing training outcomes on in-sample predictions via $y = \beta \cdot \hat{y}$ (no intercept)---recovers the lost scale. Because the regression passes through the origin, this targets a pure scale correction, appropriate here since both $y$ and $\hat{y}$ are standardized. In the empirical application, this calibration is applied to the competition variant (LGB+) only: at each expanding-window training step, $\beta$ is estimated on the in-sample fitted values and applied to the out-of-sample predictions for that window. The calibration coefficient is thus updated recursively as new data arrive, and because it is estimated from training data alone, no look-ahead bias is introduced.

\paragraph{Some Interpretability.} In both variants, a practical benefit of the additive structure is that the final prediction decomposes linearly into its linear and tree components: $\hat{y} = f^{\text{lin}}(X) + f^{\text{tree}}(X)$. This enables a channel-specific variable importance decomposition for both variants (Section~\ref{sec:interpretability}). It also opens the door to a dual representation of forecasts as weighted averages of training outcomes, with the observation weights themselves decomposing into linear and tree contributions---a decomposition developed here for the alternating variant via the AXIL framework \citep{geertsema2023axil,gouletcoulombe2024dual} and feasible but not yet implemented for the competition variant (Section~\ref{sec:dual}).

\paragraph{Lineage.} The starting point is two-stage residualization: fit a linear model first, then apply a tree ensemble to the residuals in the spirit of generalized additive models  \citep{hastie2017generalized}. This simple composition---used, among other places, as the AR+RF benchmark in \citet{gouletcoulombe2020macroeconomic}---has an obvious limitation: the linear component is estimated upfront, at full strength, absorbing as much signal as it can before the trees ever see the data. A natural fix is to scale the linear fit down, applying a learning rate so the trees retain residual signal to work with. But once the linear step is shrunk, the question becomes: why stop at one? If a single small linear correction helps, a sequence of small corrections should help more; and if they are interspersed with tree updates under the same learning rate, one arrives at the integrated loop that defines LGB+.

\paragraph{Connection to $\ell_1$ regularization.} The greedy univariate linear updates have a well-known theoretical counterpart. Applying them repeatedly---selecting the most correlated predictor and nudging its coefficient by a small step $\eta$---traces the path of forward stagewise regression, which \citet{efron2004least} showed approximates the LASSO solution, with the effective regularization level governed by the step size and the number of updates: smaller steps or earlier stopping correspond to stronger $\ell_1$ penalization. The linear channel of LGB+ thus builds up a sparse, shrunk linear predictor without explicit penalty tuning: sparsity and shrinkage emerge automatically from the greedy selection mechanism and early stopping, not from solving a penalized program.

\subsection{A More Economical Alternative: LGB$^{\texttt{A}}$+}\label{sec:alternating}

The competition-based design evaluates two candidates at every step, adding non-trivial per-step overhead relative to standard boosting. When computational budget is a concern, or when the practitioner prefers a simpler scheme, an alternative is to \textit{alternate deterministically} between blocks of tree updates and single linear corrections. Specifically, after every $T$ consecutive tree updates, the algorithm computes residuals, identifies the predictor most correlated with those residuals, fits a univariate OLS regression, and applies the linear update with learning rate $\eta_{\text{lin}}$. This yields a fixed tree/linear schedule rather than a data-driven one.

The alternating variant, denoted LGB$^{\texttt{A}}$+ (Algorithm~\ref{alg:lgbplus_alt}), removes the need for OOB evaluation at each step and eliminates one hyperparameter (the candidate fraction $\rho$), at the cost of fixing the tree/linear ratio ex ante and introducing a separate linear learning rate $\eta_{\text{lin}}$ in its place. A secondary benefit is robustness: with macroeconomic training samples of $N \approx 250$ observations, OOB hold-out sets at each individual step are small enough that the step-by-step winner selection in LGB+ can be noisy, occasionally pivoting toward a candidate that happened to fare better on an unrepresentative subsample. The fixed schedule of LGB$^{\texttt{A}}$+ sidesteps this variance entirely, which may partly explain why it wins a larger share of simulation configurations overall despite the competition variant having access to a richer, data-driven allocation mechanism. In the simulation cross-check, both variants perform comparably across all DGPs, suggesting that the core insight of incorporating linear updates into the boosting loop matters more than the specific scheduling mechanism. In the empirical application, both variants are included in the model comparison.

\begin{algorithm}[t]
\SetAlgoLined
\DontPrintSemicolon
      \vspace{0.35em}
\SetKwInput{KwData}{Input}
\SetKwInput{KwResult}{Output}
\KwData{Data $\{(X_i, y_i)\}_{i=1}^N$, cycles $M$, trees/cycle $T$, rates $\eta_{\text{tree}}, \eta_{\text{lin}}$}
\KwResult{Prediction function $F_M(\cdot)$}
% \\[1.8em]  % was silently dropped in original SSRN compile (TeX "no line here to end")
      \vspace{0.65em}
\algostep{\color{dpd}Initialize:} $F_0(\cdot) \gets \bar{y}$\;
% \\[0.5em]  % was silently dropped in original SSRN compile (TeX "no line here to end")
\For{$m = 1, \ldots, M$}{
  % \\[2.89em]  % was silently dropped in original SSRN compile (TeX "no line here to end")
      \vspace{0.15em}
  \tcp{\algostep{\color{dpd}Tree Block}}
    \vspace{0.065em}
  \begin{itemize} \itemsep 0.03em
      \item[]   Compute residuals: $r_i \gets y_i - F_{m-1}(X_i)$
      \item[]   Fit $T$ trees $\{h_t\}_{t=1}^T$ sequentially to $\{r_i\}$
      \item[]   Update function   $\tilde{F}_m(\cdot) \gets F_{m-1}(\cdot) + \eta_{\text{tree}} \sum_{t=1}^T h_t(\cdot)$
  \vspace{-0.25em}

  \end{itemize}
  \tcp{\algostep{\color{dpd} Linear Update}}
  \vspace{-0.85em}
  \begin{itemize} \itemsep 0.03em
      \item[] Compute residuals: $r'_i \gets y_i - \tilde{F}_m(X_i)$
      \item[] Find best linear predictor: $k^* \gets \arg\max_k |\mathrm{Corr}(X_{\cdot,k}, \boldsymbol{r}')|$
      \item[] Fit $\boldsymbol{r}' = \hat{\alpha} + \hat{\beta}\, X_{k^*}$ via OLS
      \item[] Update function
  $F_m(\cdot) \gets \tilde{F}_m(\cdot) + \eta_{\text{lin}}(\hat{\alpha} + \hat{\beta}\, X_{\cdot,k^*})$
    \end{itemize}
      \vspace{-0.65em}
}
\Return{$F_M(\cdot)$}
      \vspace{0.35em}
\caption{{\fontfamily{phv}\selectfont\textbf{LGB$^{\texttt{A}}$+ (Alternating)}}}\label{alg:lgbplus_alt}
\end{algorithm}

For practitioners, the main free parameters are $T$ and $\eta_{\text{lin}}$. $T$ governs how frequently linear corrections are injected: smaller values produce a denser linear schedule, while larger values let trees accumulate longer before each correction. Values of $T$ between 5 and 20 work well across the macroeconomic applications considered here, with $\eta_{\text{lin}}$ set equal to or slightly below the tree learning rate as a starting point. In practice, both variants are viable defaults. LGB+ lets the data determine the tree/linear mix step by step, while LGB$^{\texttt{A}}$+ is faster and avoids the noise inherent in per-step OOB selection at small sample sizes.

Which variant should a practitioner choose? At equal computational cost, the simulation and empirical results suggest it matters less than one might expect: the core gain comes from incorporating linear updates at all, not from the specific scheduling mechanism. That said, each variant has a distinct advantage. LGB$^{\texttt{A}}$+ forces implicit model averaging between linear and nonlinear components by construction. LGB+ is more adaptive, but whether that adaptiveness pays off out of sample is not guaranteed. Intermediate designs are conceivable, and learning or adapting the tree--linear schedule over time remains a promising direction for future work.

\subsection{Simulation Cross-Check}\label{sec:crosscheck}

Before turning to the empirical application, I verify that both variants (LGB+ and LGB$^{\texttt{A}}$+) perform well across a range of data-generating processes. I conduct a Monte Carlo experiment spanning eight DGPs, from purely linear to highly nonlinear, with mixed cases in between. Results are averaged over 10 replications using a fixed test set of 1000 observations. Full details, including mathematical definitions of all DGPs, are provided in Appendix~\ref{app:simulation}; here I summarize the key findings.

\begin{figure}[htbp]
  \caption{Simulation Cross-Check: Model Performance Across Data Generating Processes}
  \vspace*{-0.4em}
  \label{fig:simulation}
  \centering
  \includegraphics[width=0.99\textwidth]{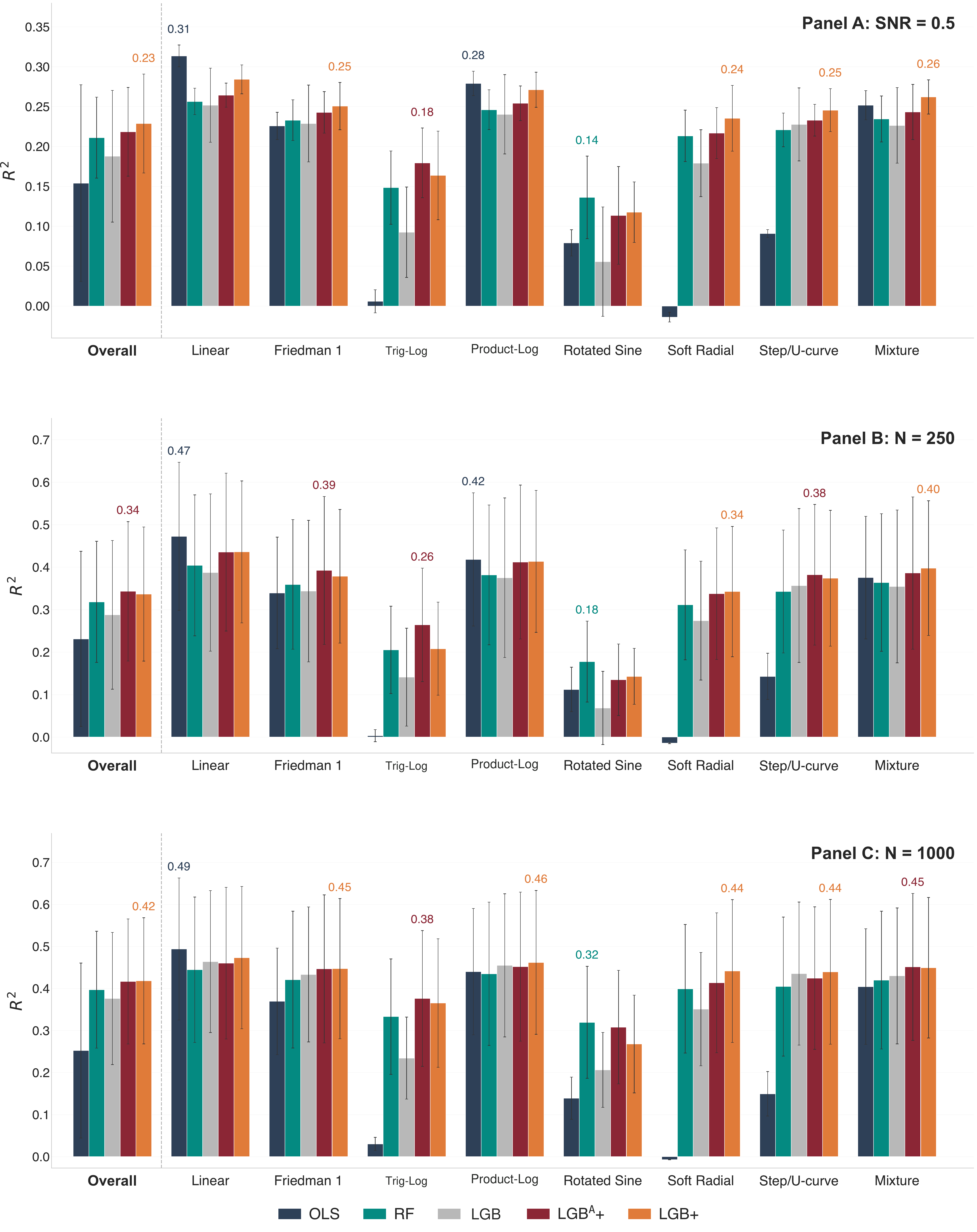}
  \vspace{-0.2cm}
  \begin{minipage}{\textwidth}
    \fontfamily{phv}\selectfont\footnotesize
    \textit{Notes}: Average out-of-sample $R^2$ across experimental conditions. Panel A: SNR$=0.5$, averaged over sample sizes. Panel B: $N=250$, averaged over SNRs. Panel C: $N=1000$, averaged over SNRs. Error bars represent $\pm 1$ standard deviation. See Table~\ref{tab:simulation_results} in Appendix~\ref{app:simulation} for detailed results.
  \end{minipage}
\end{figure}

I consider training sample sizes $N \in \{250, 500, 1000\}$ and signal-to-noise ratios $\text{SNR} \in \{0.5, 1, 2\}$, representative of typical macroeconomic forecasting environments. Figure~\ref{fig:simulation} displays results for three illustrative cases: Panel A shows results at low SNR (averaged over sample sizes), where LGB+ is expected to have an edge; Panels B and C contrast small ($N=250$) versus large ($N=1000$) samples, illustrating how this edge erodes as nonparametric methods catch up with more data. Across the full range of DGPs, from purely linear environments where OLS visibly dominates to highly nonlinear settings where OLS struggles greatly, both LGB+ variants fare well. Both variants perform comparably, suggesting that the core insight---incorporating linear updates into the boosting loop---matters more than the specific scheduling mechanism (see Table~\ref{tab:simulation_results} for detailed configuration-level results).

Several patterns emerge from the results. First, when the DGP is purely \textit{linear}, LGB+ trails OLS the least among all nonparametric methods. This is because the linear channel can capture the true DGP directly; tree updates, facing residuals with no exploitable nonlinear structure, contribute little under OOB selection or early stopping, so the effective model collapses toward a regularized linear predictor. Second, when the DGP is \textit{mixed}, as in the Mixture DGP which combines a linear component with a Friedman 1 nonlinear part, LGB+ is the only method that beats OLS in the small-sample, low-SNR configurations most representative of macroeconomic settings (though RF and LightGBM close the gap at larger $N$ or higher SNR). Third, when the DGP is highly \textit{nonlinear}, LGB+ performs essentially as well as pure tree-based methods; the linear component does not hurt even when useless. This is fairly intuitive: if the linear updates are orthogonal to the true signal and applied in modest amounts, the estimated coefficients will hover around zero and contribute little to predictions. Moreover, the greedy feature selection implies that, under pure noise, different predictors will be selected across iterations, effectively diversifying away the risk of overfitting, an argument formalized in \citet{gouletcoulombe2024bag}. Fourth, when the DGP is \textit{extremely} nonlinear, as in the Rotated Sine case featuring high-frequency oscillations, a purely nonlinear method like Random Forest performs best, though LGB+ remains competitive. 

As expected, increasing the sample size or the signal-to-noise ratio erodes LGB+'s advantage as fully nonparametric methods eventually catch up. But for settings that resemble macroeconomic data (small samples, low signal-to-noise), as illustrated in Panels A and B of Figure~\ref{fig:simulation}, LGB+ clears the simulation cross-check and is ready for a road test.

\section{Macroeconomic Forecasting}\label{sec:empirical}

I conduct a comprehensive macroeconomic forecasting exercise comparing classical econometric models, modern ML algorithms, a transformer-based foundation model, and the Survey of Professional Forecasters. The forecasting setup and data build closely on \citet{gouletcoulombe2026risk}---which develops the risk-adjusted evaluation framework used throughout---but the scope here is broader: that paper evaluates forecast reliability across models; this one puts LGB+ center stage.

\paragraph{Data and Forecasting Setup.} I forecast six U.S. macroeconomic variables at the quarterly frequency: headline CPI inflation, GDP growth, the unemployment rate, housing starts (log growth rate), industrial production, and the term spread (10-year minus 3-month Treasury yield). For five of these variables (inflation, GDP, unemployment, housing starts, and industrial production), the Survey of Professional Forecasters (SPF) provides benchmark forecasts, enabling direct comparison between ML models and professional judgment. Predictors are drawn from the FRED-QD database \citep{mccracken2020fred}, transformed to induce stationarity, and augmented with four lags plus moving averages of order 2, 4, and 8 following the MARX transformation \citep{gouletcoulombe2024transformation}. Four principal components are extracted from the transformed predictor panel and included with four lags each, following evidence in \citet{gouletcoulombe2024transformation} that PCA-based factors can improve forecasting performance when the transformation of the underlying data is chosen carefully. All series are standardized to zero mean and unit variance over the training sample. To account for the ragged edge of quarterly data, 37 slowly released FRED-QD variables (output, productivity, compensation, and Federal Reserve balance sheet items) are shifted forward by one quarter so that only information realistically available at forecast time enters the model. I estimate direct forecasts at horizons $h = 1, 2, 4$ quarters ahead, with separate models for each horizon.

The out-of-sample evaluation proceeds on an expanding window across two distinct subsamples, with models re-estimated every 8 quarters (2 years). The \textit{pre-COVID subsample} (2007Q2--2019Q4) covers the Great Financial Crisis, the subsequent recovery, and the pre-pandemic expansion, with training data beginning in 1961Q2. The \textit{post-COVID subsample} starts in 2022Q1 for GDP, unemployment, and industrial production, and in 2021Q1 for inflation, housing starts, and the term spread (all ending 2025Q1). The later start for the first group excludes the immediate post-pandemic period where extreme base effects distort forecast evaluation. Comparing performance across these regimes---one characterized by low inflation and gradual recovery, the other by unprecedented volatility and rapid policy shifts---allows assessing whether ML gains are stable or driven by specific episodes.

\paragraph{Models.} I compare both LGB+ variants (LGB+ and LGB$^{\texttt{A}}$+) against a suite of benchmarks including: AR(4) (the naive benchmark any forecaster must beat); Factor-Augmented AR \citep{stock2002forecasting}; Ridge Regression (RR) and Kernel Ridge Regression (KRR); Random Forest \citep{breiman2001random}; standard LightGBM \citep{ke2017lightgbm}; feed-forward neural networks (NN); and TabPFN, a transformer-based foundation model \citep{hollmann2022tabpfn}. For inflation, I additionally include the Hemisphere Neural Network (HNN) from \citet{goulet2025neural}. Details on model specifications and hyperparameter tuning are provided in Appendix~\ref{app:models}.

\paragraph{Results.} Table~\ref{tab:summary_rmse} summarizes forecast performance using RMSE relative to the AR benchmark for each variable and horizon, separately for the pre-COVID and post-COVID subsamples. Tables~\ref{tab:gdp_h1}--\ref{tab:dSpread_h4} in the Appendix provide a complementary perspective: for each variable and horizon, they report detailed risk-adjusted metrics computed \textit{over time}, characterizing the risk-return tradeoff of using each model's forecasts and balancing accuracy with the risk of significant failure in any given period.

% RMSE Summary Table
\input{03_tables/summary_table.tex}

As expected, LGB+ performs well pre-COVID, particularly at short horizons where it can leverage lagged relationships. The strongest results appear for unemployment at $h=1$ (0.67 relative to AR) and industrial production at $h=1$ (0.88 for LGB$^{\texttt{A}}$+). Housing starts also improve at $h=1$ (0.97 for LGB+), while GDP and the term spread see more modest gains. The hybrid approach consistently outperforms plain LGB, demonstrating that the linear component adds value beyond pure tree-based boosting.

The post-COVID period presents a more challenging environment. Most models that performed well before COVID struggle afterward, and vice versa. The exception is the SPF, which---excluding inflation---maintains solid performance across both periods. This is consistent with the broader literature: during regime changes driven by policy shifts, survey forecasters' qualitative judgment and real-time information provide an edge that purely statistical methods lack. Interestingly, Kernel Ridge Regression (KRR), which underperformed pre-COVID, emerges as the strongest ML method post-COVID, particularly for inflation. Among tree-based methods, LGB+ maintains its edge for housing starts, while plain LGB and Random Forest suffer catastrophic failures on unemployment and industrial production---variables where the AR benchmark itself collapses (a denominator effect). Post-COVID relative RMSE figures for unemployment and industrial production should be read as comparative rankings among models rather than evidence of high absolute forecast quality: when the AR benchmark itself degrades, any model avoiding the worst mistakes benefits mechanically from a small denominator.

Overall, for targets where short-horizon autoregressive dynamics are known to matter---unemployment, industrial production, housing---LGB+ delivers as anticipated. At longer horizons where such linear structure becomes less relevant, the advantage naturally fades. The question is whether this conjecture holds: does the linear component of LGB+ actually leverage the ``right'' predictors? The next section investigates.

\subsection{Linear--Nonlinear Decomposition and Variable Importance}
\label{sec:interpretability}

The additive structure of LGB+ offers avenues for interpretability that go beyond what standard tree-based methods provide.

\paragraph{Linear-Nonlinear Decomposition of Predictions.} Because the final prediction is the sum of a linear part and a tree part, $\hat{y} = f^{\text{lin}}(X) + f^{\text{tree}}(X)$, one can regroup all linear updates and all tree updates to obtain an exact decomposition of each forecast into its linear and nonlinear contributions. This is not an approximation: it follows directly from the additive structure of the boosting procedure. For any given test observation, one can inspect how much of the predicted value comes from linear relationships (captured by the greedy univariate OLS updates) and how much from nonlinear patterns (captured by the trees). The top panels of the dashboard figures below display this decomposition as stacked shaded regions over time.

\paragraph{Variable Importance.} Beyond decomposing the prediction itself, one can assess which variables matter and through which channel. I measure variable importance using permutation-based methods following \citet{goulet2025neural}: for each variable group $g$---aggregating all lags and MARX transformations of the same underlying series---I randomly permute its values, re-compute predictions, and measure the relative increase in MSE:
\begin{equation}
{\text{VI}}_g = \frac{\text{MSE}_{\text{permuted}} - \text{MSE}_{\text{baseline}}}{\text{MSE}_{\text{baseline}}} \times 100.
\end{equation}
Because predictions decompose into two additive components, importance can be computed separately for each channel: when computing $\text{VI}_g^{\text{linear}}$, I permute group $g$ only in the linear predictions while keeping its contribution to the tree component fixed, and vice versa for $\text{VI}_g^{\text{trees}}$. When $g$ is permuted in both channels simultaneously, the MSE algebra produces a cross-term. Let $\pi_g$ denote a random permutation of group $g$'s values across observations, and define $\Delta_{g,i}^{\text{lin}} = f^{\text{lin}}(X_i) - f^{\text{lin}}(X_{\pi_g,i})$ and $\Delta_{g,i}^{\text{tree}} = f^{\text{tree}}(X_i) - f^{\text{tree}}(X_{\pi_g,i})$ as the prediction changes induced by permuting group $g$ in each channel. The full importance decomposes as:
\begin{equation}
{\text{VI}}_g = {\text{VI}}_g^{\text{linear}} +  {\text{VI}}_g^{\text{trees}}  + {C_g},
\end{equation}
where $C_g = \frac{200}{n\,\text{MSE}_{\text{base}}} \sum_{i} \Delta_{g,i}^{\text{lin}}\,\Delta_{g,i}^{\text{tree}}$ is the cross-term, on the same scale as $\text{VI}_g$. $C_g$ is a synergy term: it is positive when the two channels' prediction changes point in the same direction across observations, and negative when they draw opposing implications from the same variable. This decomposition is displayed in the dashboard figures using three colors: teal (linear), red (trees), and gray (cross-term).

A caveat is in order. The linear/nonlinear split is indicative rather than exact---trees are not guaranteed to capture only nonlinear patterns. However, if a relationship is linear, the linear updates should capture it, especially in the competition variant where linear and tree candidates compete head-to-head at each step, leaving trees to focus on genuinely nonlinear structure.

% ============================================================================
% DASHBOARD FIGURES - Reordered: UR h1, UR h2, INDPRO, HOUST h1, dSpread h1
% HOUST h2 and dSpread h2 moved to appendix
% ============================================================================

\begin{figure}[t!]
  \caption{Unemployment Rate ($h=1$), {\fontfamily{phv}\selectfont\textbf{LGB+}}}
  \vspace*{-0.3em}
  \label{fig:dashboard_ur}
  \centering
  \includegraphics[width=\textwidth]{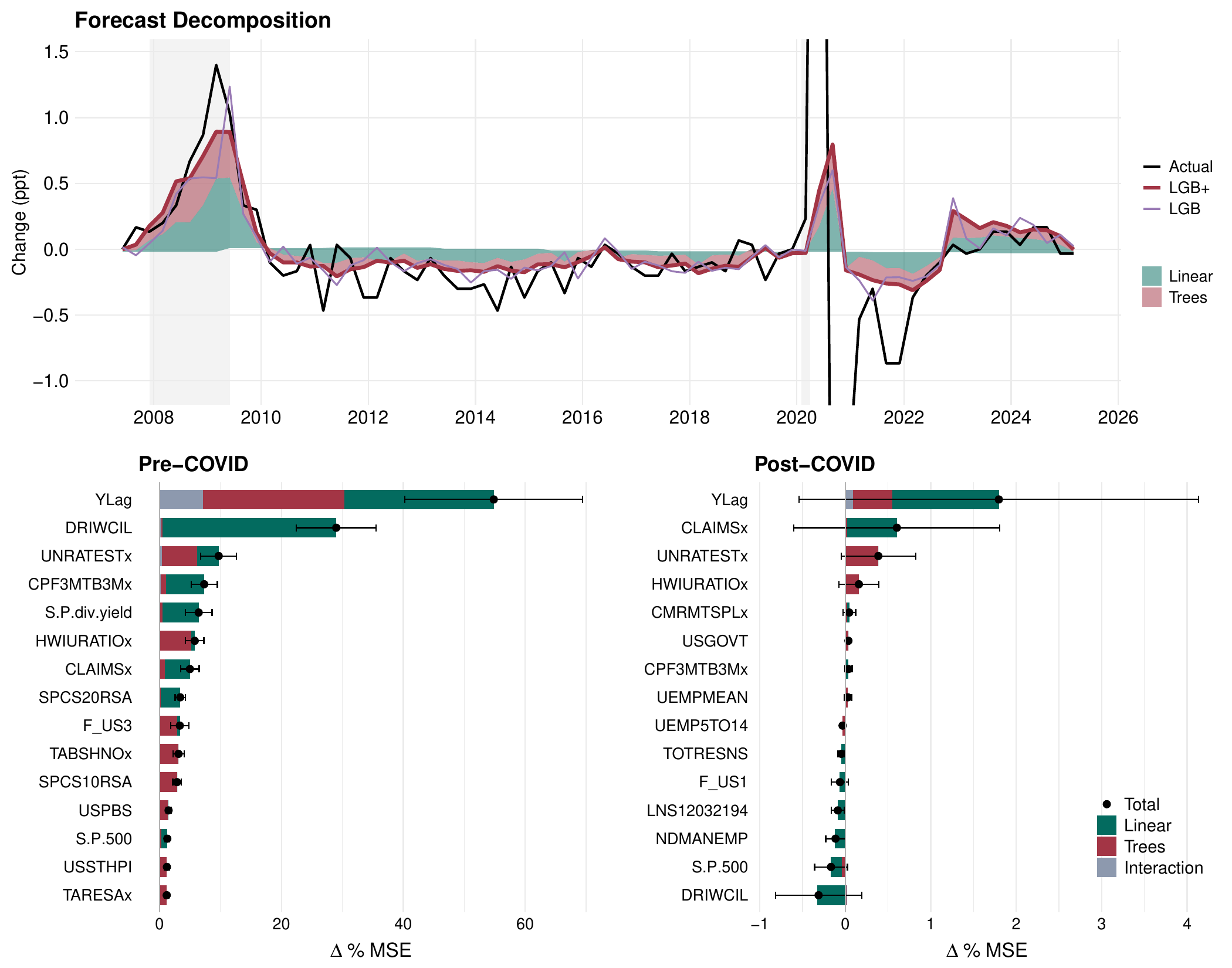}
  \vspace{0.3em}
  \parbox{\textwidth}{\fontfamily{phv}\selectfont\scriptsize
    \textit{Notes:} Top panel shows actual values (black), forecast (red), and LGB forecast (purple dashed). Shaded regions decompose the forecast into linear (teal), tree-based (red), and interaction (gray) components. Gray bands indicate NBER recession dates. Bottom panels show permutation-based variable importance for pre-COVID (left) and post-COVID (right) periods.
  }
\end{figure}

\begin{figure}[t!]
  \caption{Unemployment Rate ($h=2$), {\fontfamily{phv}\selectfont\textbf{LGB+}}}
  \vspace*{-0.3em}
  \label{fig:dashboard_ur_h2}
  \centering
  \includegraphics[width=\textwidth]{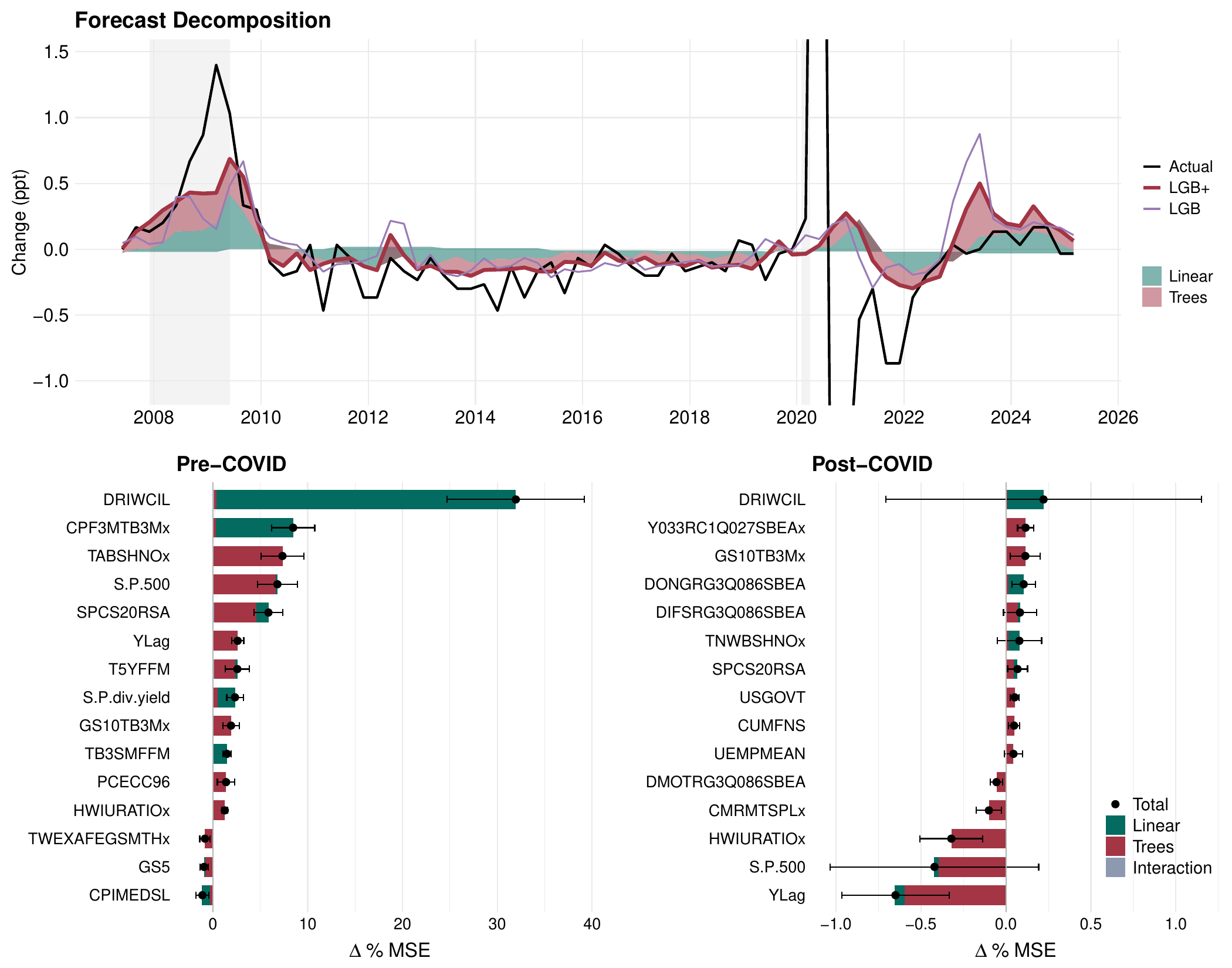}
  \vspace{0.3em}
  \parbox{\textwidth}{\fontfamily{phv}\selectfont\scriptsize
    \textit{Notes:} Top panel shows actual values (black), forecast (red), and LGB forecast (purple dashed). Shaded regions decompose the forecast into linear (teal), tree-based (red), and interaction (gray) components. Gray bands indicate NBER recession dates. Bottom panels show permutation-based variable importance for pre-COVID (left) and post-COVID (right) periods.
  }
\end{figure}

\begin{figure}[t!]
  \caption{Industrial Production ($h=1$), {\fontfamily{phv}\selectfont\textbf{LGB$^{\texttt{A}}$+}}}
  \vspace*{-0.3em}
  \label{fig:dashboard_indpro}
  \centering
  \includegraphics[width=\textwidth]{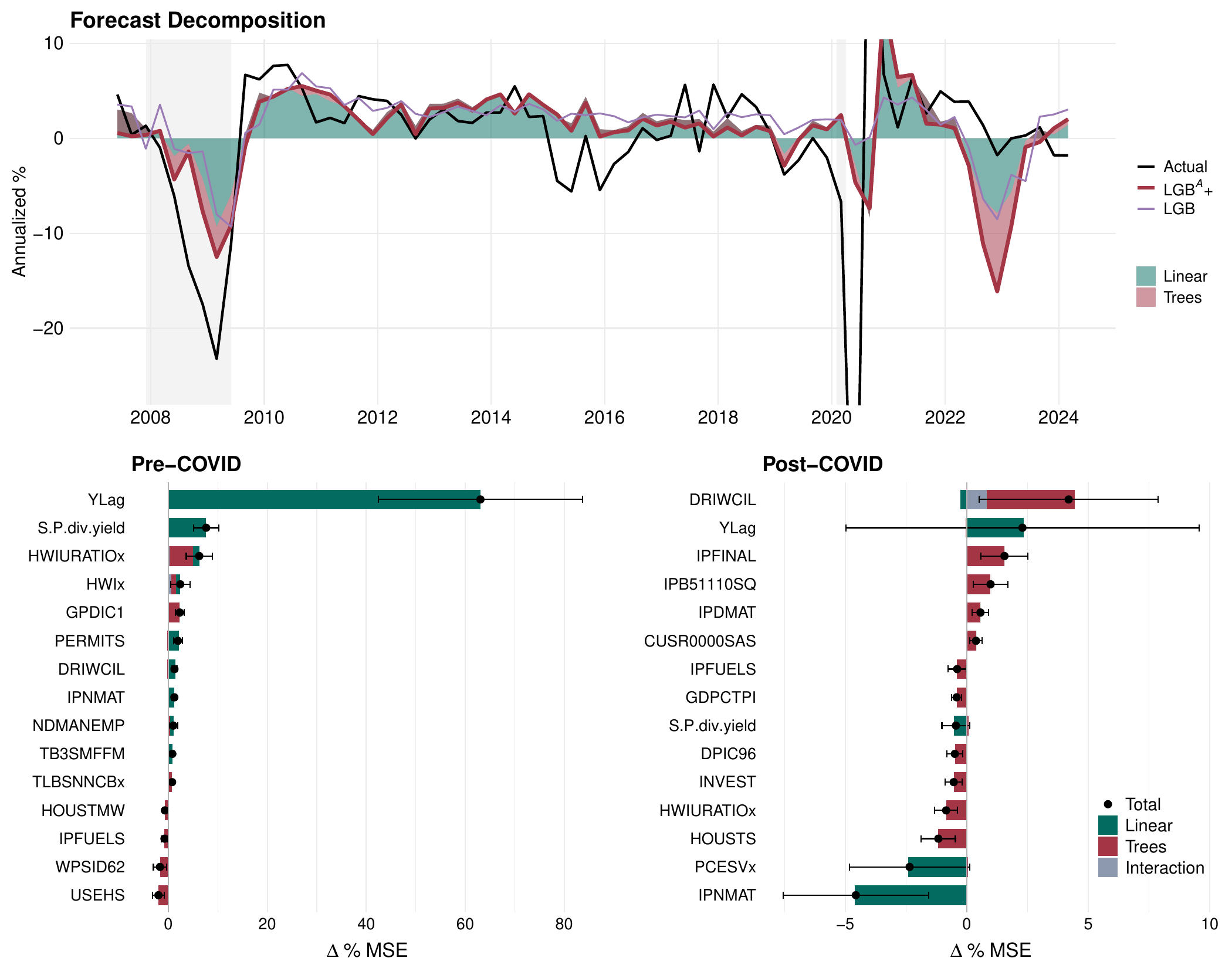}
  \vspace{0.3em}
  \parbox{\textwidth}{\fontfamily{phv}\selectfont\scriptsize
    \textit{Notes:} Top panel shows actual values (black), forecast (red), and LGB forecast (purple dashed). Shaded regions decompose the forecast into linear (teal), tree-based (red), and interaction (gray) components. Gray bands indicate NBER recession dates. Bottom panels show permutation-based variable importance for pre-COVID (left) and post-COVID (right) periods.
  }
\end{figure}

\begin{figure}[t!]
  \caption{Housing Starts ($h=1$), {\fontfamily{phv}\selectfont\textbf{LGB$^{\texttt{A}}$+}}}
  \vspace*{-0.3em}
  \label{fig:dashboard_houst_h1}
  \centering
  \includegraphics[width=\textwidth]{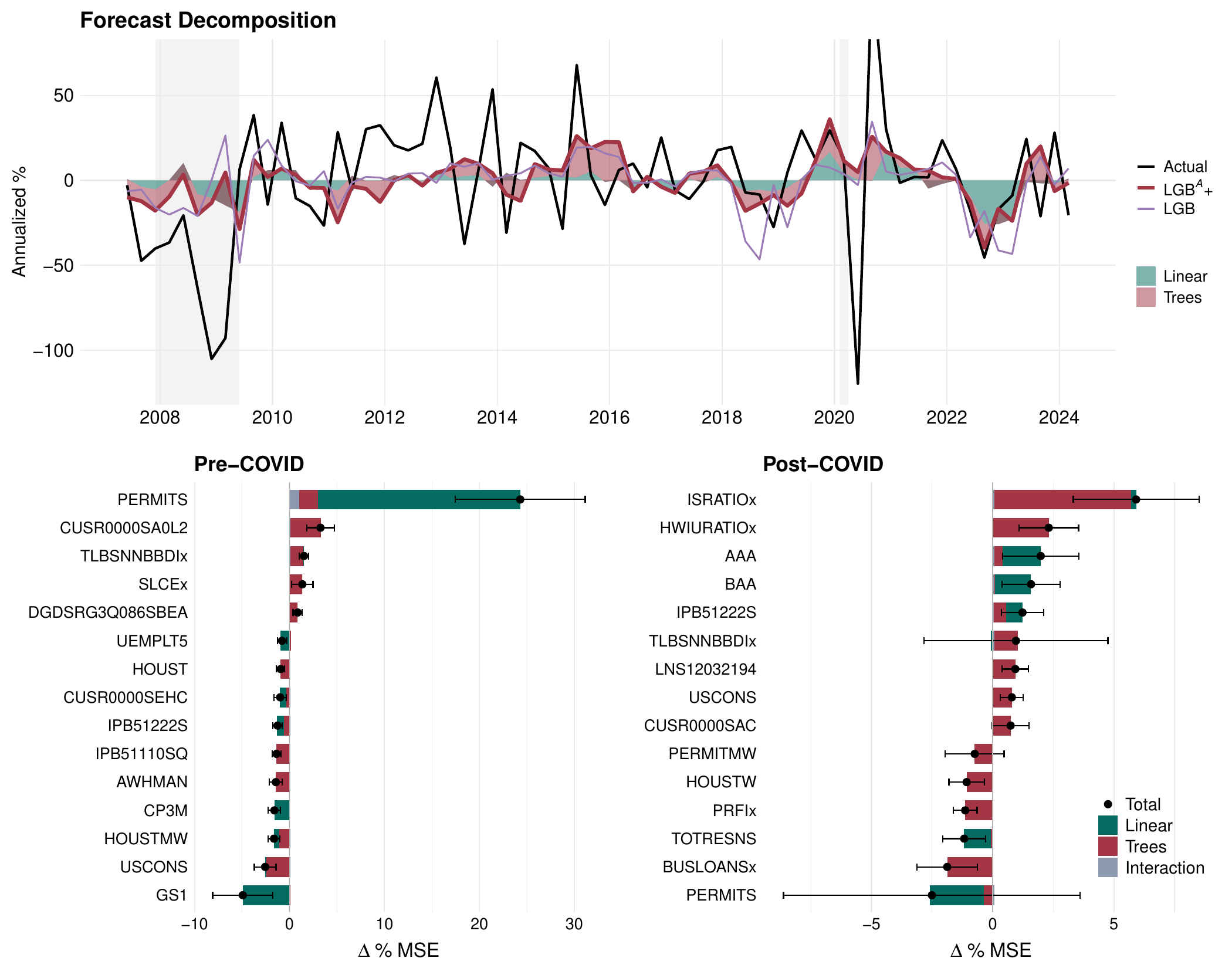}
  \vspace{0.3em}
  \parbox{\textwidth}{\fontfamily{phv}\selectfont\scriptsize
    \textit{Notes:} Top panel shows actual values (black), forecast (red), and LGB forecast (purple dashed). Shaded regions decompose the forecast into linear (teal), tree-based (red), and interaction (gray) components. Gray bands indicate NBER recession dates. Bottom panels show permutation-based variable importance for pre-COVID (left) and post-COVID (right) periods.
  }
\end{figure}

\begin{figure}[t!]
  \caption{$\Delta$Spread ($h=1$), {\fontfamily{phv}\selectfont\textbf{LGB$^{\texttt{A}}$+}}}
  \vspace*{-0.3em}
  \label{fig:dashboard_dSpread_h1}
  \centering
  \includegraphics[width=\textwidth]{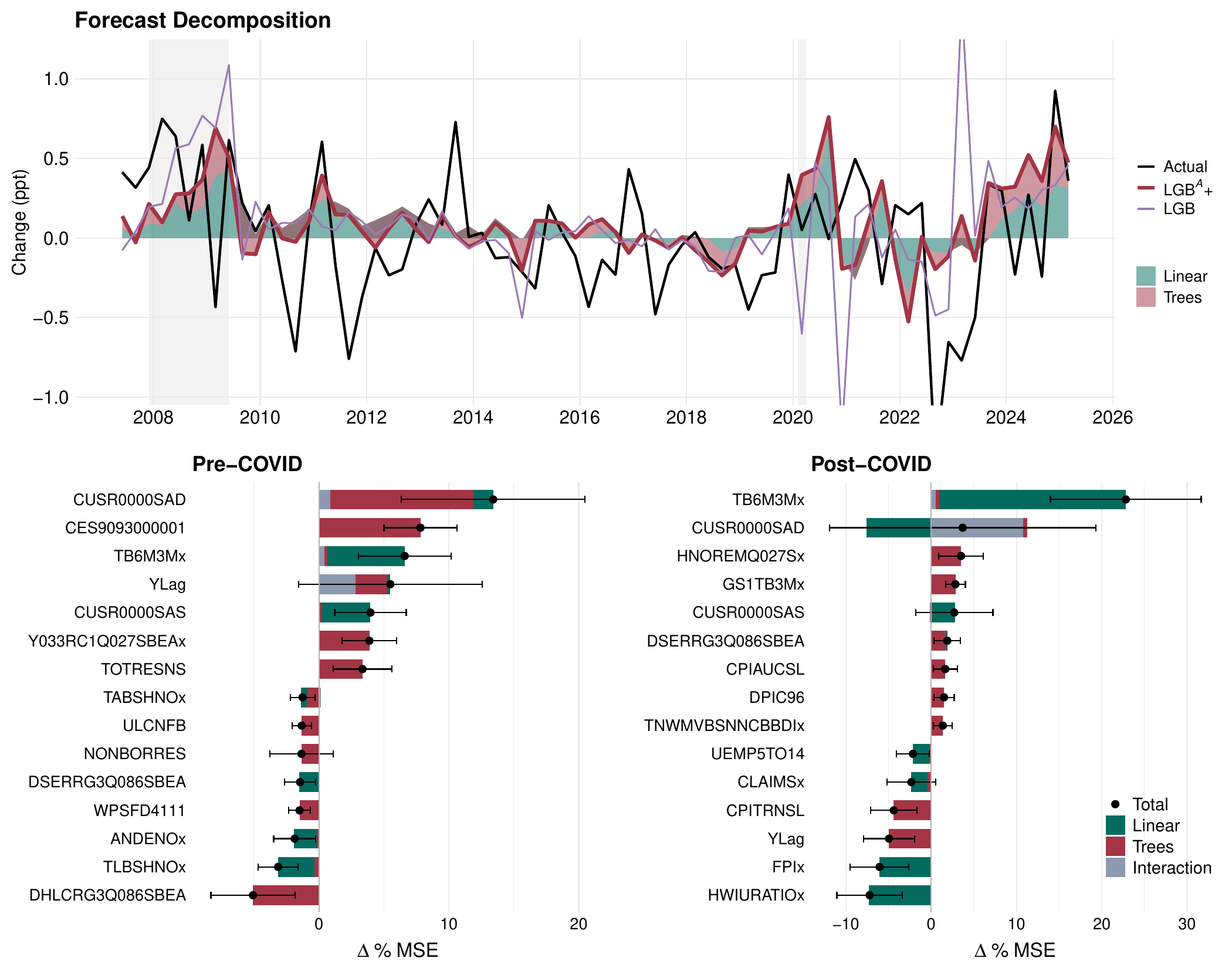}
  \vspace{0.3em}
  \parbox{\textwidth}{\fontfamily{phv}\selectfont\scriptsize
    \textit{Notes:} Top panel shows actual values (black), forecast (red), and LGB forecast (purple dashed). Shaded regions decompose the forecast into linear (teal), tree-based (red), and interaction (gray) components. Gray bands indicate NBER recession dates. Bottom panels show permutation-based variable importance for pre-COVID (left) and post-COVID (right) periods.
  }
\end{figure}

Figures~\ref{fig:dashboard_ur}--\ref{fig:dashboard_dSpread_h1} present forecast decompositions for selected variables, illustrating how LGB+ allocates predictive content between linear and nonlinear channels. For each target, I show the variant (LGB+ or LGB$^{\texttt{A}}$+) that performs best for that particular forecasting problem.

\paragraph{Unemployment Rate ($h=1$).}

Figure~\ref{fig:dashboard_ur} illustrates the decomposition for unemployment rate forecasting at $h=1$, where LGB+ delivers its strongest gains (RMSE ratio 0.67). Pre-COVID, the linear component dominates, with lags of the dependent variable (\mn{YLag}) by far the most important predictor. Notably, \mn{YLag} enters both linearly and through trees, consistent with asymmetric autoregressive dynamics: unemployment rises sharply during recessions but declines gradually during expansions, a nonlinearity that trees can capture. \mn{DRIWCIL} (bank lending willingness, from the Senior Loan Officer Opinion Survey) enters purely through the linear channel. \mn{CLAIMSx} (initial unemployment claims) also contributes linearly---by construction a near-accounting relationship, since claims filings mechanically precede measured unemployment. The S\&P~500 dividend yield and short-term unemployment (\mn{UNRATESTx}, duration $<$27 weeks) contribute through both channels. Post-COVID, \mn{YLag} and \mn{CLAIMSx} dominate, with the latter entering linearly in both periods. The key comparison is with plain LGB (the purple dashed line): it forecast a notably larger spike in unemployment that never materialized. LGB+ also overshoots, but less so---it tempers the overreaction to recession signals that affected the pure tree-based approach.

\paragraph{Unemployment Rate ($h=2$).}

Figure~\ref{fig:dashboard_ur_h2} extends the unemployment analysis to $h=2$, where LGB+ achieves an RMSE ratio of 0.71 pre-COVID. The variable importance profile shifts from predominantly autoregressive (at $h=1$) toward more forward-looking financial indicators. Pre-COVID, \mn{DRIWCIL} dominates through the linear channel, followed by the commercial paper--Treasury spread (\mn{CPF3MTB3Mx}) and household financial assets (\mn{TABSHNOx}). The S\&P~500 enters through the tree component---consistent with the view that the stock market ``has predicted nine of the last five recessions''---while the Case-Shiller home price index (\mn{SPCS20RSA}) contributes through both linear and nonlinear channels. \mn{YLag} retains a modest role but now enters nonlinearly, in contrast with $h=1$ where it was overwhelmingly linear---suggesting that purely linear autoregressive persistence is primarily a short-run phenomenon. Post-COVID, performance deteriorates (RMSE ratio 1.69): \mn{DRIWCIL} switches to the tree channel, and many variables exhibit negative importance, suggesting that historically reliable signals actively mislead in this regime. Additional results at $h=2$ for housing starts and the term spread are provided in Appendix~\ref{app:dashboard_extra}.

\paragraph{Industrial Production ($h=1$).}

Figure~\ref{fig:dashboard_indpro} illustrates the case of industrial production, which aligns well with the central premise: linear autoregressive relationships matter---and if nonlinearities can be built on top of that, forecasts improve further. Pre-COVID, the linear component dominates overwhelmingly, driven by lagged industrial production (\mn{YLag}), which accounts for roughly 60\% of MSE reduction almost entirely through the linear channel. All other variables---the S\&P~500 dividend yield, the help-wanted/unemployment-vacancy ratio (\mn{HWIURATIOx}), building permits, and \mn{DRIWCIL}---contribute modestly through trees.

Post-COVID, the importance structure shifts. \mn{DRIWCIL} (bank lending willingness) replaces \mn{YLag} as the top variable, operating primarily through trees. Several sub-indices of industrial production (\mn{IPFINAL}, \mn{IPB51110SQ}, \mn{IPDMAT}) also contribute nonlinearly. Critically, the linear channel becomes actively harmful: \mn{IPNMAT} (nondurable goods materials) and \mn{PCESVx} (personal consumption services) have negative variable importance, meaning their linear contributions increase forecast error. The wide confidence bands on \mn{YLag}'s post-COVID importance are also informative: its contribution is highly period-dependent, helping in some quarters and hurting in others, signaling that the autoregressive relationship has become unstable. These variables signal a recession that did not materialize following the 2022 rate hikes---nondurable materials production historically contracts sharply ahead of recessions, but in a post-pandemic economy with unprecedented supply chain dynamics, this historically reliable signal proved misleading. This pattern recurs across real activity targets: among all models considered, only the Survey of Professional Forecasters avoids calling a recession that did not occur.

\paragraph{Housing Starts ($h=1$).}

Figure~\ref{fig:dashboard_houst_h1} presents the case of housing starts. Pre-COVID, building permits (\mn{PERMITS}) is by far the most important variable, contributing primarily through the linear channel---by construction a near-accounting relationship, since permits mechanically precede starts. Trees and a small interaction term add on top of this dominant linear contribution. Post-COVID, the picture changes dramatically: \mn{PERMITS} flips to negative importance, meaning its inclusion actively hurts the forecast. Instead, the inventory-to-sales ratio (\mn{ISRATIOx}) and the help-wanted/unemployment-vacancy ratio (\mn{HWIURATIOx}) emerge as the dominant positive contributors, both operating through the tree channel. Corporate bond rates (\mn{AAA}, \mn{BAA}) also contribute through both linear and tree channels, consistent with the interest-rate-driven slowdown in housing following the 2022 tightening cycle. Remarkably, the model captures with precision this slowdown, drawing on nonlinear channels to identify the rate-sensitivity of housing demand. Additional results at $h=2$ are provided in Appendix~\ref{app:dashboard_extra}.

The near-accounting relationships identified here---claims preceding unemployment, permits preceding housing starts---are likely even more prevalent at higher frequencies, where many indicator-to-aggregate mappings are nearly mechanical and partly linear. Monthly forecasting and nowcasting settings, where such relationships abound, represent a natural application of the hybrid approach; indeed, \citet{chinn2023nowcasting} find that partly linear models improve nowcasting performance for world trade.

\paragraph{$\Delta$Spread ($h=1$).}

Figure~\ref{fig:dashboard_dSpread_h1} presents the term spread ($\Delta$(10Y--3M)). Pre-COVID, no method beats the AR benchmark convincingly: LGB$^{\texttt{A}}$+ matches it exactly (RMSE ratio 1.00) while Random Forests edge slightly ahead (0.96). Post-COVID, the picture improves markedly---LGB+ achieves 0.80, with neural networks and TabPFN both at 0.82, all well below the AR. Variable importance reveals that the short-term interest rate spread (\mn{TB6M3Mx}, 6-month minus 3-month Treasury) dominates both periods through the linear channel, with CPI durables (\mn{CUSR0000SAD}) contributing primarily through the interaction term. Post-COVID, the 10-year rate (\mn{GS10}) and a housing-related financial flow variable (\mn{HNOREMQ027Sx}) gain prominence through the tree channel. Forecast decompositions for housing starts and the term spread at $h=2$, where the linear--nonlinear allocation shifts meaningfully relative to $h=1$, are discussed in Appendix~\ref{app:dashboard_extra}.

\subsection{Dual Interpretation of Specific Forecasts}\label{sec:dual}

\begin{figure}[t!]
  \caption{Dual Interpretation, Unemployment Rate ($h=1$), 2009Q2. {\fontfamily{phv}\selectfont\textbf{LGB$^{\texttt{A}}$+}}}
  \vspace*{-0.3em}
  \label{fig:dual_ur}
  \centering
  \includegraphics[width=\textwidth]{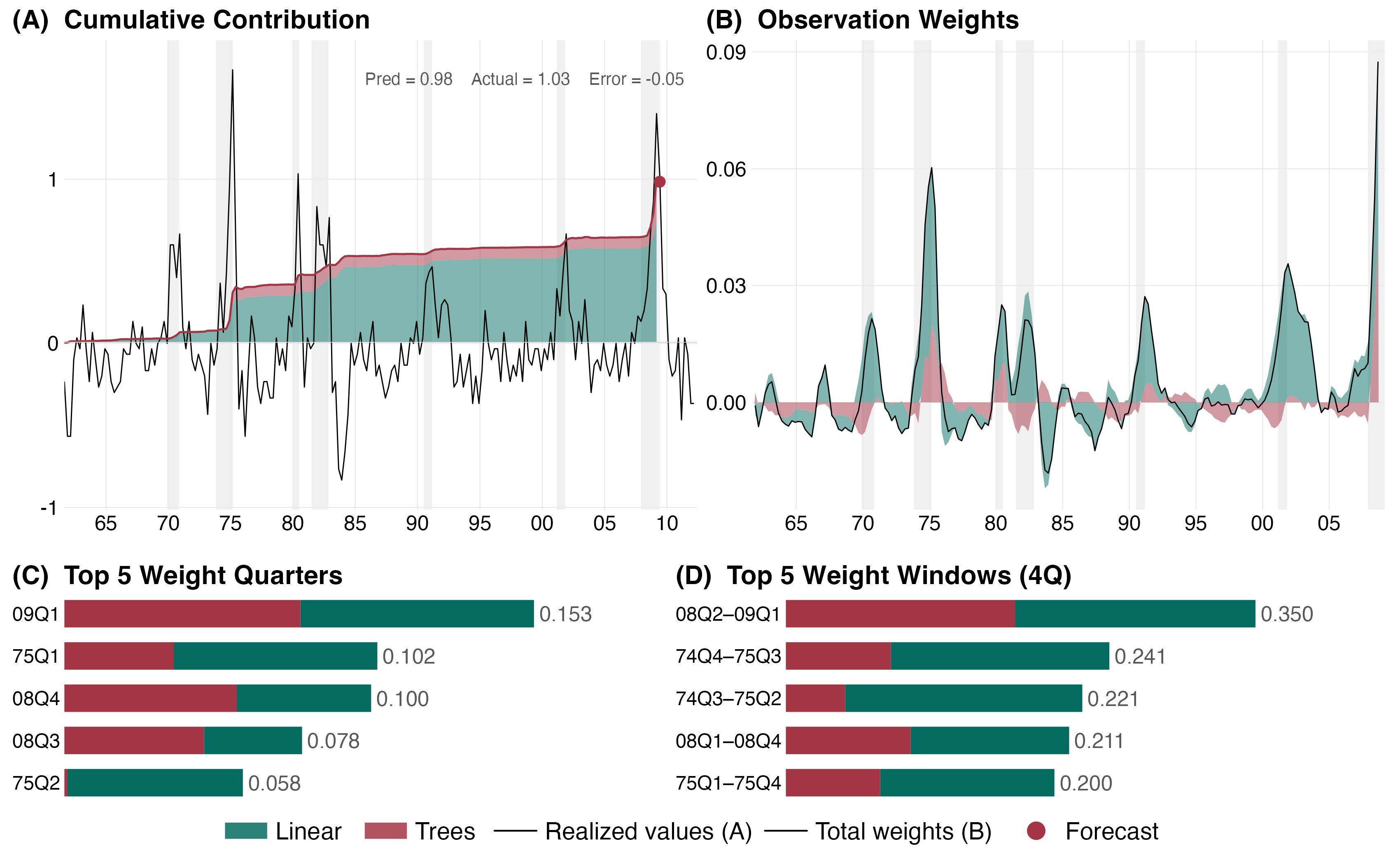}
  \vspace{0.3em}
  \parbox{\textwidth}{\fontfamily{phv}\selectfont\scriptsize
    \textit{Notes:} Observation weights for the LGB$^{\texttt{A}}$+ forecast of unemployment near the trough of the Great Recession. Weights are smoothed using a 4-quarter centered moving average. Panels (C) and (D) show the top-weighted individual quarters and 4-quarter windows, decomposed into linear (teal) and tree-based (red) contributions.
  }
\end{figure}

\begin{figure}[t!]
  \caption{Dual Interpretation, $\Delta$Spread ($h=1$), 2024Q4. {\fontfamily{phv}\selectfont\textbf{LGB$^{\texttt{A}}$+}}}
  \vspace*{-0.3em}
  \label{fig:dual_dSpread}
  \centering
  \includegraphics[width=\textwidth]{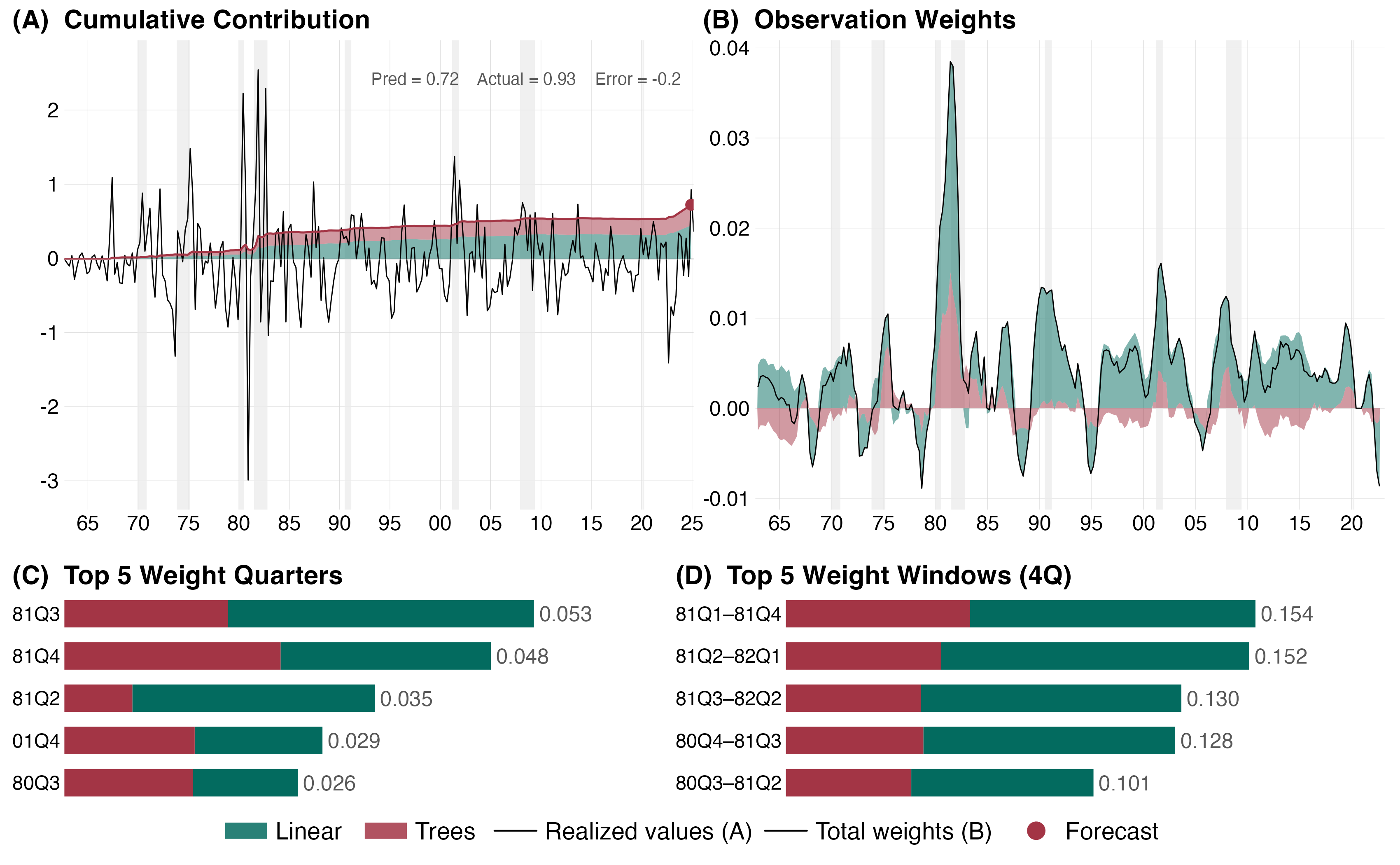}
  \vspace{0.3em}
  \parbox{\textwidth}{\fontfamily{phv}\selectfont\scriptsize
    \textit{Notes:} Observation weights for the LGB$^{\texttt{A}}$+ forecast of the change in the term spread (10Y--3M) at 2024Q4, when the yield curve steepened sharply following the Fed's September rate cut. COVID quarters (2020Q1--2021Q2) are excluded from training and receive zero weight. Weights are smoothed using a 4-quarter centered moving average. Panels (C) and (D) show the top-weighted individual quarters and 4-quarter windows, decomposed into linear (teal) and tree-based (red) contributions.
  }
\end{figure}

\begin{figure}[t!]
  \caption{Dual Interpretation, Housing Starts ($h=1$), 2022Q3. {\fontfamily{phv}\selectfont\textbf{LGB$^{\texttt{A}}$+}}}
  \vspace*{-0.3em}
  \label{fig:dual_houst}
  \centering
  \includegraphics[width=\textwidth]{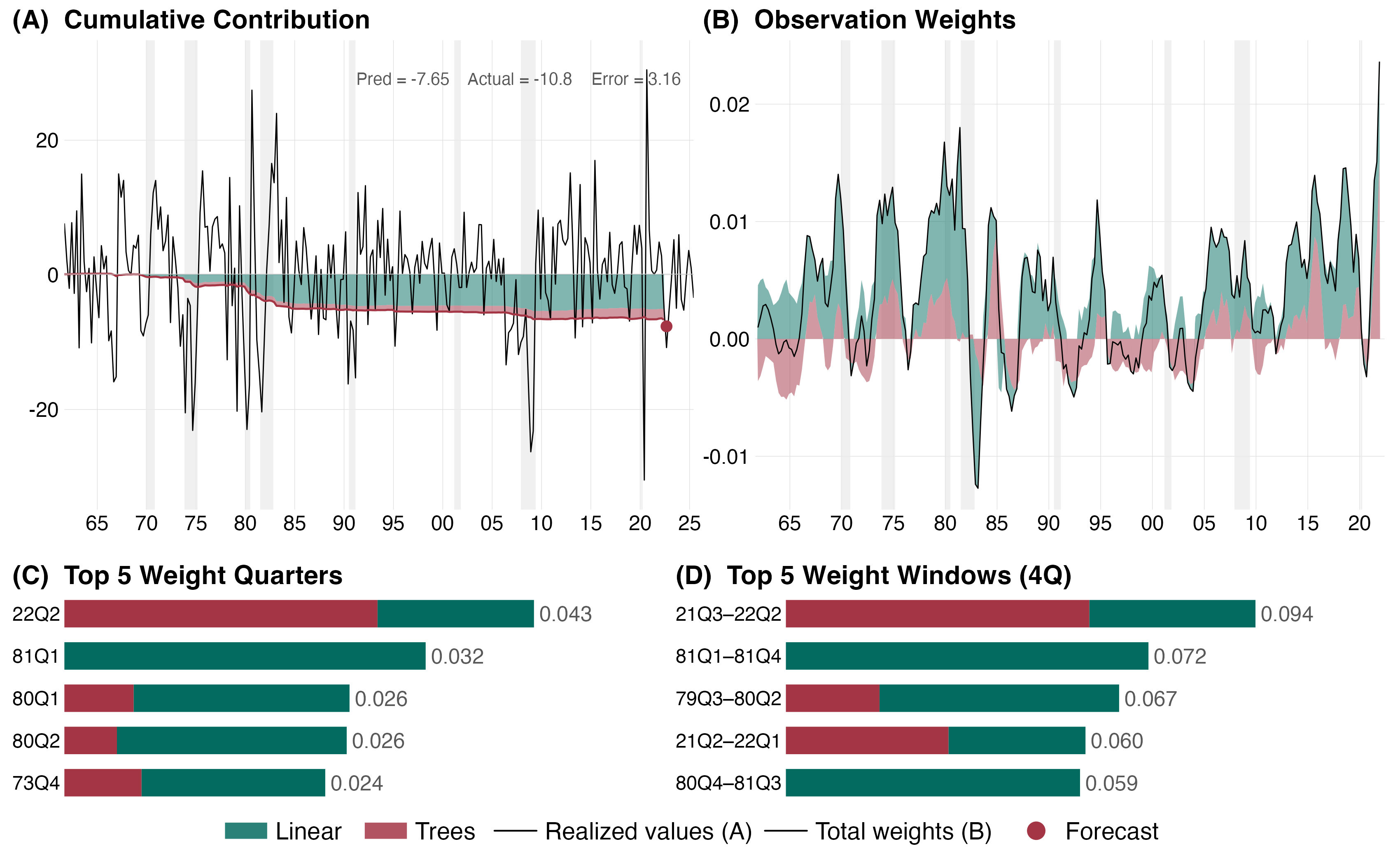}
  \vspace{0.3em}
  \parbox{\textwidth}{\fontfamily{phv}\selectfont\scriptsize
    \textit{Notes:} Observation weights for the LGB$^{\texttt{A}}$+ forecast of housing starts at the onset of the 2022 interest rate hiking cycle. Weights are smoothed using a 4-quarter centered moving average. Panels (C) and (D) show the top-weighted individual quarters and 4-quarter windows, decomposed into linear (teal) and tree-based (red) contributions.
  }
\end{figure}

The permutation importance decomposition tells us which variables matter and through which channel. A complementary question is: which historical precedents are leveraged to construct the forecast? \citet{gouletcoulombe2024dual} show that the vast majority of machine learning predictions can be written as a weighted average of training outcomes; gradient boosting and least squares are among the methods for which this representation holds. Directly leveraging the AXIL algorithm of \citet{geertsema2023axil}, I decompose the observation weights of the alternating variant (LGB$^{\texttt{A}}$+) into linear and tree contributions:
\begin{equation}
\hat{y}_j = \boldsymbol{w}_j \boldsymbol{y} = \sum_{i=1}^{N} w_{ji}\, y_i,
\end{equation}
where $w_{ji}$ captures how much training observation $i$ influences the forecast for test point $j$, and $\boldsymbol{w}_j$ is the $1 \times N$ vector of data portfolio weights. Because LGB$^{\texttt{A}}$+ is additive in its linear and tree components, these observation weights decompose accordingly:
\begin{equation}
w_{ji} = w_{ji}^{\text{lin}} + w_{ji}^{\text{tree}},
\end{equation}
where $w_{ji}^{\text{lin}}$ reflects the contribution through the linear updates and $w_{ji}^{\text{tree}}$ through the tree updates. This decomposition reveals not just \textit{which} historical episodes the model draws upon, but \textit{how}: through smooth, linear proximity or through tree-based proximity. Figures~\ref{fig:dual_ur}--\ref{fig:dual_houst} illustrate this decomposition for three selected forecast dates, chosen to highlight economically interesting episodes.

Figure~\ref{fig:dual_ur} shows the weight decomposition for the unemployment forecast at June 2009---the NBER trough of the Great Recession. The forecast is accurate (predicted: 0.98, actual: 1.03). The model leverages all past recessions with a clear hierarchy: the most recent year (2008Q2--2009Q1) receives the largest 4-quarter weight (0.350), predominantly through the linear channel---effectively an implicit Sahm-type rule recognizing that rapid unemployment acceleration tends to persist. The second-most-weighted window is 1974Q4--1975Q3 (0.241), when unemployment nearly doubled following the oil embargo and the Fed's tightening under Burns, with both linear and nonlinear channels contributing.

% [DELETED: Two dSpread dual figures (Dec 2022 + Dec 2023) and their discussion paragraphs.
%  These were based on a buggy target construction (second difference instead of first difference).
%  Replaced with a single corrected dual plot for Dec 2024.]

Figure~\ref{fig:dual_dSpread} shows the dual decomposition for the term spread forecast at December 2024 (predicted: 0.72, actual: 0.93), when the spread widened sharply as the Fed began easing after two years of tightening. The top 4-quarter windows are 1981Q1--1981Q4 (0.154) and 1981Q2--1982Q1 (0.152), both from the Volcker-era twin recessions, loaded heavily on the tree component. This indicates the model recognizes nonlinear similarities in the predictor configuration---a rapid reversal from deep inversion---rather than simply matching lagged spread levels. All five top individual quarters (81Q3, 81Q4, 81Q2, 01Q4, 80Q3) also come from high-volatility rate-cycle episodes. The pattern is consistent with the variable importance panel: post-COVID, the 10-year rate and short-term interest rate spreads drive the forecast through both linear and nonlinear channels, and the model's ``nearest neighbors'' in the training data are precisely the episodes of aggressive monetary policy reversal that produce similar predictor configurations.

Figure~\ref{fig:dual_houst} shows the housing starts weights for September 2022, when mortgage rates had roughly doubled in under a year. The top individual quarter is 2022Q2 (0.043), loaded heavily on the tree channel---the model relies on the most recent rate-shock conditions. The second-ranked quarter is 1981Q1 (0.032), loaded almost entirely on the linear channel, pointing to the Volcker-era rate shocks as the closest historical analogue. The 4-quarter window ranking tells a similar story: 2021Q3--2022Q2 (0.094) ranks first, capturing the immediate tightening episode, followed by 1981Q1--1981Q4 (0.072)---a period when 30-year mortgage rates approached 18\% and housing activity contracted sharply. The 1970s and early-1980s recessions also receive substantial weight, all episodes where monetary tightening weighed heavily on housing.

Nearly absent is the 2007--08 housing crisis, which receives negligible weight---economically sensible since that collapse was driven by subprime credit risk, not financing costs. A visible band of negative weights around 1983 is also informative: housing starts rebounded roughly 60\% as Volcker eased, a recovery episode representing the \emph{opposite} of September 2022. The importance decomposition in Figure~\ref{fig:dashboard_houst_h1} corroborates this: labor market indicators such as the help-wanted index and the unemployment--vacancy ratio are the dominant positive contributors through the tree channel, with corporate bond rates (AAA, BAA) contributing through both linear and nonlinear channels.

\section{Conclusion}

LGB+ is a family of hybrid boosting procedures that interleave greedy univariate linear updates with standard tree updates within the boosting loop. The competition variant selects the better candidate at each step via out-of-bag evaluation; the alternating variant (LGB$^{\texttt{A}}$+) achieves comparable results on a fixed schedule, at lower computational cost and without step-by-step OOB evaluation. Both are fast, require minimal tuning beyond standard LightGBM hyperparameters, and are robust across a wide range of data-generating processes---from purely linear to highly nonlinear---with no systematic failures at either extreme. In a quarterly U.S.\ macroeconomic forecasting exercise against a competitive benchmark suite, both variants hold their own, with particular gains at short horizons where autoregressive structure is most relevant. The additive prediction structure also yields interpretability tools not available to standard gradient boosting: a three-way variable importance decomposition (linear, tree, and interaction channels) and, for the alternating variant, a dual representation that expresses each forecast as a historically grounded weighted average of past outcomes, with weights that themselves decompose into linear and tree contributions.

The most natural applications for LGB+ are settings where linear relationships are expected to be pervasive but nonlinear subtleties are suspected---precisely the situations where plain tree-based methods struggle to beat an AR benchmark. Because LGB+ incorporates linear updates from the start, it may effectively begin from an AR-like baseline and builds nonlinear refinements on top, rather than having to rediscover autoregressive structure with splits. Nowcasting is an obvious candidate: many indicators bear near-accounting linear relationships to target variables yet to be observed, while nonlinear regime effects around turning points, policy shifts, or supply shocks remain relevant. High-frequency short-run forecasting is another natural fit. Beyond macro, applications in finance suggest themselves: realized volatility and credit spreads both have pronounced linear components (persistence, mean reversion) alongside nonlinear tail dynamics. More broadly, the hybrid structure documented here is likely just a starting point: time-varying tree/linear schedules and online adaptation of the competition threshold are natural extensions. For now, the practical recommendation is straightforward: run both variants, since they are complementary in their tradeoffs, and expect that at minimum they will carve out a distinct profile from the rest of the ML benchmark set.

\clearpage
\setlength\bibsep{5pt}

\bibliographystyle{apalike}
 
\setstretch{0.75}

% PERSON USING, CHANGE THIS 
% FORMAT '/Users[\dboxpath]/Dropbox/...'
\def\dboxpath{/karinklieber} 
\def\dboxpath{/UQAM} % FOR PHIL : /UQAM
 \bibliography{references}

\clearpage
 
\appendix
%\appendixpage
%\addappheadtotoc
\newcounter{saveeqn}
\setcounter{saveeqn}{\value{section}}
\renewcommand{\theequation}{\mbox{\Alph{saveeqn}.\arabic{equation}}} \setcounter{saveeqn}{1}
\setcounter{equation}{0}
\setstretch{1.25}
 
%\pagebreak

%%%%%%%%%%%%%%%%%%%%%%%%%%%%%%%%%%%%%%%%%%%%%%%%%%%%%%%%%%%%

\section{Model Specifications}\label{app:models}

\begin{enumerate}[leftmargin=6em, labelwidth=5em, labelsep=0.5em, align=left]
   \item[\texttt{\fontfamily{phv}\selectfont \textbf{\phantom{Tab}AR(4)}:}] Fourth-order autoregression on the target variable.

    \item[\texttt{\fontfamily{phv}\selectfont \textbf{\phantom{Ta}FAAR}:}] Factor-Augmented AR \citep{stock2002forecasting}. Includes four lags of the target and two lags of four principal components extracted from the predictor panel.

    \item[\texttt{\fontfamily{phv}\selectfont \textbf{\phantom{TabP}RR}:}] Ridge Regression. $L_2$-penalized linear regression on the full predictor set; $\lambda$ selected by cross-validation.

    \item[\texttt{\fontfamily{phv}\selectfont \textbf{\phantom{Tab}KRR}:}] Kernel Ridge Regression. Gaussian and Laplacian kernels; bandwidth $\sigma$ and $\lambda$ selected by cross-validation.

    \item[\texttt{\fontfamily{phv}\selectfont \textbf{\phantom{TabP}RF}:}] Random Forest \citep{breiman2001random}. 500 trees, 75\% row subsampling, minimum node size 5, random feature subsets.

    \item[\texttt{\fontfamily{phv}\selectfont \textbf{\phantom{Tab}LGB}:}] LightGBM \citep{ke2017lightgbm}. Histogram-based gradient boosting; learning rate, depth, and sampling fractions selected by cross-validation with early stopping.

    \item[\texttt{\fontfamily{phv}\selectfont \textbf{\phantom{Ta}LGB+}:}] At each boosting step, a tree-based and a linear update compete; the winner is selected via out-of-bag validation.

    \item[\texttt{\fontfamily{phv}\selectfont \textbf{LGB$^{\texttt{A}}$+}:}] The alternating variant of LGB+. A more computationally economical version that alternates tree ensembles with linear corrections in a fixed pattern each boosting cycle.

    \item[\texttt{\fontfamily{phv}\selectfont \textbf{\phantom{TabP}NN}:}] Feed-forward neural network with three hidden layers (400 neurons each), ReLU activations, and dropout regularization (rate 0.2); trained via Adam optimizer with early stopping.

    \item[\texttt{\fontfamily{phv}\selectfont \textbf{\phantom{Tab}HNN}:}] Hemisphere Neural Network \citep{goulet2025neural}. A constrained neural architecture designed for inflation forecasting, with separate hemispheres for inflation expectations, output gap, and commodity prices. Used for inflation only.

    \item[\texttt{\fontfamily{phv}\selectfont \textbf{TabPFN}:}] Transformer-based foundation model \citep{hollmann2022tabpfn} pre-trained on synthetic tabular datasets. Performs in-context learning: ingests training data as context and produces predictions in a single forward pass.

    \item[\texttt{\fontfamily{phv}\selectfont \textbf{\phantom{Tab}SPF}:}] Survey of Professional Forecasters \citep{stark2010realistic,engelberg2022manskie}. Median forecast from the Philadelphia Fed survey. Available for CPI inflation, GDP growth, housing starts, unemployment, and industrial production.
\end{enumerate}

\section{Detailed Simulation Results}
\label{app:simulation}

\subsection{Simulation Design}

The simulation study considers eight data-generating processes (DGPs), spanning purely linear to highly nonlinear settings. For each DGP, I generate $P=6$ predictors drawn uniformly from $[0,1]$ and construct the outcome as $y = f(\boldsymbol{x}) + \varepsilon$, where the signal function $f(\cdot)$ varies by DGP and $\varepsilon$ is Gaussian noise calibrated to achieve the target signal-to-noise ratio (SNR). Results are averaged over 10 Monte Carlo replications, with a fixed test set of $N_{\text{test}}=1000$ observations.

\paragraph{Linear DGPs.}
\begin{itemize}[leftmargin=2em, itemsep=0.2em]
    \item \textsc{Linear}: $f(\boldsymbol{x}) = 2x_1 - x_2 + 3x_3 + 1.5x_4 + 0.5x_5$ (centered at 0.5). Pure linear relationship.
\end{itemize}

\paragraph{Partially Linear DGPs.}
\begin{itemize}[leftmargin=2em, itemsep=0.2em]
    \item \textsc{Friedman 1}: $f(\boldsymbol{x}) = 10\sin(\pi x_1 x_2) + 20(x_3 - 0.5)^2 + 10x_4 + 5x_5$. Contains substantial linear terms ($10x_4 + 5x_5$) alongside nonlinear components.
    \item \textsc{Product-Log}: $f(\boldsymbol{x}) = x_1 x_2 + \log(x_3 + x_4 + 2)$. Multiplicative interaction plus smooth logarithmic transformation.
    \item \textsc{Mixture}: $f(\boldsymbol{x}) = \text{Linear}(x_1, x_2) + \text{Nonlinear}(x_3, x_4, x_5)$, where the linear part is $2x_1 + 1.5x_2$ and the nonlinear part is $3\sin(\pi x_3 x_4) + 2(x_5 - 0.5)^2$, combined with equal variance contributions.
\end{itemize}

\paragraph{Nonlinear DGPs.}
\begin{itemize}[leftmargin=2em, itemsep=0.2em]
    \item \textsc{Trig-Log}: $f(\boldsymbol{x}) = \sin(\pi(x_1 + x_2 + x_3)) + \log(1 + x_4^2)$. Trigonometric plus logarithmic; no linear structure.
    \item \textsc{Soft Radial}: $f(\boldsymbol{x}) = 1/(1 + 5\|\boldsymbol{x} - 0.5\|^2)$. Radial basis function centered at $(0.5, \ldots, 0.5)$.
    \item \textsc{Step/U-curve}: Combination of step functions and U-curves: $f(\boldsymbol{x}) = 3\cdot\mathbf{1}_{0.3 < x_1 < 0.7} + 2\cdot\mathbf{1}_{x_2 > 0.5} - \mathbf{1}_{x_2 < 0.2} + 4(x_3 - 0.5)^2 + 3|x_4 - 0.5| + 2\cdot\mathbf{1}_{0.25 < x_5 < 0.75,\, x_6 > 0.5}$. Pure discontinuities and non-smooth curvature; tests whether linear updates hurt.
\end{itemize}

\paragraph{Highly Nonlinear DGPs.}
\begin{itemize}[leftmargin=2em, itemsep=0.2em]
    \item \textsc{Rotated Sine}: $f(\boldsymbol{x}) = \sin(3(x_1 + x_2 + x_3 + x_4))$. High-frequency oscillation in a rotated coordinate system; difficult for all methods.
\end{itemize}

\definecolor{tableShade}{gray}{0.95}

% Simulation table - Version 2 (compact single-page)
\input{03_tables/simulation_table_v2.tex}

% % Original simulation table
% \input{03_tables/simulation_table.tex}
%
% \clearpage
% % Alternative Version 1: Full longtable with dpd2 colors and RowAlt shading
% \input{03_tables/simulation_table_v1.tex}
%
% \clearpage
% % Alternative Version 3: Minimal changes - dpd2 colors, phv font, RowAlt
% \input{03_tables/simulation_table_v3.tex}

% --- GDP (SPF available) ---
\input{03_tables/gdp__h1.tex}

\input{03_tables/gdp__h2.tex}

\input{03_tables/gdp__h4.tex}

% --- Unemployment Rate (SPF available) ---
\input{03_tables/ur__h1.tex}

\input{03_tables/ur__h2.tex}

\input{03_tables/ur__h4.tex}

% --- Inflation (SPF available) ---
\input{03_tables/infl__h1.tex}

\input{03_tables/infl__h2.tex}

\input{03_tables/infl__h4.tex}

% --- Housing Starts (SPF available) ---
\input{03_tables/houst__h1.tex}

\input{03_tables/houst__h2.tex}

\input{03_tables/houst__h4.tex}

% --- Industrial Production (no SPF) ---
\input{03_tables/indpro__h1.tex}

\input{03_tables/indpro__h2.tex}

\input{03_tables/indpro__h4.tex}

% --- $\Delta$Spread (no SPF) ---
\input{03_tables/dSpread__h1.tex}

\input{03_tables/dSpread__h2.tex}

\input{03_tables/dSpread__h4.tex}

%% Standalone time series plots removed per Phase 5

\section{Additional Forecast Decompositions}\label{app:dashboard_extra}

This appendix presents forecast decomposition dashboards for targets and horizons not shown in the main text. In the main text, I focus on $h=1$ for housing starts and the term spread. The $h=2$ results below provide a complementary view of how the linear--nonlinear allocation evolves as the forecast horizon lengthens.

\begin{figure}[t!]
  \caption{Housing Starts ($h=2$), {\fontfamily{phv}\selectfont\textbf{LGB$^{\texttt{A}}$+}}}
  \vspace*{-0.3em}
  \label{fig:dashboard_houst_h2}
  \centering
  \includegraphics[width=\textwidth]{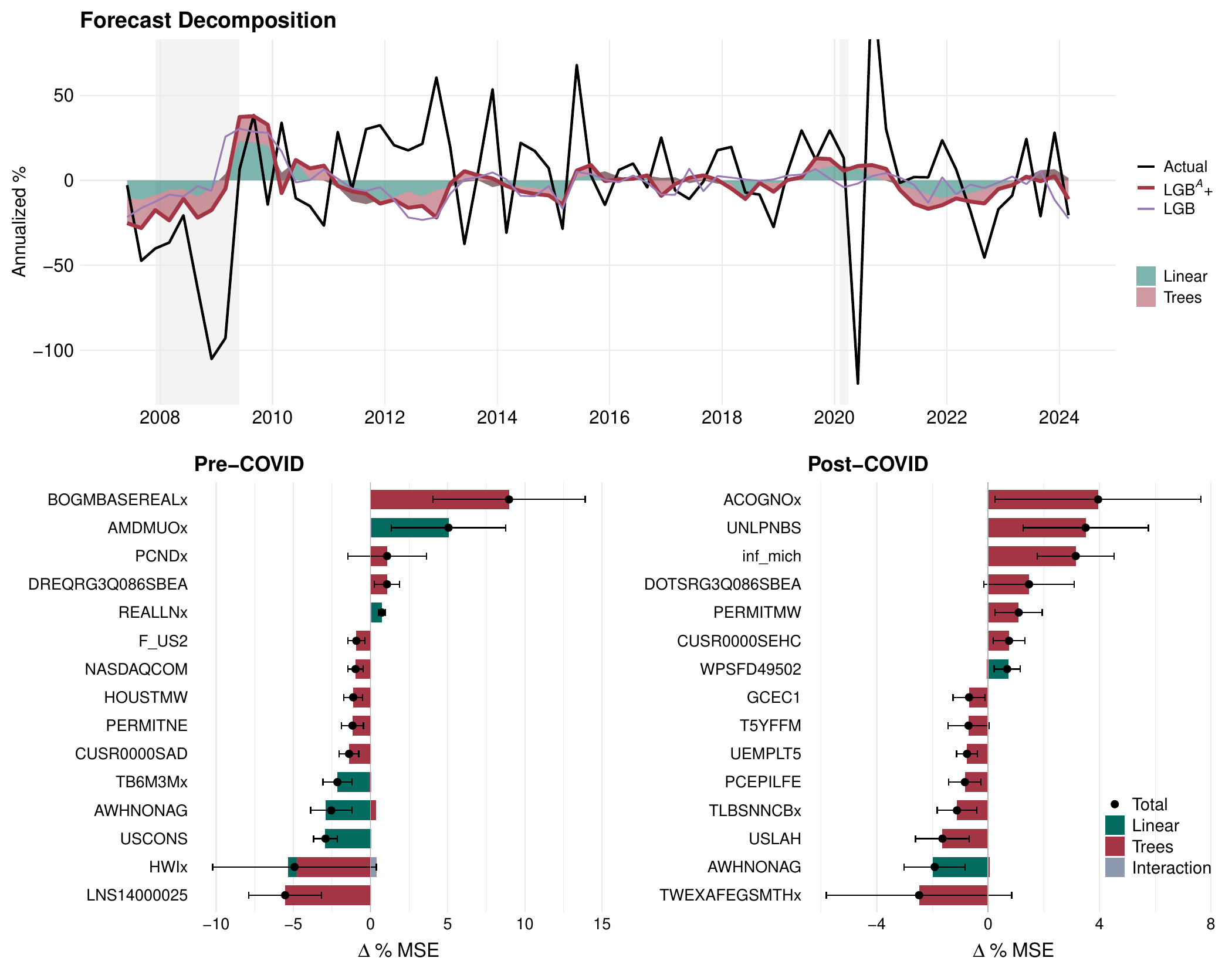}
  \vspace{0.3em}
  \parbox{\textwidth}{\fontfamily{phv}\selectfont\scriptsize
    \textit{Notes:} Top panel shows actual values (black), forecast (red), and LGB forecast (purple dashed). Shaded regions decompose the forecast into linear (teal), tree-based (red), and interaction (gray) components. Gray bands indicate NBER recession dates. Bottom panels show permutation-based variable importance for pre-COVID (left) and post-COVID (right) periods.
  }
\end{figure}

\paragraph{Housing Starts ($h=2$).}

At $h=2$, the variable importance profile differs across periods. Pre-COVID, the monetary base (\mn{BOGMBASEREALx}) and the real M2 aggregate (\mn{AMDMUOx}) dominate---\mn{BOGMBASEREALx} with substantial contributions from both channels, \mn{AMDMUOx} primarily through the linear channel---alongside \mn{PCNDx} (personal consumption on nondurables). Post-COVID, the composition shifts toward leading indicators: \mn{ACOGNOx} (new orders for capital goods) and \mn{UNLPNBS} (unit nonlabor payments, nonfarm business) rank highest, with the University of Michigan consumer sentiment index (\mn{uf\_mich}) entering through the nonlinear channel. The tree component gains importance at $h=2$ relative to $h=1$, though overall variable importance magnitudes are smaller, consistent with the weaker forecastability of housing starts at longer horizons.

\begin{figure}[t!]
  \caption{$\Delta$Spread ($h=2$), {\fontfamily{phv}\selectfont\textbf{LGB$^{\texttt{A}}$+}}}
  \vspace*{-0.3em}
  \label{fig:dashboard_dSpread_h2}
  \centering
  \includegraphics[width=\textwidth]{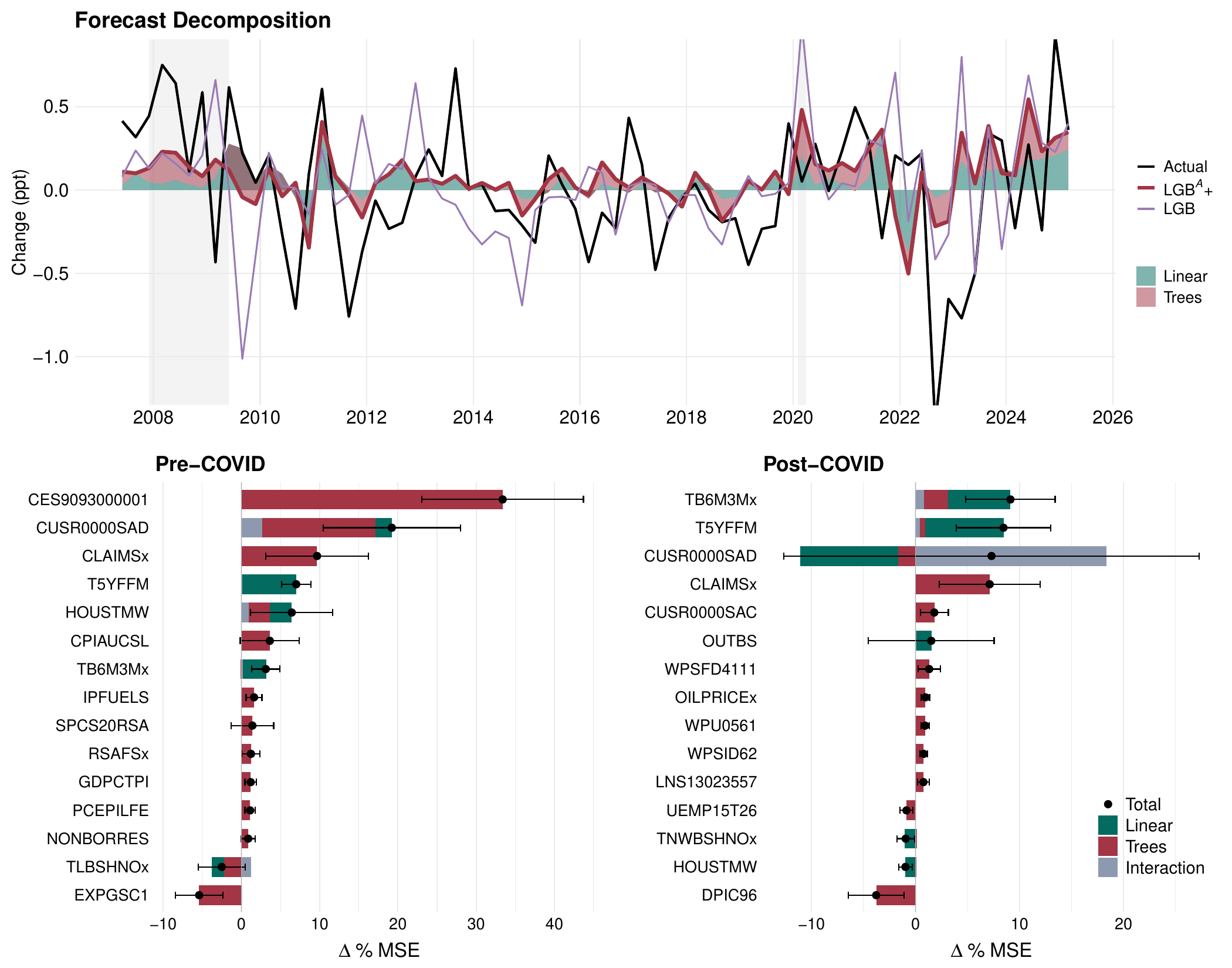}
  \vspace{0.3em}
  \parbox{\textwidth}{\fontfamily{phv}\selectfont\scriptsize
    \textit{Notes:} Top panel shows actual values (black), forecast (red), and LGB forecast (purple dashed). Shaded regions decompose the forecast into linear (teal), tree-based (red), and interaction (gray) components. Gray bands indicate NBER recession dates. Bottom panels show permutation-based variable importance for pre-COVID (left) and post-COVID (right) periods.
  }
\end{figure}

\paragraph{$\Delta$Spread ($h=2$).}

At $h=2$, LGB$^{\texttt{A}}$+ achieves an RMSE ratio of 0.89 pre-COVID; post-COVID, the competition variant LGB+ reaches 0.89 while LGB$^{\texttt{A}}$+ attains 0.96. The variable importance profile shifts markedly relative to $h=1$. Pre-COVID, government employment (\mn{CES9093000001}) dominates through the tree channel, with CPI durables (\mn{CUSR0000SAD}) second via a mix of linear and nonlinear contributions. Post-COVID, \mn{TB6M3Mx} takes the lead through the tree channel, followed by \mn{T5YFFM} (5-year Treasury minus Fed Funds Rate) through both channels, and \mn{CUSR0000SAD} through a large interaction term. The tree component gains overall importance at $h=2$ relative to $h=1$, suggesting that nonlinear regime identification becomes more valuable as the forecast horizon lengthens.

\end{document}

%% file: 03_tables/summary_table.tex
% Requires in preamble:
% \definecolor{ForestGreen}{RGB}{34,139,34}
% \definecolor{dpd2}{rgb}{0.0, 0.035, 0.38}
% \definecolor{RowAlt}{gray}{0.975}

\begin{table}[t!]
\vspace*{-0.25em}
  \centering
  \caption{\normalsize Summary of Forecast Performance: RMSE (relative to AR)}
  \label{tab:summary_rmse}
\vspace*{-0.5em}
  {\fontfamily{phv}\selectfont
  \resizebox{\textwidth}{!}{%
  \scriptsize
  \setlength{\tabcolsep}{0.3em}
  \renewcommand{\arraystretch}{1.1}
  \begin{tabular}{l r r r r r r r r c r r r r r r r r}
    \toprule

    & \multicolumn{8}{c}{\textcolor{dpd2}{\textbf{Pre-COVID (2007Q2 {--} 2019Q4)}}}
    && \multicolumn{8}{c}{\textcolor{dpd2}{\textbf{Post-COVID$^\dagger$ ({--} 2025Q1)}}}
    \\
    \cmidrule(lr){2-9}
    \cmidrule(lr){11-18}
    & RF & LGB & LGB$^{\texttt{A}}$+ & LGB+ & KRR & NN & SPF & TPFN
    && RF & LGB & LGB$^{\texttt{A}}$+ & LGB+ & KRR & NN & SPF & TPFN \\

    \midrule
    \addlinespace[0.3em]
    \multicolumn{18}{l}{\textcolor{dpd2}{\textbf{Unemployment}}} \\
    \addlinespace[0.2em]

      \quad $h$=1 & 0.86 & 0.86 & {\textbf{0.68}} & {\color{ForestGreen}\textbf{0.67}} & 1.29 & 0.94 & 1.07 & 0.84 && 1.43 & 1.51 & 1.41 & 1.39 & 3.11 & 1.41 & {\textbf{1.15}} & {\color{ForestGreen}\textbf{0.86}} \\
      \rowcolor{RowAlt}
      \quad $h$=2 & 0.95 & 0.92 & {\textbf{0.84}} & {\color{ForestGreen}\textbf{0.71}} & 1.32 & 0.90 & 0.94 & 0.87 && 1.89 & 2.91 & 1.31 & 1.69 & 2.56 & 1.29 & {\color{ForestGreen}\textbf{0.92}} & {\textbf{1.22}} \\
      \quad $h$=4 & 0.90 & {\textbf{0.85}} & 0.99 & 0.87 & 1.30 & 0.91 & 0.91 & {\color{ForestGreen}\textbf{0.72}} && 1.45 & 1.46 & 1.16 & 1.56 & 2.00 & {\textbf{0.82}} & {\color{ForestGreen}\textbf{0.49}} & 1.08 \\
    \addlinespace[0.3em]
    \multicolumn{18}{l}{\textcolor{dpd2}{\textbf{GDP}}} \\
    \addlinespace[0.2em]

      \quad $h$=1 & 0.93 & 1.02 & 0.91 & 0.91 & {\color{ForestGreen}\textbf{0.84}} & {\textbf{0.89}} & 0.90 & 0.90 && {\textbf{1.06}} & 1.40 & 1.13 & 1.07 & 1.11 & 1.14 & 1.20 & {\color{ForestGreen}\textbf{1.02}} \\
      \rowcolor{RowAlt}
      \quad $h$=2 & {\color{ForestGreen}\textbf{0.93}} & 0.98 & 1.09 & 1.10 & 1.04 & 1.08 & {\textbf{0.93}} & 0.99 && 1.21 & 1.29 & 1.25 & 1.17 & {\textbf{1.15}} & 1.28 & 1.31 & {\color{ForestGreen}\textbf{1.06}} \\
      \quad $h$=4 & 1.00 & 1.02 & 1.14 & 1.04 & 0.94 & 0.95 & {\textbf{0.93}} & {\color{ForestGreen}\textbf{0.92}} && 0.65 & 0.88 & {\textbf{0.56}} & 0.61 & 0.67 & 0.60 & 0.76 & {\color{ForestGreen}\textbf{0.53}} \\
    \addlinespace[0.3em]
    \multicolumn{18}{l}{\textcolor{dpd2}{\textbf{Ind.\ Prod.}}} \\
    \addlinespace[0.2em]

      \quad $h$=1 & 1.01 & 1.01 & {\color{ForestGreen}\textbf{0.88}} & 1.01 & 1.06 & {\textbf{0.99}} & 1.21 & 1.12 && 1.06 & 1.28 & 1.69 & {\textbf{0.97}} & 1.36 & 1.61 & {\color{ForestGreen}\textbf{0.89}} & 1.37 \\
      \rowcolor{RowAlt}
      \quad $h$=2 & 0.99 & {\textbf{0.94}} & 1.02 & 1.05 & 1.02 & {\color{ForestGreen}\textbf{0.91}} & 1.05 & 1.03 && 0.95 & 2.00 & 1.61 & {\textbf{0.94}} & 1.22 & 1.41 & {\color{ForestGreen}\textbf{0.78}} & 1.85 \\
      \quad $h$=4 & {\textbf{1.00}} & 1.04 & 1.07 & 1.07 & {\color{ForestGreen}\textbf{0.94}} & 1.00 & 1.02 & 1.01 && 1.12 & 0.92 & 1.59 & 0.97 & {\color{ForestGreen}\textbf{0.58}} & 1.13 & {\textbf{0.68}} & 1.32 \\
    \addlinespace[0.3em]
    \multicolumn{18}{l}{\textcolor{dpd2}{\textbf{Housing}}} \\
    \addlinespace[0.2em]

      \quad $h$=1 & 1.02 & 1.08 & {\textbf{1.00}} & {\color{ForestGreen}\textbf{0.97}} & 1.06 & 1.12 & 1.02 & 1.03 && 0.91 & 0.95 & {\color{ForestGreen}\textbf{0.83}} & 0.96 & 0.94 & 1.13 & {\textbf{0.84}} & 1.11 \\
      \rowcolor{RowAlt}
      \quad $h$=2 & 1.03 & 1.03 & 1.04 & 1.04 & {\color{ForestGreen}\textbf{1.02}} & 1.11 & 1.03 & {\textbf{1.02}} && {\color{ForestGreen}\textbf{0.96}} & 1.10 & 1.15 & 1.13 & 0.98 & 1.39 & 0.99 & {\textbf{0.97}} \\
      \quad $h$=4 & 1.00 & 1.03 & 1.02 & 1.05 & {\color{ForestGreen}\textbf{0.98}} & 1.04 & 1.04 & {\textbf{0.98}} && 1.13 & 1.44 & 1.45 & 1.27 & {\color{ForestGreen}\textbf{1.00}} & 1.39 & 1.15 & {\textbf{1.10}} \\
    \addlinespace[0.3em]
    \multicolumn{18}{l}{\textcolor{dpd2}{\textbf{Inflation} (NN $\rightarrow$ HNN)}} \\
    \addlinespace[0.2em]

      \quad $h$=1 & 0.96 & 1.06 & 0.92 & 0.90 & 1.03 & {\textbf{0.89}} & {\color{ForestGreen}\textbf{0.86}} & 1.02 && 1.18 & 1.58 & 1.14 & 1.03 & {\color{ForestGreen}\textbf{0.94}} & {\textbf{0.96}} & 1.59 & 1.16 \\
      \rowcolor{RowAlt}
      \quad $h$=2 & {\textbf{0.83}} & 1.01 & 0.95 & 0.91 & 0.94 & 0.86 & {\color{ForestGreen}\textbf{0.76}} & 0.95 && 1.16 & 1.16 & 1.05 & 1.02 & {\color{ForestGreen}\textbf{0.70}} & 0.94 & 1.29 & {\textbf{0.92}} \\
      \quad $h$=4 & 0.90 & 0.87 & 0.87 & 0.91 & 0.88 & {\textbf{0.84}} & {\color{ForestGreen}\textbf{0.80}} & 1.12 && 1.22 & 1.25 & 1.20 & 1.18 & {\color{ForestGreen}\textbf{0.76}} & 1.24 & {\textbf{1.00}} & 1.01 \\
    \addlinespace[0.3em]
    \multicolumn{18}{l}{\textcolor{dpd2}{\textbf{$\Delta$Spread}}} \\
    \addlinespace[0.2em]

      \quad $h$=1 & {\color{ForestGreen}\textbf{0.96}} & 1.06 & {\textbf{1.00}} & 1.06 & 1.07 & 1.20 & -- & 1.05 && 0.98 & 1.21 & 0.98 & {\color{ForestGreen}\textbf{0.80}} & 0.90 & {\textbf{0.82}} & -- & 0.82 \\
      \rowcolor{RowAlt}
      \quad $h$=2 & 0.93 & 1.16 & {\color{ForestGreen}\textbf{0.89}} & 1.05 & 0.91 & 0.95 & -- & {\textbf{0.89}} && 1.00 & 1.04 & 0.96 & {\color{ForestGreen}\textbf{0.89}} & {\textbf{0.90}} & 0.98 & -- & 0.90 \\
      \quad $h$=4 & {\color{ForestGreen}\textbf{0.95}} & 1.02 & 1.30 & 1.21 & {\textbf{0.95}} & 1.08 & -- & 0.99 && 1.19 & 1.20 & 1.12 & {\color{ForestGreen}\textbf{0.95}} & 1.02 & {\textbf{0.96}} & -- & 1.14 \\

    \addlinespace[0.2em]
    \bottomrule
  \end{tabular}%
  }

  \vspace{0.35em}
  \parbox{\linewidth}{\fontfamily{phv}\selectfont\scriptsize
    \textit{Notes}: RMSE relative to AR benchmark. LGB$^{\texttt{A}}+$ = alternating variant; LGB+ = competition variant. NN/HNN column shows HNN for inflation, NN otherwise. SPF available only for unemployment, GDP, inflation, and housing starts. $\Delta$Spread has no SPF. $^\dagger$Post-COVID evaluation starts 2022Q1 for GDP, unemployment, and industrial production; 2021Q1 for inflation, housing starts, and $\Delta$Spread.

    Best: \textcolor{ForestGreen}{\textbf{bold green}}; second-best: \textbf{bold}.
  }
  }

\end{table}

%% file: 03_tables/simulation_table_v2.tex
% Version 2: Ultra-compact single-page version with all N values
% Updated: LGB+ = competition, LGB^A+ = alternating
% Requires in preamble:
% \definecolor{ForestGreen}{RGB}{34,139,34}
% \definecolor{dpd2}{rgb}{0.0, 0.035, 0.38}
% \definecolor{RowAlt}{gray}{0.975}

\begin{table}[t!]
\vspace*{-0.5em}
  \centering
  \caption{Model Performance (Out-of-Sample $R^2$) Across Data Generating Processes}
  \label{tab:simulation_results}
\vspace*{-0.75em}
  {\fontfamily{phv}\selectfont
  \resizebox{\textwidth}{!}{%
  \tiny
  \setlength{\tabcolsep}{0.22em}
  \renewcommand{\arraystretch}{1.22}
  \begin{tabular}{l r ccccc c r ccccc c r ccccc}
    \toprule

    & \multicolumn{6}{c}{\textcolor{dpd2}{\textbf{$N=250$}}}
    && \multicolumn{6}{c}{\textcolor{dpd2}{\textbf{$N=500$}}}
    && \multicolumn{6}{c}{\textcolor{dpd2}{\textbf{$N=1000$}}}
    \\
    \cmidrule(lr){2-7}
    \cmidrule(lr){9-14}
    \cmidrule(lr){16-21}
    \textcolor{dpd2}{\textbf{DGP}} & \textcolor{dpd2}{\textbf{SNR}} & OLS & RF & LGB & LGB+ & LGB$^{\texttt{A}}$+
    && \textcolor{dpd2}{\textbf{SNR}} & OLS & RF & LGB & LGB+ & LGB$^{\texttt{A}}$+
    && \textcolor{dpd2}{\textbf{SNR}} & OLS & RF & LGB & LGB+ & LGB$^{\texttt{A}}$+ \\

    \midrule
    \addlinespace[0.15em]
    \multicolumn{21}{l}{\textcolor{dpd2}{\textbf{Linear}}} \\
    \addlinespace[0.1em]

      \rowcolor{RowAlt}
      \quad Linear & 0.5 & {\color{ForestGreen}\textbf{.32}} & .25 & .22 & .27 & \textbf{.27} && 0.5 & {\color{ForestGreen}\textbf{.31}} & .26 & .25 & .28 & \textbf{.28} && 0.5 & {\color{ForestGreen}\textbf{.34}} & .28 & .30 & .30 & \textbf{.31} \\
      \rowcolor{RowAlt}
        & 1 & {\color{ForestGreen}\textbf{.49}} & .42 & .40 & .43 & \textbf{.45} && 1 & {\color{ForestGreen}\textbf{.48}} & .43 & .43 & .45 & \textbf{.45} && 1 & {\color{ForestGreen}\textbf{.50}} & .45 & .47 & .47 & \textbf{.48} \\
      \rowcolor{RowAlt}
        & 2 & {\color{ForestGreen}\textbf{.66}} & .58 & .58 & .60 & \textbf{.63} && 2 & {\color{ForestGreen}\textbf{.66}} & .60 & .61 & .62 & \textbf{.62} && 2 & {\color{ForestGreen}\textbf{.67}} & .62 & .64 & .64 & \textbf{.65} \\

    \addlinespace[0.15em]
    \multicolumn{21}{l}{\textcolor{dpd2}{\textbf{Partially Linear}}} \\
    \addlinespace[0.1em]

      \quad {\color{blue}Product-Log} & 0.5 & {\color{ForestGreen}\textbf{.28}} & .23 & .20 & .25 & \textbf{.25} && 0.5 & {\color{ForestGreen}\textbf{.29}} & .25 & .25 & .27 & \textbf{.28} && 0.5 & \textbf{.29} & .27 & .28 & .29 & {\color{ForestGreen}\textbf{.29}} \\
        & 1 & {\color{ForestGreen}\textbf{.43}} & .39 & .39 & .41 & \textbf{.42} && 1 & .44 & .42 & .43 & \textbf{.44} & {\color{ForestGreen}\textbf{.45}} && 1 & .44 & .44 & .45 & \textbf{.46} & {\color{ForestGreen}\textbf{.46}} \\
        & 2 & \textbf{.58} & .56 & .57 & .58 & {\color{ForestGreen}\textbf{.59}} && 2 & .59 & .59 & .60 & \textbf{.61} & {\color{ForestGreen}\textbf{.62}} && 2 & .59 & .61 & .63 & \textbf{.63} & {\color{ForestGreen}\textbf{.64}} \\
      \rowcolor{RowAlt}
      \quad Mixture & 0.5 & {\color{ForestGreen}\textbf{.25}} & .22 & .20 & .24 & \textbf{.24} && 0.5 & \textbf{.26} & .25 & .24 & .26 & {\color{ForestGreen}\textbf{.26}} && 0.5 & .27 & .26 & .27 & \textbf{.28} & {\color{ForestGreen}\textbf{.29}} \\
      \rowcolor{RowAlt}
        & 1 & .39 & .38 & .37 & \textbf{.40} & {\color{ForestGreen}\textbf{.40}} && 1 & .40 & .41 & .40 & \textbf{.42} & {\color{ForestGreen}\textbf{.43}} && 1 & .41 & .43 & .43 & \textbf{.45} & {\color{ForestGreen}\textbf{.45}} \\
      \rowcolor{RowAlt}
        & 2 & .53 & .54 & .54 & \textbf{.56} & {\color{ForestGreen}\textbf{.56}} && 2 & .54 & .57 & .57 & \textbf{.59} & {\color{ForestGreen}\textbf{.60}} && 2 & .55 & .59 & .60 & \textbf{.62} & {\color{ForestGreen}\textbf{.62}} \\
      \quad Friedman 1 & 0.5 & \textbf{.23} & .22 & .19 & .22 & {\color{ForestGreen}\textbf{.23}} && 0.5 & .23 & .23 & .23 & \textbf{.25} & {\color{ForestGreen}\textbf{.26}} && 0.5 & .24 & .26 & .27 & \textbf{.28} & {\color{ForestGreen}\textbf{.28}} \\
        & 1 & .36 & .37 & .35 & \textbf{.37} & {\color{ForestGreen}\textbf{.39}} && 1 & .35 & .39 & .39 & \textbf{.41} & {\color{ForestGreen}\textbf{.42}} && 1 & .37 & .42 & .43 & \textbf{.45} & {\color{ForestGreen}\textbf{.45}} \\
        & 2 & .49 & .52 & .52 & \textbf{.53} & {\color{ForestGreen}\textbf{.55}} && 2 & .48 & .55 & .56 & \textbf{.57} & {\color{ForestGreen}\textbf{.59}} && 2 & .50 & .59 & .59 & \textbf{.61} & {\color{ForestGreen}\textbf{.62}} \\

    \addlinespace[0.15em]
    \multicolumn{21}{l}{\textcolor{dpd2}{\textbf{Nonlinear}}} \\
    \addlinespace[0.1em]

      \rowcolor{RowAlt}
      \quad {\color{blue}Trig-Log} & 0.5 & .00 & \textbf{.11} & .04 & .10 & {\color{ForestGreen}\textbf{.13}} && 0.5 & .01 & .14 & .10 & \textbf{.17} & {\color{ForestGreen}\textbf{.18}} && 0.5 & .02 & .20 & .15 & \textbf{.21} & {\color{ForestGreen}\textbf{.21}} \\
      \rowcolor{RowAlt}
        & 1 & .02 & .20 & .14 & \textbf{.20} & {\color{ForestGreen}\textbf{.25}} && 1 & .03 & .26 & .20 & \textbf{.31} & {\color{ForestGreen}\textbf{.31}} && 1 & .04 & .34 & .25 & {\color{ForestGreen}\textbf{.37}} & .34 \\
      \rowcolor{RowAlt}
        & 2 & .03 & .30 & .25 & \textbf{.32} & {\color{ForestGreen}\textbf{.37}} && 2 & .05 & .38 & .31 & \textbf{.47} & {\color{ForestGreen}\textbf{.45}} && 2 & .06 & .47 & .35 & {\color{ForestGreen}\textbf{.52}} & .48 \\
      \quad Soft Radial & 0.5 & $-$.02 & \textbf{.19} & .14 & .19 & {\color{ForestGreen}\textbf{.20}} && 0.5 & $-$.01 & .22 & .19 & \textbf{.24} & {\color{ForestGreen}\textbf{.26}} && 0.5 & $-$.01 & .25 & .23 & \textbf{.27} & {\color{ForestGreen}\textbf{.28}} \\
        & 1 & $-$.02 & .32 & .28 & \textbf{.34} & {\color{ForestGreen}\textbf{.36}} && 1 & $-$.01 & .36 & .32 & \textbf{.41} & {\color{ForestGreen}\textbf{.41}} && 1 & $-$.01 & .40 & .36 & \textbf{.44} & {\color{ForestGreen}\textbf{.44}} \\
        & 2 & $-$.02 & .44 & .42 & \textbf{.50} & {\color{ForestGreen}\textbf{.51}} && 2 & $-$.02 & .50 & .46 & \textbf{.57} & {\color{ForestGreen}\textbf{.58}} && 2 & $-$.01 & .56 & .50 & {\color{ForestGreen}\textbf{.61}} & .60 \\
      \rowcolor{RowAlt}
      \quad Step/U-curve & 0.5 & .08 & .19 & .18 & \textbf{.21} & {\color{ForestGreen}\textbf{.22}} && 0.5 & .09 & .23 & .24 & \textbf{.25} & {\color{ForestGreen}\textbf{.26}} && 0.5 & .10 & .24 & .26 & \textbf{.27} & {\color{ForestGreen}\textbf{.28}} \\
      \rowcolor{RowAlt}
        & 1 & .14 & .34 & .35 & \textbf{.37} & {\color{ForestGreen}\textbf{.38}} && 1 & .14 & .38 & .41 & \textbf{.41} & {\color{ForestGreen}\textbf{.42}} && 1 & .15 & .40 & .43 & \textbf{.44} & {\color{ForestGreen}\textbf{.45}} \\
      \rowcolor{RowAlt}
        & 2 & .19 & .49 & .53 & \textbf{.54} & {\color{ForestGreen}\textbf{.55}} && 2 & .20 & .54 & .58 & \textbf{.58} & {\color{ForestGreen}\textbf{.60}} && 2 & .20 & .57 & .61 & \textbf{.61} & {\color{ForestGreen}\textbf{.62}} \\

    \addlinespace[0.15em]
    \multicolumn{21}{l}{\textcolor{dpd2}{\textbf{Highly Nonlinear}}} \\
    \addlinespace[0.1em]

      \quad Rotated Sine & 0.5 & .08 & {\color{ForestGreen}\textbf{.10}} & .01 & \textbf{.08} & .07 && 0.5 & .09 & {\color{ForestGreen}\textbf{.14}} & .07 & \textbf{.12} & .09 && 0.5 & .09 & {\color{ForestGreen}\textbf{.19}} & .11 & \textbf{.15} & .12 \\
        & 1 & .13 & {\color{ForestGreen}\textbf{.19}} & .10 & \textbf{.14} & .13 && 1 & .14 & {\color{ForestGreen}\textbf{.25}} & .15 & \textbf{.21} & .18 && 1 & .13 & {\color{ForestGreen}\textbf{.32}} & .20 & \textbf{.26} & .21 \\
        & 2 & .18 & {\color{ForestGreen}\textbf{.29}} & .19 & \textbf{.21} & .20 && 2 & .19 & {\color{ForestGreen}\textbf{.37}} & .25 & \textbf{.32} & .26 && 2 & .18 & {\color{ForestGreen}\textbf{.45}} & .29 & \textbf{.39} & .31 \\

    \addlinespace[0.1em]
    \bottomrule
  \end{tabular}%
  }

  \vspace{0.35em}
  \parbox{\textwidth}{\fontfamily{phv}\selectfont\scriptsize
    \textit{Notes}: Out-of-sample $R^2$ averaged over 10 replications. Test set size: $N_{\text{test}}=1000$. LGB = standard LightGBM. LGB+ = competition variant (without calibration); LGB$^{\texttt{A}}$+ = alternating variant.
    Best: \textcolor{ForestGreen}{\textbf{green bold}}; second-best: \textbf{bold}.
  }
  }

\end{table}

%% file: 03_tables/gdp__h1.tex
% ------------------------------------------------------------
%  EW VERSION - THREE-PANEL TABLE (SE / AE / Classical)
%  Template: dpd2 color + RowAlt alternating rows
% ------------------------------------------------------------

\begin{landscape}
\begin{table}[t!]
  \centering
  \caption{\normalsize GDP ($h=1$)}
  \vspace*{-0.65em}
  \label{tab:gdp_h1}

  {\fontfamily{phv}\selectfont
  \resizebox{\linewidth}{!}{%
  \scriptsize
  \setlength{\tabcolsep}{0.3em}
  \renewcommand{\arraystretch}{1.65}
  \begin{tabular}{l r r r r r r r r r r c r r r r r r r r r r}
    \toprule

    & \multicolumn{10}{c}{\textcolor{dpd2}{\textbf{2007Q2 -- 2019Q4}}}
    && \multicolumn{10}{c}{\textcolor{dpd2}{\textbf{2022Q1 -- 2025Q1}}}
    \\
    \cmidrule(lr){2-11}
    \cmidrule(lr){13-22}
    & FAAR & RF & LGB & LGB$^{\texttt{A}}$+ & LGB+ & KRR & NN & RR & SPF & TabPFN 
    && FAAR & RF & LGB & LGB$^{\texttt{A}}$+ & LGB+ & KRR & NN & RR & SPF & TabPFN \\

    \midrule
    \addlinespace[0.5em]
    \multicolumn{22}{l}{\textcolor{dpd2}{\textbf{Panel A: Squared Error}}} \\
    \addlinespace[0.3em]

      Return
      & -0.01 & 0.13 & -0.05 & 0.17 & 0.17 & {\color{ForestGreen}\textbf{0.29}} & {\textbf{0.21}} & 0.17 & 0.19 & 0.20 &  & -1.21 & -0.12 & -0.96 & -0.29 & -0.14 & -0.23 & -0.31 & {\color{ForestGreen}\textbf{0.11}} & -0.45 & {\textbf{-0.05}} \\
      \rowcolor{RowAlt}
      Sharpe
      & -0.01 & 0.24 & -0.05 & 0.27 & 0.29 & {\textbf{0.44}} & 0.30 & 0.39 & {\color{ForestGreen}\textbf{0.47}} & 0.35 &  & -0.83 & -0.27 & -0.95 & -0.56 & -0.66 & -0.37 & -0.40 & {\color{ForestGreen}\textbf{0.31}} & -0.73 & {\textbf{-0.11}} \\
      Sortino
      & -0.02 & 0.50 & -0.09 & 0.74 & 1.11 & {\color{ForestGreen}\textbf{2.85}} & 1.09 & 1.01 & {\textbf{1.80}} & 1.18 &  & -0.82 & -0.38 & -0.95 & -0.64 & -0.75 & -0.43 & -0.44 & {\color{ForestGreen}\textbf{0.62}} & -0.79 & {\textbf{-0.14}} \\
      \rowcolor{RowAlt}
      Omega
      & 0.98 & 1.69 & 0.89 & 1.87 & 2.13 & {\color{ForestGreen}\textbf{4.03}} & 2.22 & 2.61 & {\textbf{3.78}} & 2.40 &  & 0.19 & 0.69 & 0.25 & 0.45 & 0.41 & 0.57 & 0.50 & {\color{ForestGreen}\textbf{1.57}} & 0.38 & {\textbf{0.86}} \\
      MaxDD
      & -13.63 & -0.49 & -12.20 & -0.61 & -0.34 & {\color{ForestGreen}\textbf{-0.11}} & -0.22 & -0.24 & {\textbf{-0.13}} & -0.26 &  & -17.99 & -4.21 & -14.92 & -5.47 & {\textbf{-1.45}} & -6.20 & -7.81 & {\color{ForestGreen}\textbf{-0.67}} & -8.06 & -3.59 \\
      \rowcolor{RowAlt}
      Edge
      & {\textbf{0.85}} & 0.00 & {\color{ForestGreen}\textbf{0.93}} & 0.17 & 0.00 & 0.02 & 0.10 & 0.09 & 0.08 & 0.12 &  & 0.05 & 0.00 & {\textbf{0.11}} & 0.00 & 0.00 & {\color{ForestGreen}\textbf{0.12}} & 0.00 & 0.00 & 0.08 & 0.05 \\

    \addlinespace[0.5em]
    \midrule
    \addlinespace[0.5em]
    \multicolumn{22}{l}{\textcolor{dpd2}{\textbf{Panel B: Absolute Error}}} \\
    \addlinespace[0.3em]

      Return
      & -0.12 & 0.01 & -0.12 & -0.01 & 0.02 & {\textbf{0.11}} & 0.04 & {\color{ForestGreen}\textbf{0.12}} & 0.07 & 0.04 &  & -0.63 & -0.23 & -0.66 & -0.27 & -0.13 & -0.29 & -0.31 & {\color{ForestGreen}\textbf{-0.10}} & -0.34 & {\textbf{-0.11}} \\
      \rowcolor{RowAlt}
      Sharpe
      & -0.25 & 0.05 & -0.28 & -0.03 & 0.08 & 0.40 & 0.13 & {\color{ForestGreen}\textbf{0.60}} & {\textbf{0.42}} & 0.16 &  & -0.92 & -0.61 & -1.17 & -0.64 & -0.58 & -0.60 & -0.59 & {\textbf{-0.34}} & -0.68 & {\color{ForestGreen}\textbf{-0.25}} \\
      Sortino
      & -0.32 & 0.07 & -0.34 & -0.04 & 0.13 & {\textbf{0.88}} & 0.23 & {\color{ForestGreen}\textbf{1.25}} & 0.76 & 0.24 &  & -0.89 & -0.69 & -1.10 & -0.73 & -0.66 & -0.65 & -0.62 & {\textbf{-0.41}} & -0.80 & {\color{ForestGreen}\textbf{-0.32}} \\
      \rowcolor{RowAlt}
      Omega
      & 0.72 & 1.07 & 0.66 & 0.96 & 1.12 & 1.81 & 1.23 & {\color{ForestGreen}\textbf{2.27}} & {\textbf{1.84}} & 1.24 &  & 0.20 & 0.45 & 0.21 & 0.44 & 0.45 & 0.43 & 0.39 & {\textbf{0.64}} & 0.45 & {\color{ForestGreen}\textbf{0.71}} \\
      MaxDD
      & -10.57 & -3.48 & -6.95 & -3.39 & -0.69 & {\textbf{-0.26}} & -0.52 & {\color{ForestGreen}\textbf{-0.23}} & -0.27 & -0.90 &  & -8.91 & -4.62 & -9.41 & -5.01 & {\color{ForestGreen}\textbf{-2.24}} & -5.64 & -6.09 & {\textbf{-3.28}} & -6.31 & -3.95 \\
      \rowcolor{RowAlt}
      Edge
      & {\color{ForestGreen}\textbf{1.10}} & 0.05 & {\textbf{0.47}} & 0.12 & 0.02 & 0.12 & 0.27 & 0.42 & 0.16 & 0.23 &  & 0.22 & 0.00 & 0.05 & 0.00 & 0.00 & 0.13 & 0.00 & 0.00 & {\color{ForestGreen}\textbf{0.34}} & {\textbf{0.29}} \\

    \addlinespace[0.5em]
    \midrule
    \addlinespace[0.5em]
    \multicolumn{22}{l}{\textcolor{dpd2}{\textbf{Panel C: Classical Forecast Accuracy}}} \\
    \addlinespace[0.3em]

      RMSE
      & 1.00 & 0.93 & 1.02 & 0.91 & 0.91 & {\color{ForestGreen}\textbf{0.84}} & {\textbf{0.89}} & 0.91 & 0.90 & 0.90 &  & 1.49 & 1.06 & 1.40 & 1.13 & 1.07 & 1.11 & 1.14 & {\color{ForestGreen}\textbf{0.95}} & 1.20 & {\textbf{1.02}} \\
      \rowcolor{RowAlt}
      MAE
      & 1.12 & 0.99 & 1.12 & 1.01 & 0.98 & {\textbf{0.89}} & 0.96 & {\color{ForestGreen}\textbf{0.88}} & 0.93 & 0.96 &  & 1.63 & 1.23 & 1.66 & 1.27 & 1.13 & 1.29 & 1.31 & {\color{ForestGreen}\textbf{1.10}} & 1.34 & {\textbf{1.11}} \\
      $\rho(1)$
      & 0.39 & 0.23 & {\color{ForestGreen}\textbf{0.07}} & {\textbf{0.11}} & 0.22 & 0.15 & 0.15 & 0.32 & 0.25 & 0.19 &  & 0.65 & 0.46 & 0.53 & 0.43 & 0.42 & 0.45 & 0.49 & {\textbf{0.39}} & 0.56 & {\color{ForestGreen}\textbf{0.38}} \\
      \rowcolor{RowAlt}
      DM $t$-stat
      & -0.03 & 1.04 & -0.23 & 0.92 & 1.03 & 1.41 & 0.95 & {\textbf{1.48}} & {\color{ForestGreen}\textbf{1.56}} & 1.24 &  & -1.26 & -0.50 & -1.84 & -1.01 & -1.19 & -0.55 & -0.59 & {\color{ForestGreen}\textbf{0.48}} & -1.32 & {\textbf{-0.19}} \\

    \addlinespace[0.3em]
    \bottomrule
  \end{tabular}%
  }

  \vspace{0.25em}
  \parbox{\linewidth}{\scriptsize
    \textit{Notes}: Panels A--B report risk-adjusted metrics; Panel C reports classical forecast accuracy metrics. $\rho(1)$ = first-order autocorrelation of errors.
    Best: \textcolor{ForestGreen}{\textbf{bold green}}; second-best: \textbf{bold}.
  }
  }

\end{table}
\end{landscape}

%% file: 03_tables/gdp__h2.tex
% ------------------------------------------------------------
%  EW VERSION - THREE-PANEL TABLE (SE / AE / Classical)
%  Template: dpd2 color + RowAlt alternating rows
% ------------------------------------------------------------

\begin{landscape}
\begin{table}[t!]
  \centering
  \caption{\normalsize GDP ($h=2$)}
  \vspace*{-0.65em}
  \label{tab:gdp_h2}

  {\fontfamily{phv}\selectfont
  \resizebox{\linewidth}{!}{%
  \scriptsize
  \setlength{\tabcolsep}{0.3em}
  \renewcommand{\arraystretch}{1.65}
  \begin{tabular}{l r r r r r r r r r r c r r r r r r r r r r}
    \toprule

    & \multicolumn{10}{c}{\textcolor{dpd2}{\textbf{2007Q2 -- 2019Q4}}}
    && \multicolumn{10}{c}{\textcolor{dpd2}{\textbf{2022Q1 -- 2025Q1}}}
    \\
    \cmidrule(lr){2-11}
    \cmidrule(lr){13-22}
    & FAAR & RF & LGB & LGB$^{\texttt{A}}$+ & LGB+ & KRR & NN & RR & SPF & TabPFN 
    && FAAR & RF & LGB & LGB$^{\texttt{A}}$+ & LGB+ & KRR & NN & RR & SPF & TabPFN \\

    \midrule
    \addlinespace[0.5em]
    \multicolumn{22}{l}{\textcolor{dpd2}{\textbf{Panel A: Squared Error}}} \\
    \addlinespace[0.3em]

      Return
      & -1.06 & {\textbf{0.13}} & 0.03 & -0.19 & -0.22 & -0.09 & -0.16 & 0.05 & {\color{ForestGreen}\textbf{0.13}} & 0.03 &  & -0.90 & -0.47 & -0.67 & -0.57 & -0.37 & -0.32 & -0.64 & {\color{ForestGreen}\textbf{0.13}} & -0.73 & {\textbf{-0.13}} \\
      \rowcolor{RowAlt}
      Sharpe
      & -0.80 & {\textbf{0.28}} & 0.04 & -0.25 & -0.31 & -0.25 & -0.40 & 0.22 & {\color{ForestGreen}\textbf{0.42}} & 0.11 &  & -1.06 & -0.95 & -0.86 & -0.79 & -0.76 & -0.39 & -0.60 & {\color{ForestGreen}\textbf{0.31}} & -0.84 & {\textbf{-0.20}} \\
      Sortino
      & -0.80 & {\textbf{0.76}} & 0.10 & -0.28 & -0.33 & -0.30 & -0.44 & 0.43 & {\color{ForestGreen}\textbf{1.46}} & 0.16 &  & -1.05 & -0.92 & -0.83 & -0.82 & -0.79 & -0.50 & -0.66 & {\color{ForestGreen}\textbf{0.62}} & -0.86 & {\textbf{-0.26}} \\
      \rowcolor{RowAlt}
      Omega
      & 0.23 & {\textbf{2.02}} & 1.11 & 0.56 & 0.43 & 0.65 & 0.47 & 1.51 & {\color{ForestGreen}\textbf{2.95}} & 1.19 &  & 0.21 & 0.20 & 0.20 & 0.32 & 0.30 & 0.63 & 0.43 & {\color{ForestGreen}\textbf{1.52}} & 0.29 & {\textbf{0.78}} \\
      MaxDD
      & -59.81 & {\color{ForestGreen}\textbf{-0.31}} & -9.86 & -16.08 & -15.15 & -6.95 & -10.71 & -0.57 & {\textbf{-0.55}} & -0.74 &  & -12.61 & -6.54 & -10.78 & -9.63 & {\textbf{-3.65}} & -9.93 & -13.15 & {\color{ForestGreen}\textbf{-0.71}} & -9.10 & -6.39 \\
      \rowcolor{RowAlt}
      Edge
      & {\textbf{0.90}} & 0.07 & {\color{ForestGreen}\textbf{1.30}} & 0.06 & 0.01 & 0.01 & 0.03 & 0.02 & 0.08 & 0.00 &  & {\color{ForestGreen}\textbf{0.35}} & 0.00 & 0.00 & 0.00 & 0.01 & {\textbf{0.28}} & 0.00 & 0.00 & 0.06 & 0.00 \\

    \addlinespace[0.5em]
    \midrule
    \addlinespace[0.5em]
    \multicolumn{22}{l}{\textcolor{dpd2}{\textbf{Panel B: Absolute Error}}} \\
    \addlinespace[0.3em]

      Return
      & -0.57 & {\textbf{0.04}} & -0.09 & -0.08 & -0.10 & -0.04 & -0.09 & 0.03 & {\color{ForestGreen}\textbf{0.09}} & 0.02 &  & -0.54 & -0.35 & -0.53 & -0.43 & {\textbf{-0.20}} & -0.34 & -0.46 & {\color{ForestGreen}\textbf{-0.07}} & -0.36 & -0.21 \\
      \rowcolor{RowAlt}
      Sharpe
      & -0.85 & {\textbf{0.19}} & -0.24 & -0.21 & -0.29 & -0.15 & -0.30 & 0.19 & {\color{ForestGreen}\textbf{0.52}} & 0.12 &  & -0.97 & -0.98 & -1.34 & -0.98 & -0.73 & -0.60 & -0.75 & {\color{ForestGreen}\textbf{-0.26}} & -0.71 & {\textbf{-0.48}} \\
      Sortino
      & -0.87 & 0.29 & -0.32 & -0.24 & -0.32 & -0.21 & -0.40 & {\textbf{0.30}} & {\color{ForestGreen}\textbf{0.86}} & 0.17 &  & -1.01 & -0.96 & -1.16 & -0.97 & -0.79 & -0.72 & -0.80 & {\color{ForestGreen}\textbf{-0.36}} & -0.79 & {\textbf{-0.59}} \\
      \rowcolor{RowAlt}
      Omega
      & 0.31 & {\textbf{1.31}} & 0.72 & 0.72 & 0.60 & 0.82 & 0.67 & 1.27 & {\color{ForestGreen}\textbf{1.91}} & 1.16 &  & 0.27 & 0.23 & 0.11 & 0.29 & 0.38 & 0.51 & 0.39 & {\color{ForestGreen}\textbf{0.74}} & 0.42 & {\textbf{0.57}} \\
      MaxDD
      & -33.45 & {\textbf{-0.68}} & -6.90 & -8.72 & -7.47 & -3.97 & -6.16 & -0.72 & {\color{ForestGreen}\textbf{-0.31}} & -0.93 &  & -8.25 & -5.03 & -7.74 & -7.10 & {\textbf{-3.18}} & -7.73 & -8.70 & {\color{ForestGreen}\textbf{-3.15}} & -5.94 & -5.50 \\
      \rowcolor{RowAlt}
      Edge
      & {\color{ForestGreen}\textbf{1.01}} & 0.09 & {\textbf{0.51}} & 0.24 & 0.09 & 0.07 & 0.19 & 0.03 & 0.25 & 0.01 &  & {\color{ForestGreen}\textbf{0.73}} & 0.00 & 0.00 & 0.00 & 0.10 & 0.20 & 0.00 & 0.00 & {\textbf{0.25}} & 0.03 \\

    \addlinespace[0.5em]
    \midrule
    \addlinespace[0.5em]
    \multicolumn{22}{l}{\textcolor{dpd2}{\textbf{Panel C: Classical Forecast Accuracy}}} \\
    \addlinespace[0.3em]

      RMSE
      & 1.43 & {\textbf{0.93}} & 0.98 & 1.09 & 1.10 & 1.04 & 1.08 & 0.97 & {\color{ForestGreen}\textbf{0.93}} & 0.99 &  & 1.38 & 1.21 & 1.29 & 1.25 & 1.17 & 1.15 & 1.28 & {\color{ForestGreen}\textbf{0.93}} & 1.31 & {\textbf{1.06}} \\
      \rowcolor{RowAlt}
      MAE
      & 1.57 & {\textbf{0.96}} & 1.09 & 1.08 & 1.10 & 1.04 & 1.09 & 0.97 & {\color{ForestGreen}\textbf{0.91}} & 0.98 &  & 1.54 & 1.35 & 1.53 & 1.43 & {\textbf{1.20}} & 1.34 & 1.46 & {\color{ForestGreen}\textbf{1.07}} & 1.36 & 1.21 \\
      $\rho(1)$
      & {\textbf{0.09}} & 0.32 & 0.30 & {\color{ForestGreen}\textbf{0.08}} & 0.14 & 0.49 & 0.45 & 0.47 & 0.42 & 0.41 &  & {\color{ForestGreen}\textbf{0.29}} & 0.58 & 0.50 & 0.52 & 0.47 & 0.56 & 0.57 & {\textbf{0.43}} & 0.57 & 0.48 \\
      \rowcolor{RowAlt}
      DM $t$-stat
      & -2.30 & {\textbf{1.00}} & 0.15 & -0.69 & -0.84 & -0.77 & -1.49 & 0.78 & {\color{ForestGreen}\textbf{1.65}} & 0.49 &  & -1.93 & -1.38 & -1.21 & -1.09 & -1.38 & -0.60 & -0.92 & {\color{ForestGreen}\textbf{0.45}} & -1.40 & {\textbf{-0.29}} \\

    \addlinespace[0.3em]
    \bottomrule
  \end{tabular}%
  }

  \vspace{0.25em}
  \parbox{\linewidth}{\scriptsize
    \textit{Notes}: Panels A--B report risk-adjusted metrics; Panel C reports classical forecast accuracy metrics. $\rho(1)$ = first-order autocorrelation of errors.
    Best: \textcolor{ForestGreen}{\textbf{bold green}}; second-best: \textbf{bold}.
  }
  }

\end{table}
\end{landscape}

%% file: 03_tables/gdp__h4.tex
% ------------------------------------------------------------
%  EW VERSION - THREE-PANEL TABLE (SE / AE / Classical)
%  Template: dpd2 color + RowAlt alternating rows
% ------------------------------------------------------------

\begin{landscape}
\begin{table}[t!]
  \centering
  \caption{\normalsize GDP ($h=4$)}
  \vspace*{-0.65em}
  \label{tab:gdp_h4}

  {\fontfamily{phv}\selectfont
  \resizebox{\linewidth}{!}{%
  \scriptsize
  \setlength{\tabcolsep}{0.3em}
  \renewcommand{\arraystretch}{1.65}
  \begin{tabular}{l r r r r r r r r r r c r r r r r r r r r r}
    \toprule

    & \multicolumn{10}{c}{\textcolor{dpd2}{\textbf{2007Q2 -- 2019Q4}}}
    && \multicolumn{10}{c}{\textcolor{dpd2}{\textbf{2022Q1 -- 2025Q1}}}
    \\
    \cmidrule(lr){2-11}
    \cmidrule(lr){13-22}
    & FAAR & RF & LGB & LGB$^{\texttt{A}}$+ & LGB+ & KRR & NN & RR & SPF & TabPFN 
    && FAAR & RF & LGB & LGB$^{\texttt{A}}$+ & LGB+ & KRR & NN & RR & SPF & TabPFN \\

    \midrule
    \addlinespace[0.5em]
    \multicolumn{22}{l}{\textcolor{dpd2}{\textbf{Panel A: Squared Error}}} \\
    \addlinespace[0.3em]

      Return
      & -0.45 & 0.01 & -0.04 & -0.29 & -0.09 & 0.11 & 0.10 & 0.05 & {\textbf{0.14}} & {\color{ForestGreen}\textbf{0.16}} &  & -0.06 & 0.57 & 0.23 & {\textbf{0.69}} & 0.62 & 0.56 & 0.64 & 0.57 & 0.43 & {\color{ForestGreen}\textbf{0.72}} \\
      \rowcolor{RowAlt}
      Sharpe
      & -0.75 & 0.04 & -0.15 & -0.80 & -0.50 & {\textbf{0.46}} & 0.21 & 0.24 & {\color{ForestGreen}\textbf{0.64}} & 0.44 &  & -0.53 & 0.46 & 0.17 & 0.50 & {\textbf{0.57}} & 0.49 & 0.46 & 0.49 & 0.46 & {\color{ForestGreen}\textbf{0.57}} \\
      Sortino
      & -0.72 & 0.06 & -0.21 & -0.78 & -0.54 & {\textbf{1.58}} & 0.44 & 0.51 & {\color{ForestGreen}\textbf{1.97}} & 1.28 &  & -0.58 & 2.67 & 0.32 & 4.50 & {\color{ForestGreen}\textbf{8.50}} & 4.30 & 2.47 & 4.84 & 2.25 & {\textbf{6.78}} \\
      \rowcolor{RowAlt}
      Omega
      & 0.16 & 1.07 & 0.78 & 0.25 & 0.43 & {\textbf{2.65}} & 1.56 & 1.60 & {\color{ForestGreen}\textbf{3.65}} & 2.51 &  & 0.38 & 3.21 & 1.43 & 4.23 & {\color{ForestGreen}\textbf{8.55}} & 4.53 & 3.29 & 5.16 & 3.15 & {\textbf{6.41}} \\
      MaxDD
      & -24.38 & -0.91 & -5.64 & -14.63 & -5.15 & {\color{ForestGreen}\textbf{-0.12}} & -0.48 & -0.51 & {\textbf{-0.19}} & -0.21 &  & -0.68 & -0.29 & -0.71 & -0.21 & {\color{ForestGreen}\textbf{-0.10}} & -0.20 & -0.28 & -0.17 & -0.23 & {\textbf{-0.10}} \\
      \rowcolor{RowAlt}
      Edge
      & 0.04 & 0.00 & 0.30 & 0.09 & 0.04 & 0.01 & {\color{ForestGreen}\textbf{1.18}} & 0.00 & {\textbf{1.10}} & 0.55 &  & 0.02 & {\textbf{0.13}} & 0.00 & {\color{ForestGreen}\textbf{0.20}} & 0.03 & 0.00 & 0.00 & 0.00 & 0.13 & 0.00 \\

    \addlinespace[0.5em]
    \midrule
    \addlinespace[0.5em]
    \multicolumn{22}{l}{\textcolor{dpd2}{\textbf{Panel B: Absolute Error}}} \\
    \addlinespace[0.3em]

      Return
      & -0.33 & 0.03 & -0.05 & -0.16 & -0.03 & {\textbf{0.07}} & -0.02 & 0.02 & {\color{ForestGreen}\textbf{0.10}} & 0.06 &  & -0.05 & 0.06 & -0.09 & 0.16 & {\textbf{0.20}} & 0.02 & 0.12 & 0.04 & 0.09 & {\color{ForestGreen}\textbf{0.22}} \\
      \rowcolor{RowAlt}
      Sharpe
      & -0.97 & 0.14 & -0.19 & -0.57 & -0.18 & {\textbf{0.42}} & -0.06 & 0.15 & {\color{ForestGreen}\textbf{0.72}} & 0.25 &  & -0.34 & 0.09 & -0.13 & 0.20 & {\color{ForestGreen}\textbf{0.43}} & 0.04 & 0.15 & 0.07 & 0.20 & {\textbf{0.36}} \\
      Sortino
      & -0.91 & 0.21 & -0.24 & -0.64 & -0.24 & {\textbf{0.71}} & -0.08 & 0.22 & {\color{ForestGreen}\textbf{1.41}} & 0.39 &  & -0.41 & 0.17 & -0.19 & 0.54 & {\color{ForestGreen}\textbf{1.31}} & 0.09 & 0.34 & 0.17 & 0.34 & {\textbf{1.02}} \\
      \rowcolor{RowAlt}
      Omega
      & 0.20 & 1.20 & 0.78 & 0.47 & 0.79 & {\textbf{1.70}} & 0.92 & 1.21 & {\color{ForestGreen}\textbf{2.46}} & 1.40 &  & 0.63 & 1.13 & 0.83 & 1.38 & {\color{ForestGreen}\textbf{2.14}} & 1.06 & 1.26 & 1.12 & 1.29 & {\textbf{1.76}} \\
      MaxDD
      & -17.34 & -0.75 & -5.11 & -7.84 & -2.65 & {\textbf{-0.29}} & -0.98 & -0.70 & {\color{ForestGreen}\textbf{-0.17}} & -0.46 &  & -1.20 & -4.59 & -5.82 & -0.78 & {\color{ForestGreen}\textbf{-0.42}} & -0.92 & -0.92 & -0.78 & -0.68 & {\textbf{-0.55}} \\
      \rowcolor{RowAlt}
      Edge
      & 0.14 & 0.06 & 0.39 & 0.30 & 0.23 & 0.02 & 0.28 & 0.01 & {\color{ForestGreen}\textbf{1.23}} & {\textbf{0.47}} &  & 0.24 & 0.21 & 0.00 & {\textbf{0.31}} & 0.12 & 0.00 & 0.00 & 0.00 & {\color{ForestGreen}\textbf{0.58}} & 0.01 \\

    \addlinespace[0.5em]
    \midrule
    \addlinespace[0.5em]
    \multicolumn{22}{l}{\textcolor{dpd2}{\textbf{Panel C: Classical Forecast Accuracy}}} \\
    \addlinespace[0.3em]

      RMSE
      & 1.20 & 1.00 & 1.02 & 1.14 & 1.04 & 0.94 & 0.95 & 0.97 & {\textbf{0.93}} & {\color{ForestGreen}\textbf{0.92}} &  & 1.03 & 0.65 & 0.88 & {\textbf{0.56}} & 0.61 & 0.67 & 0.60 & 0.66 & 0.76 & {\color{ForestGreen}\textbf{0.53}} \\
      \rowcolor{RowAlt}
      MAE
      & 1.33 & 0.97 & 1.05 & 1.16 & 1.03 & {\textbf{0.93}} & 1.02 & 0.98 & {\color{ForestGreen}\textbf{0.90}} & 0.94 &  & 1.05 & 0.94 & 1.09 & 0.84 & {\textbf{0.80}} & 0.98 & 0.88 & 0.96 & 0.91 & {\color{ForestGreen}\textbf{0.78}} \\
      $\rho(1)$
      & 0.57 & 0.53 & 0.52 & 0.59 & 0.58 & 0.52 & {\textbf{0.50}} & 0.55 & 0.53 & {\color{ForestGreen}\textbf{0.36}} &  & {\color{ForestGreen}\textbf{-0.01}} & 0.31 & 0.29 & 0.44 & {\textbf{0.20}} & 0.41 & 0.47 & 0.42 & 0.34 & 0.29 \\
      \rowcolor{RowAlt}
      DM $t$-stat
      & -2.02 & 0.16 & -0.55 & -1.99 & -1.90 & {\textbf{1.63}} & 0.54 & 0.90 & {\color{ForestGreen}\textbf{2.57}} & 1.22 &  & -0.97 & 0.83 & 0.30 & 0.90 & {\textbf{1.03}} & 0.89 & 0.82 & 0.88 & 0.85 & {\color{ForestGreen}\textbf{1.03}} \\

    \addlinespace[0.3em]
    \bottomrule
  \end{tabular}%
  }

  \vspace{0.25em}
  \parbox{\linewidth}{\scriptsize
    \textit{Notes}: Panels A--B report risk-adjusted metrics; Panel C reports classical forecast accuracy metrics. $\rho(1)$ = first-order autocorrelation of errors.
    Best: \textcolor{ForestGreen}{\textbf{bold green}}; second-best: \textbf{bold}.
  }
  }

\end{table}
\end{landscape}

%% file: 03_tables/ur__h1.tex
% ------------------------------------------------------------
%  EW VERSION - THREE-PANEL TABLE (SE / AE / Classical)
%  Template: dpd2 color + RowAlt alternating rows
% ------------------------------------------------------------

\begin{landscape}
\begin{table}[t!]
  \centering
  \caption{\normalsize Unemployment Rate ($h=1$)}
  \vspace*{-0.65em}
  \label{tab:ur_h1}

  {\fontfamily{phv}\selectfont
  \resizebox{\linewidth}{!}{%
  \scriptsize
  \setlength{\tabcolsep}{0.3em}
  \renewcommand{\arraystretch}{1.65}
  \begin{tabular}{l r r r r r r r r r r c r r r r r r r r r r}
    \toprule

    & \multicolumn{10}{c}{\textcolor{dpd2}{\textbf{2007Q2 -- 2019Q4}}}
    && \multicolumn{10}{c}{\textcolor{dpd2}{\textbf{2022Q1 -- 2025Q1}}}
    \\
    \cmidrule(lr){2-11}
    \cmidrule(lr){13-22}
    & FAAR & RF & LGB & LGB$^{\texttt{A}}$+ & LGB+ & KRR & NN & RR & SPF & TabPFN 
    && FAAR & RF & LGB & LGB$^{\texttt{A}}$+ & LGB+ & KRR & NN & RR & SPF & TabPFN \\

    \midrule
    \addlinespace[0.5em]
    \multicolumn{22}{l}{\textcolor{dpd2}{\textbf{Panel A: Squared Error}}} \\
    \addlinespace[0.3em]

      Return
      & 0.21 & 0.27 & 0.26 & {\textbf{0.54}} & {\color{ForestGreen}\textbf{0.55}} & -0.67 & 0.11 & -0.07 & -0.14 & 0.29 &  & -2.77 & -1.05 & -1.27 & -0.99 & -0.94 & -8.64 & -1.00 & {\textbf{-0.30}} & -0.31 & {\color{ForestGreen}\textbf{0.26}} \\
      \rowcolor{RowAlt}
      Sharpe
      & 0.23 & {\color{ForestGreen}\textbf{0.80}} & 0.57 & 0.68 & 0.69 & -0.38 & 0.20 & -0.17 & -0.19 & {\textbf{0.70}} &  & -0.75 & -0.66 & -0.54 & -0.79 & -0.56 & -1.22 & -0.56 & {\textbf{-0.33}} & -0.33 & {\color{ForestGreen}\textbf{0.31}} \\
      Sortino
      & 0.47 & 2.84 & 1.48 & {\textbf{6.38}} & {\color{ForestGreen}\textbf{7.73}} & -0.43 & 0.33 & -0.21 & -0.20 & 1.89 &  & -0.73 & -0.68 & -0.55 & -0.83 & -0.59 & -1.07 & -0.58 & {\textbf{-0.38}} & -0.40 & {\color{ForestGreen}\textbf{0.45}} \\
      \rowcolor{RowAlt}
      Omega
      & 1.70 & 4.80 & 3.04 & {\textbf{10.50}} & {\color{ForestGreen}\textbf{11.48}} & 0.43 & 1.46 & 0.76 & 0.66 & 3.81 &  & 0.20 & 0.29 & 0.29 & 0.33 & 0.38 & 0.02 & 0.37 & {\textbf{0.61}} & 0.60 & {\color{ForestGreen}\textbf{1.55}} \\
      MaxDD
      & -0.36 & -0.09 & -0.31 & {\textbf{-0.05}} & {\color{ForestGreen}\textbf{-0.03}} & -37.30 & -0.45 & -7.26 & -16.69 & -0.18 &  & -44.83 & -15.18 & -18.74 & -16.96 & -18.92 & -65.90 & -17.79 & {\textbf{-8.29}} & -9.04 & {\color{ForestGreen}\textbf{-0.94}} \\
      \rowcolor{RowAlt}
      Edge
      & 0.09 & 0.03 & 0.12 & 0.05 & {\color{ForestGreen}\textbf{0.55}} & {\textbf{0.27}} & 0.08 & 0.03 & 0.23 & 0.17 &  & {\textbf{0.04}} & 0.01 & 0.03 & 0.02 & 0.00 & 0.00 & 0.04 & 0.00 & 0.03 & {\color{ForestGreen}\textbf{0.18}} \\

    \addlinespace[0.5em]
    \midrule
    \addlinespace[0.5em]
    \multicolumn{22}{l}{\textcolor{dpd2}{\textbf{Panel B: Absolute Error}}} \\
    \addlinespace[0.3em]

      Return
      & 0.07 & 0.26 & 0.21 & {\color{ForestGreen}\textbf{0.31}} & {\textbf{0.30}} & -0.23 & 0.12 & 0.02 & 0.05 & 0.24 &  & -0.52 & -0.32 & -0.28 & -0.53 & -0.30 & -1.87 & -0.41 & -0.17 & {\textbf{-0.09}} & {\color{ForestGreen}\textbf{0.13}} \\
      \rowcolor{RowAlt}
      Sharpe
      & 0.20 & {\textbf{1.12}} & 0.72 & {\color{ForestGreen}\textbf{1.14}} & 1.06 & -0.38 & 0.37 & 0.06 & 0.13 & 0.90 &  & -0.47 & -0.46 & -0.33 & -0.88 & -0.42 & -1.48 & -0.63 & -0.37 & {\textbf{-0.15}} & {\color{ForestGreen}\textbf{0.28}} \\
      Sortino
      & 0.30 & 2.94 & 1.42 & {\textbf{3.42}} & {\color{ForestGreen}\textbf{3.71}} & -0.44 & 0.62 & 0.08 & 0.17 & 2.01 &  & -0.50 & -0.53 & -0.37 & -0.93 & -0.48 & -1.23 & -0.69 & -0.46 & {\textbf{-0.20}} & {\color{ForestGreen}\textbf{0.41}} \\
      \rowcolor{RowAlt}
      Omega
      & 1.34 & 4.46 & 2.63 & {\textbf{5.00}} & {\color{ForestGreen}\textbf{5.15}} & 0.56 & 1.71 & 1.08 & 1.22 & 3.42 &  & 0.48 & 0.51 & 0.59 & 0.36 & 0.56 & 0.07 & 0.42 & 0.61 & {\textbf{0.82}} & {\color{ForestGreen}\textbf{1.41}} \\
      MaxDD
      & -0.52 & {\textbf{-0.11}} & -0.49 & {\color{ForestGreen}\textbf{-0.11}} & -0.12 & -13.22 & -0.56 & -3.09 & -5.22 & -0.17 &  & -12.65 & -5.92 & -5.11 & -9.03 & -7.86 & -17.89 & -7.03 & {\textbf{-4.41}} & -4.90 & {\color{ForestGreen}\textbf{-0.83}} \\
      \rowcolor{RowAlt}
      Edge
      & 0.11 & 0.16 & 0.35 & 0.05 & 0.35 & {\textbf{0.40}} & 0.21 & 0.11 & {\color{ForestGreen}\textbf{0.48}} & 0.37 &  & {\textbf{0.29}} & 0.10 & 0.20 & 0.12 & 0.02 & 0.00 & 0.07 & 0.00 & {\color{ForestGreen}\textbf{0.31}} & 0.13 \\

    \addlinespace[0.5em]
    \midrule
    \addlinespace[0.5em]
    \multicolumn{22}{l}{\textcolor{dpd2}{\textbf{Panel C: Classical Forecast Accuracy}}} \\
    \addlinespace[0.3em]

      RMSE
      & 0.89 & 0.86 & 0.86 & {\textbf{0.68}} & {\color{ForestGreen}\textbf{0.67}} & 1.29 & 0.94 & 1.04 & 1.07 & 0.84 &  & 1.94 & 1.43 & 1.51 & 1.41 & 1.39 & 3.11 & 1.41 & {\textbf{1.14}} & 1.15 & {\color{ForestGreen}\textbf{0.86}} \\
      \rowcolor{RowAlt}
      MAE
      & 0.93 & 0.74 & 0.79 & {\color{ForestGreen}\textbf{0.69}} & {\textbf{0.70}} & 1.23 & 0.88 & 0.98 & 0.95 & 0.76 &  & 1.52 & 1.32 & 1.28 & 1.53 & 1.30 & 2.87 & 1.41 & 1.17 & {\textbf{1.09}} & {\color{ForestGreen}\textbf{0.87}} \\
      $\rho(1)$
      & 0.06 & 0.27 & {\textbf{0.02}} & 0.08 & 0.04 & -0.12 & {\color{ForestGreen}\textbf{-0.01}} & 0.52 & 0.60 & 0.21 &  & 0.72 & 0.29 & {\textbf{0.09}} & 0.66 & 0.56 & 0.39 & 0.21 & 0.43 & 0.40 & {\color{ForestGreen}\textbf{0.06}} \\
      \rowcolor{RowAlt}
      DM $t$-stat
      & 0.82 & {\color{ForestGreen}\textbf{2.53}} & 1.73 & {\textbf{1.91}} & 1.86 & -1.18 & 0.55 & -0.55 & -0.51 & 1.88 &  & -1.04 & -1.05 & -1.00 & -1.51 & -0.80 & -2.09 & -0.88 & {\textbf{-0.55}} & -0.60 & {\color{ForestGreen}\textbf{0.66}} \\

    \addlinespace[0.3em]
    \bottomrule
  \end{tabular}%
  }

  \vspace{0.25em}
  \parbox{\linewidth}{\scriptsize
    \textit{Notes}: Panels A--B report risk-adjusted metrics; Panel C reports classical forecast accuracy metrics. $\rho(1)$ = first-order autocorrelation of errors.
    Best: \textcolor{ForestGreen}{\textbf{bold green}}; second-best: \textbf{bold}.
  }
  }

\end{table}
\end{landscape}

%% file: 03_tables/ur__h2.tex
% ------------------------------------------------------------
%  EW VERSION - THREE-PANEL TABLE (SE / AE / Classical)
%  Template: dpd2 color + RowAlt alternating rows
% ------------------------------------------------------------

\begin{landscape}
\begin{table}[t!]
  \centering
  \caption{\normalsize Unemployment Rate ($h=2$)}
  \vspace*{-0.65em}
  \label{tab:ur_h2}

  {\fontfamily{phv}\selectfont
  \resizebox{\linewidth}{!}{%
  \scriptsize
  \setlength{\tabcolsep}{0.3em}
  \renewcommand{\arraystretch}{1.65}
  \begin{tabular}{l r r r r r r r r r r c r r r r r r r r r r}
    \toprule

    & \multicolumn{10}{c}{\textcolor{dpd2}{\textbf{2007Q2 -- 2019Q4}}}
    && \multicolumn{10}{c}{\textcolor{dpd2}{\textbf{2022Q1 -- 2025Q1}}}
    \\
    \cmidrule(lr){2-11}
    \cmidrule(lr){13-22}
    & FAAR & RF & LGB & LGB$^{\texttt{A}}$+ & LGB+ & KRR & NN & RR & SPF & TabPFN 
    && FAAR & RF & LGB & LGB$^{\texttt{A}}$+ & LGB+ & KRR & NN & RR & SPF & TabPFN \\

    \midrule
    \addlinespace[0.5em]
    \multicolumn{22}{l}{\textcolor{dpd2}{\textbf{Panel A: Squared Error}}} \\
    \addlinespace[0.3em]

      Return
      & 0.12 & 0.09 & 0.15 & {\textbf{0.29}} & {\color{ForestGreen}\textbf{0.50}} & -0.74 & 0.19 & 0.02 & 0.13 & 0.25 &  & -2.87 & -2.56 & -7.49 & -0.73 & -1.86 & -5.56 & -0.67 & {\textbf{-0.04}} & {\color{ForestGreen}\textbf{0.16}} & -0.49 \\
      \rowcolor{RowAlt}
      Sharpe
      & 0.25 & 0.29 & 0.34 & {\textbf{0.68}} & {\color{ForestGreen}\textbf{0.82}} & -0.44 & 0.24 & 0.13 & 0.64 & 0.58 &  & -1.40 & -0.77 & -0.83 & -0.46 & -0.65 & -1.34 & -0.46 & {\textbf{-0.06}} & {\color{ForestGreen}\textbf{0.20}} & -0.60 \\
      Sortino
      & 0.47 & 0.52 & 0.65 & {\textbf{5.06}} & {\color{ForestGreen}\textbf{10.98}} & -0.51 & 0.61 & 0.22 & 1.73 & 1.71 &  & -1.19 & -0.75 & -0.79 & -0.49 & -0.65 & -1.14 & -0.51 & {\textbf{-0.07}} & {\color{ForestGreen}\textbf{0.29}} & -0.69 \\
      \rowcolor{RowAlt}
      Omega
      & 1.56 & 1.67 & 1.80 & {\textbf{6.82}} & {\color{ForestGreen}\textbf{14.31}} & 0.39 & 1.70 & 1.28 & 3.09 & 3.67 &  & 0.10 & 0.16 & 0.08 & 0.45 & 0.24 & 0.01 & 0.52 & {\textbf{0.93}} & {\color{ForestGreen}\textbf{1.31}} & 0.46 \\
      MaxDD
      & -0.43 & -0.70 & -0.32 & {\textbf{-0.09}} & {\color{ForestGreen}\textbf{-0.02}} & -57.01 & -0.50 & -0.67 & -0.54 & -0.21 &  & -41.63 & -35.13 & -101.22 & -14.60 & -28.98 & -56.13 & -16.05 & {\textbf{-4.82}} & {\color{ForestGreen}\textbf{-0.90}} & -9.57 \\
      \rowcolor{RowAlt}
      Edge
      & 0.12 & 0.01 & 0.03 & 0.00 & {\textbf{0.32}} & {\color{ForestGreen}\textbf{0.56}} & 0.02 & 0.01 & 0.09 & 0.00 &  & 0.02 & 0.00 & 0.01 & 0.01 & {\textbf{0.03}} & 0.00 & {\color{ForestGreen}\textbf{0.06}} & 0.03 & 0.02 & 0.00 \\

    \addlinespace[0.5em]
    \midrule
    \addlinespace[0.5em]
    \multicolumn{22}{l}{\textcolor{dpd2}{\textbf{Panel B: Absolute Error}}} \\
    \addlinespace[0.3em]

      Return
      & -0.02 & 0.10 & 0.09 & 0.16 & {\color{ForestGreen}\textbf{0.31}} & -0.42 & 0.02 & 0.01 & 0.10 & {\textbf{0.17}} &  & -0.98 & -0.63 & -1.17 & -0.21 & -0.49 & -1.60 & -0.20 & {\textbf{0.03}} & {\color{ForestGreen}\textbf{0.11}} & -0.31 \\
      \rowcolor{RowAlt}
      Sharpe
      & -0.05 & 0.41 & 0.29 & {\textbf{0.69}} & {\color{ForestGreen}\textbf{1.07}} & -0.58 & 0.06 & 0.06 & 0.55 & 0.69 &  & -1.38 & -0.67 & -0.74 & -0.31 & -0.55 & -1.89 & -0.28 & {\textbf{0.07}} & {\color{ForestGreen}\textbf{0.20}} & -0.61 \\
      Sortino
      & -0.06 & 0.60 & 0.43 & {\textbf{1.44}} & {\color{ForestGreen}\textbf{3.15}} & -0.63 & 0.09 & 0.08 & 1.08 & 1.24 &  & -1.20 & -0.70 & -0.74 & -0.37 & -0.60 & -1.40 & -0.35 & {\textbf{0.10}} & {\color{ForestGreen}\textbf{0.30}} & -0.72 \\
      \rowcolor{RowAlt}
      Omega
      & 0.94 & 1.75 & 1.45 & {\textbf{2.62}} & {\color{ForestGreen}\textbf{4.67}} & 0.40 & 1.10 & 1.09 & 2.03 & 2.52 &  & 0.18 & 0.35 & 0.27 & 0.64 & 0.43 & 0.02 & 0.72 & {\textbf{1.10}} & {\color{ForestGreen}\textbf{1.30}} & 0.46 \\
      MaxDD
      & -6.34 & -0.88 & -0.90 & {\textbf{-0.42}} & {\color{ForestGreen}\textbf{-0.11}} & -28.56 & -0.94 & -2.97 & -0.50 & -0.49 &  & -15.68 & -10.05 & -18.33 & -5.82 & -9.47 & -17.89 & -7.66 & {\textbf{-3.32}} & {\color{ForestGreen}\textbf{-0.91}} & -6.00 \\
      \rowcolor{RowAlt}
      Edge
      & 0.31 & 0.09 & 0.13 & 0.04 & {\textbf{0.47}} & {\color{ForestGreen}\textbf{0.47}} & 0.08 & 0.07 & 0.19 & 0.02 &  & 0.10 & 0.00 & 0.08 & 0.10 & 0.07 & 0.00 & {\color{ForestGreen}\textbf{0.33}} & 0.11 & {\textbf{0.21}} & 0.00 \\

    \addlinespace[0.5em]
    \midrule
    \addlinespace[0.5em]
    \multicolumn{22}{l}{\textcolor{dpd2}{\textbf{Panel C: Classical Forecast Accuracy}}} \\
    \addlinespace[0.3em]

      RMSE
      & 0.94 & 0.95 & 0.92 & {\textbf{0.84}} & {\color{ForestGreen}\textbf{0.71}} & 1.32 & 0.90 & 0.99 & 0.94 & 0.87 &  & 1.97 & 1.89 & 2.91 & 1.31 & 1.69 & 2.56 & 1.29 & {\textbf{1.02}} & {\color{ForestGreen}\textbf{0.92}} & 1.22 \\
      \rowcolor{RowAlt}
      MAE
      & 1.02 & 0.90 & 0.91 & 0.84 & {\color{ForestGreen}\textbf{0.69}} & 1.42 & 0.98 & 0.99 & 0.90 & {\textbf{0.83}} &  & 1.98 & 1.63 & 2.17 & 1.21 & 1.49 & 2.60 & 1.20 & {\textbf{0.97}} & {\color{ForestGreen}\textbf{0.89}} & 1.31 \\
      $\rho(1)$
      & 0.47 & 0.66 & 0.50 & 0.48 & 0.38 & {\color{ForestGreen}\textbf{-0.11}} & {\textbf{0.20}} & 0.69 & 0.63 & 0.57 &  & {\textbf{0.42}} & 0.62 & 0.67 & 0.68 & 0.65 & 0.52 & 0.52 & 0.46 & {\color{ForestGreen}\textbf{0.40}} & 0.72 \\
      \rowcolor{RowAlt}
      DM $t$-stat
      & 0.63 & 0.89 & 1.21 & 1.77 & 1.76 & -1.28 & 0.59 & 0.39 & {\color{ForestGreen}\textbf{2.16}} & {\textbf{2.13}} &  & -2.00 & -1.15 & -1.21 & -0.70 & -1.03 & -2.45 & -0.72 & {\textbf{-0.10}} & {\color{ForestGreen}\textbf{0.33}} & -1.01 \\

    \addlinespace[0.3em]
    \bottomrule
  \end{tabular}%
  }

  \vspace{0.25em}
  \parbox{\linewidth}{\scriptsize
    \textit{Notes}: Panels A--B report risk-adjusted metrics; Panel C reports classical forecast accuracy metrics. $\rho(1)$ = first-order autocorrelation of errors.
    Best: \textcolor{ForestGreen}{\textbf{bold green}}; second-best: \textbf{bold}.
  }
  }

\end{table}
\end{landscape}

%% file: 03_tables/ur__h4.tex
% ------------------------------------------------------------
%  EW VERSION - THREE-PANEL TABLE (SE / AE / Classical)
%  Template: dpd2 color + RowAlt alternating rows
% ------------------------------------------------------------

\begin{landscape}
\begin{table}[t!]
  \centering
  \caption{\normalsize Unemployment Rate ($h=4$)}
  \vspace*{-0.65em}
  \label{tab:ur_h4}

  {\fontfamily{phv}\selectfont
  \resizebox{\linewidth}{!}{%
  \scriptsize
  \setlength{\tabcolsep}{0.3em}
  \renewcommand{\arraystretch}{1.65}
  \begin{tabular}{l r r r r r r r r r r c r r r r r r r r r r}
    \toprule

    & \multicolumn{10}{c}{\textcolor{dpd2}{\textbf{2007Q2 -- 2019Q4}}}
    && \multicolumn{10}{c}{\textcolor{dpd2}{\textbf{2022Q1 -- 2025Q1}}}
    \\
    \cmidrule(lr){2-11}
    \cmidrule(lr){13-22}
    & FAAR & RF & LGB & LGB$^{\texttt{A}}$+ & LGB+ & KRR & NN & RR & SPF & TabPFN 
    && FAAR & RF & LGB & LGB$^{\texttt{A}}$+ & LGB+ & KRR & NN & RR & SPF & TabPFN \\

    \midrule
    \addlinespace[0.5em]
    \multicolumn{22}{l}{\textcolor{dpd2}{\textbf{Panel A: Squared Error}}} \\
    \addlinespace[0.3em]

      Return
      & -0.15 & 0.20 & {\textbf{0.28}} & 0.01 & 0.25 & -0.69 & 0.18 & 0.11 & 0.16 & {\color{ForestGreen}\textbf{0.48}} &  & -2.90 & -1.11 & -1.15 & -0.36 & -1.42 & -3.01 & 0.33 & {\textbf{0.65}} & {\color{ForestGreen}\textbf{0.76}} & -0.16 \\
      \rowcolor{RowAlt}
      Sharpe
      & -0.36 & {\color{ForestGreen}\textbf{1.10}} & 0.67 & 0.05 & 0.74 & -0.47 & 0.25 & 0.47 & {\textbf{0.92}} & 0.72 &  & -0.89 & -0.59 & -0.62 & -0.32 & -0.75 & -1.43 & 0.32 & {\color{ForestGreen}\textbf{0.91}} & {\textbf{0.88}} & -0.16 \\
      Sortino
      & -0.39 & {\textbf{4.82}} & 1.86 & 0.09 & 1.98 & -0.56 & 0.51 & 1.03 & 2.16 & {\color{ForestGreen}\textbf{6.14}} &  & -0.84 & -0.63 & -0.66 & -0.38 & -0.76 & -1.20 & 0.63 & {\textbf{6.01}} & {\color{ForestGreen}\textbf{6.50}} & -0.22 \\
      \rowcolor{RowAlt}
      Omega
      & 0.52 & {\textbf{8.46}} & 4.69 & 1.09 & 4.97 & 0.41 & 1.79 & 2.71 & 4.44 & {\color{ForestGreen}\textbf{10.20}} &  & 0.01 & 0.35 & 0.35 & 0.57 & 0.27 & 0.04 & 1.59 & {\color{ForestGreen}\textbf{7.70}} & {\textbf{6.87}} & 0.78 \\
      MaxDD
      & -10.33 & {\color{ForestGreen}\textbf{-0.09}} & -1.65 & -0.78 & -0.24 & -53.00 & -0.50 & -0.32 & -0.47 & {\textbf{-0.12}} &  & -34.72 & -21.34 & -21.79 & -9.81 & -24.45 & -25.90 & -0.46 & {\textbf{-0.10}} & {\color{ForestGreen}\textbf{-0.06}} & -7.39 \\
      \rowcolor{RowAlt}
      Edge
      & 0.07 & 0.01 & 0.06 & 0.00 & 0.01 & {\textbf{0.75}} & 0.01 & 0.00 & 0.08 & {\color{ForestGreen}\textbf{4.45}} &  & 0.01 & 0.00 & 0.00 & 0.00 & 0.00 & 0.00 & {\textbf{0.47}} & 0.23 & {\color{ForestGreen}\textbf{0.82}} & 0.00 \\

    \addlinespace[0.5em]
    \midrule
    \addlinespace[0.5em]
    \multicolumn{22}{l}{\textcolor{dpd2}{\textbf{Panel B: Absolute Error}}} \\
    \addlinespace[0.3em]

      Return
      & -0.13 & 0.19 & 0.19 & -0.02 & {\textbf{0.22}} & -0.44 & 0.10 & 0.07 & 0.18 & {\color{ForestGreen}\textbf{0.35}} &  & -0.89 & -0.45 & -0.47 & -0.27 & -0.61 & -1.13 & 0.15 & {\textbf{0.31}} & {\color{ForestGreen}\textbf{0.41}} & -0.14 \\
      \rowcolor{RowAlt}
      Sharpe
      & -0.39 & {\color{ForestGreen}\textbf{1.21}} & 0.89 & -0.12 & {\textbf{1.02}} & -0.70 & 0.33 & 0.50 & 0.98 & 0.98 &  & -1.44 & -0.53 & -0.57 & -0.44 & -0.74 & -1.68 & 0.22 & {\color{ForestGreen}\textbf{0.81}} & {\textbf{0.74}} & -0.23 \\
      Sortino
      & -0.45 & {\color{ForestGreen}\textbf{3.21}} & 1.59 & -0.16 & 2.45 & -0.76 & 0.53 & 0.78 & 2.14 & {\textbf{2.98}} &  & -1.21 & -0.59 & -0.62 & -0.49 & -0.76 & -1.32 & 0.36 & {\textbf{2.12}} & {\color{ForestGreen}\textbf{2.47}} & -0.29 \\
      \rowcolor{RowAlt}
      Omega
      & 0.58 & {\color{ForestGreen}\textbf{4.99}} & 3.24 & 0.86 & 4.10 & 0.41 & 1.50 & 1.93 & 3.80 & {\textbf{4.84}} &  & 0.06 & 0.45 & 0.44 & 0.49 & 0.33 & 0.08 & 1.33 & {\textbf{3.02}} & {\color{ForestGreen}\textbf{3.19}} & 0.73 \\
      MaxDD
      & -8.34 & {\color{ForestGreen}\textbf{-0.14}} & -0.79 & -4.78 & -0.31 & -29.29 & -0.54 & -0.48 & -0.63 & {\textbf{-0.26}} &  & -10.75 & -9.74 & -9.72 & -6.03 & -10.92 & -11.81 & -0.52 & {\textbf{-0.27}} & {\color{ForestGreen}\textbf{-0.15}} & -4.90 \\
      \rowcolor{RowAlt}
      Edge
      & 0.26 & 0.11 & 0.19 & 0.02 & 0.07 & {\textbf{0.58}} & 0.08 & 0.03 & 0.18 & {\color{ForestGreen}\textbf{2.95}} &  & 0.08 & 0.00 & 0.02 & 0.00 & 0.02 & 0.00 & {\color{ForestGreen}\textbf{1.24}} & 0.52 & {\textbf{1.16}} & 0.00 \\

    \addlinespace[0.5em]
    \midrule
    \addlinespace[0.5em]
    \multicolumn{22}{l}{\textcolor{dpd2}{\textbf{Panel C: Classical Forecast Accuracy}}} \\
    \addlinespace[0.3em]

      RMSE
      & 1.07 & 0.90 & {\textbf{0.85}} & 0.99 & 0.87 & 1.30 & 0.91 & 0.95 & 0.91 & {\color{ForestGreen}\textbf{0.72}} &  & 1.98 & 1.45 & 1.46 & 1.16 & 1.56 & 2.00 & 0.82 & {\textbf{0.59}} & {\color{ForestGreen}\textbf{0.49}} & 1.08 \\
      \rowcolor{RowAlt}
      MAE
      & 1.13 & 0.81 & 0.81 & 1.02 & {\textbf{0.78}} & 1.44 & 0.90 & 0.93 & 0.82 & {\color{ForestGreen}\textbf{0.65}} &  & 1.89 & 1.45 & 1.47 & 1.27 & 1.61 & 2.13 & 0.85 & {\textbf{0.69}} & {\color{ForestGreen}\textbf{0.59}} & 1.14 \\
      $\rho(1)$
      & 0.65 & 0.74 & 0.66 & 0.76 & 0.73 & {\textbf{0.59}} & 0.60 & 0.76 & 0.73 & {\color{ForestGreen}\textbf{0.41}} &  & {\textbf{-0.06}} & 1.08 & 0.93 & 0.90 & 0.94 & {\color{ForestGreen}\textbf{0.02}} & 0.39 & 0.35 & 0.20 & 0.66 \\
      \rowcolor{RowAlt}
      DM $t$-stat
      & -1.26 & {\textbf{2.83}} & 1.60 & 0.11 & 1.94 & -1.61 & 0.91 & 1.50 & {\color{ForestGreen}\textbf{3.58}} & 1.72 &  & -1.60 & -0.97 & -2.26 & -0.43 & -2.93 & -2.24 & 0.45 & {\textbf{1.26}} & {\color{ForestGreen}\textbf{1.28}} & -0.25 \\

    \addlinespace[0.3em]
    \bottomrule
  \end{tabular}%
  }

  \vspace{0.25em}
  \parbox{\linewidth}{\scriptsize
    \textit{Notes}: Panels A--B report risk-adjusted metrics; Panel C reports classical forecast accuracy metrics. $\rho(1)$ = first-order autocorrelation of errors.
    Best: \textcolor{ForestGreen}{\textbf{bold green}}; second-best: \textbf{bold}.
  }
  }

\end{table}
\end{landscape}

%% file: 03_tables/infl__h1.tex
% ------------------------------------------------------------
%  EW VERSION - THREE-PANEL TABLE (SE / AE / Classical)
%  Template: dpd2 color + RowAlt alternating rows
% ------------------------------------------------------------

\begin{landscape}
\begin{table}[t!]
  \centering
  \caption{\normalsize Inflation ($h=1$)}
  \vspace*{-0.65em}
  \label{tab:infl_h1}

  {\fontfamily{phv}\selectfont
  \resizebox{\linewidth}{!}{%
  \scriptsize
  \setlength{\tabcolsep}{0.3em}
  \renewcommand{\arraystretch}{1.65}
  \begin{tabular}{l r r r r r r r r r r r c r r r r r r r r r r r}
    \toprule

    & \multicolumn{11}{c}{\textcolor{dpd2}{\textbf{2007Q2 -- 2019Q4}}}
    && \multicolumn{11}{c}{\textcolor{dpd2}{\textbf{2021Q1 -- 2025Q1}}}
    \\
    \cmidrule(lr){2-12}
    \cmidrule(lr){14-24}
    & FAAR & RF & LGB & LGB$^{\texttt{A}}$+ & LGB+ & KRR & NN & HNN & RR & SPF & TabPFN 
    && FAAR & RF & LGB & LGB$^{\texttt{A}}$+ & LGB+ & KRR & NN & HNN & RR & SPF & TabPFN \\

    \midrule
    \addlinespace[0.5em]
    \multicolumn{24}{l}{\textcolor{dpd2}{\textbf{Panel A: Squared Error}}} \\
    \addlinespace[0.3em]

      Return
      & -0.43 & 0.07 & -0.13 & 0.15 & 0.19 & -0.05 & 0.03 & {\textbf{0.22}} & -0.11 & {\color{ForestGreen}\textbf{0.25}} & -0.05 &  & -1.04 & -0.39 & -1.48 & -0.29 & -0.07 & {\color{ForestGreen}\textbf{0.12}} & -0.66 & {\textbf{0.07}} & -1.07 & -1.54 & -0.35 \\
      \rowcolor{RowAlt}
      Sharpe
      & -0.37 & 0.12 & -0.17 & 0.17 & {\textbf{0.32}} & -0.08 & 0.03 & {\color{ForestGreen}\textbf{0.36}} & -0.13 & 0.31 & -0.04 &  & -0.47 & -0.57 & -1.02 & -0.54 & -0.17 & {\color{ForestGreen}\textbf{0.30}} & -0.48 & {\textbf{0.09}} & -0.58 & -0.87 & -0.55 \\
      Sortino
      & -0.41 & 0.25 & -0.28 & 0.39 & {\textbf{1.61}} & -0.11 & 0.05 & {\color{ForestGreen}\textbf{2.00}} & -0.18 & 1.21 & -0.10 &  & -0.48 & -0.62 & -0.94 & -0.61 & -0.27 & {\color{ForestGreen}\textbf{0.58}} & -0.50 & {\textbf{0.12}} & -0.58 & -0.84 & -0.65 \\
      \rowcolor{RowAlt}
      Omega
      & 0.33 & 1.33 & 0.67 & 1.69 & {\textbf{2.53}} & 0.79 & 1.09 & {\color{ForestGreen}\textbf{3.01}} & 0.69 & 2.38 & 0.87 &  & 0.25 & 0.37 & 0.07 & 0.40 & 0.76 & {\color{ForestGreen}\textbf{1.59}} & 0.32 & {\textbf{1.14}} & 0.26 & 0.21 & 0.45 \\
      MaxDD
      & -25.55 & -4.61 & -13.08 & {\textbf{-0.94}} & {\color{ForestGreen}\textbf{-0.34}} & -8.22 & -8.25 & -1.99 & -12.74 & -3.00 & -13.14 &  & -3.76 & -10.03 & -25.25 & -7.65 & -3.80 & {\color{ForestGreen}\textbf{-0.47}} & -15.52 & {\textbf{-0.74}} & -24.37 & -32.80 & -9.29 \\
      \rowcolor{RowAlt}
      Edge
      & 0.01 & 0.00 & 0.06 & 0.07 & 0.01 & 0.15 & 0.06 & 0.07 & 0.00 & {\textbf{0.52}} & {\color{ForestGreen}\textbf{0.86}} &  & 0.03 & 0.00 & {\textbf{0.06}} & 0.00 & 0.00 & 0.01 & 0.05 & {\color{ForestGreen}\textbf{1.83}} & 0.00 & 0.00 & 0.05 \\

    \addlinespace[0.5em]
    \midrule
    \addlinespace[0.5em]
    \multicolumn{24}{l}{\textcolor{dpd2}{\textbf{Panel B: Absolute Error}}} \\
    \addlinespace[0.3em]

      Return
      & -0.31 & -0.04 & -0.20 & -0.01 & 0.04 & -0.01 & -0.05 & {\textbf{0.04}} & -0.11 & {\color{ForestGreen}\textbf{0.09}} & -0.16 &  & -0.13 & -0.08 & -0.32 & -0.13 & 0.01 & {\textbf{0.05}} & -0.04 & {\color{ForestGreen}\textbf{0.14}} & -0.18 & -0.32 & -0.07 \\
      \rowcolor{RowAlt}
      Sharpe
      & -0.76 & -0.13 & -0.54 & -0.03 & 0.12 & -0.02 & -0.13 & {\textbf{0.17}} & -0.32 & {\color{ForestGreen}\textbf{0.21}} & -0.40 &  & -0.25 & -0.29 & -0.76 & -0.55 & 0.05 & {\textbf{0.25}} & -0.10 & {\color{ForestGreen}\textbf{0.32}} & -0.34 & -0.51 & -0.20 \\
      Sortino
      & -0.76 & -0.17 & -0.60 & -0.03 & 0.23 & -0.02 & -0.18 & {\textbf{0.27}} & -0.38 & {\color{ForestGreen}\textbf{0.37}} & -0.48 &  & -0.29 & -0.35 & -0.77 & -0.65 & 0.10 & {\textbf{0.40}} & -0.13 & {\color{ForestGreen}\textbf{0.56}} & -0.39 & -0.56 & -0.27 \\
      \rowcolor{RowAlt}
      Omega
      & 0.28 & 0.83 & 0.46 & 0.95 & 1.21 & 0.97 & 0.81 & {\textbf{1.25}} & 0.61 & {\color{ForestGreen}\textbf{1.37}} & 0.53 &  & 0.67 & 0.69 & 0.33 & 0.47 & 1.08 & {\textbf{1.35}} & 0.86 & {\color{ForestGreen}\textbf{1.48}} & 0.58 & 0.49 & 0.78 \\
      MaxDD
      & -14.62 & -4.60 & -10.57 & -3.56 & {\color{ForestGreen}\textbf{-0.97}} & -3.88 & -5.52 & {\textbf{-2.33}} & -7.21 & -3.92 & -8.83 &  & -1.30 & -3.91 & -6.55 & -3.37 & -0.66 & {\textbf{-0.45}} & -4.27 & {\color{ForestGreen}\textbf{-0.43}} & -6.95 & -10.14 & -4.03 \\
      \rowcolor{RowAlt}
      Edge
      & 0.08 & 0.01 & 0.15 & 0.18 & 0.05 & {\textbf{0.58}} & 0.10 & 0.09 & 0.01 & {\color{ForestGreen}\textbf{0.90}} & 0.24 &  & 0.15 & 0.00 & 0.20 & 0.00 & 0.00 & 0.08 & 0.19 & {\color{ForestGreen}\textbf{2.26}} & 0.00 & 0.00 & {\textbf{0.33}} \\

    \addlinespace[0.5em]
    \midrule
    \addlinespace[0.5em]
    \multicolumn{24}{l}{\textcolor{dpd2}{\textbf{Panel C: Classical Forecast Accuracy}}} \\
    \addlinespace[0.3em]

      RMSE
      & 1.19 & 0.96 & 1.06 & 0.92 & 0.90 & 1.03 & 0.99 & {\textbf{0.89}} & 1.05 & {\color{ForestGreen}\textbf{0.86}} & 1.02 &  & 1.43 & 1.18 & 1.58 & 1.14 & 1.03 & {\color{ForestGreen}\textbf{0.94}} & 1.29 & {\textbf{0.96}} & 1.44 & 1.59 & 1.16 \\
      \rowcolor{RowAlt}
      MAE
      & 1.31 & 1.04 & 1.20 & 1.01 & 0.96 & 1.01 & 1.05 & {\textbf{0.96}} & 1.11 & {\color{ForestGreen}\textbf{0.91}} & 1.16 &  & 1.13 & 1.08 & 1.32 & 1.13 & 0.99 & {\textbf{0.95}} & 1.04 & {\color{ForestGreen}\textbf{0.86}} & 1.18 & 1.32 & 1.07 \\
      $\rho(1)$
      & 0.15 & 0.20 & 0.26 & {\textbf{0.04}} & 0.10 & 0.29 & 0.08 & {\color{ForestGreen}\textbf{0.02}} & 0.16 & 0.18 & 0.21 &  & 0.32 & 0.55 & 0.46 & 0.49 & 0.45 & {\color{ForestGreen}\textbf{0.12}} & {\textbf{0.14}} & 0.16 & 0.49 & 0.82 & 0.32 \\
      \rowcolor{RowAlt}
      DM $t$-stat
      & -1.18 & 0.72 & -0.59 & 0.92 & 1.33 & -0.27 & 0.10 & {\color{ForestGreen}\textbf{1.40}} & -0.43 & {\textbf{1.38}} & -0.16 &  & -1.00 & -1.02 & -2.13 & -1.15 & -0.43 & {\color{ForestGreen}\textbf{0.62}} & -1.15 & {\textbf{0.18}} & -1.23 & -1.19 & -0.89 \\

    \addlinespace[0.3em]
    \bottomrule
  \end{tabular}%
  }

  \vspace{0.25em}
  \parbox{\linewidth}{\scriptsize
    \textit{Notes}: Panels A--B report risk-adjusted metrics; Panel C reports classical forecast accuracy metrics. $\rho(1)$ = first-order autocorrelation of errors.
    Best: \textcolor{ForestGreen}{\textbf{bold green}}; second-best: \textbf{bold}.
  }
  }

\end{table}
\end{landscape}

%% file: 03_tables/infl__h2.tex
% ------------------------------------------------------------
%  EW VERSION - THREE-PANEL TABLE (SE / AE / Classical)
%  Template: dpd2 color + RowAlt alternating rows
% ------------------------------------------------------------

\begin{landscape}
\begin{table}[t!]
  \centering
  \caption{\normalsize Inflation ($h=2$)}
  \vspace*{-0.65em}
  \label{tab:infl_h2}

  {\fontfamily{phv}\selectfont
  \resizebox{\linewidth}{!}{%
  \scriptsize
  \setlength{\tabcolsep}{0.3em}
  \renewcommand{\arraystretch}{1.65}
  \begin{tabular}{l r r r r r r r r r r r c r r r r r r r r r r r}
    \toprule

    & \multicolumn{11}{c}{\textcolor{dpd2}{\textbf{2007Q2 -- 2019Q4}}}
    && \multicolumn{11}{c}{\textcolor{dpd2}{\textbf{2021Q1 -- 2025Q1}}}
    \\
    \cmidrule(lr){2-12}
    \cmidrule(lr){14-24}
    & FAAR & RF & LGB & LGB$^{\texttt{A}}$+ & LGB+ & KRR & NN & HNN & RR & SPF & TabPFN 
    && FAAR & RF & LGB & LGB$^{\texttt{A}}$+ & LGB+ & KRR & NN & HNN & RR & SPF & TabPFN \\

    \midrule
    \addlinespace[0.5em]
    \multicolumn{24}{l}{\textcolor{dpd2}{\textbf{Panel A: Squared Error}}} \\
    \addlinespace[0.3em]

      Return
      & -0.30 & 0.31 & -0.03 & 0.09 & 0.17 & 0.11 & {\textbf{0.36}} & 0.26 & 0.17 & {\color{ForestGreen}\textbf{0.42}} & 0.10 &  & -2.91 & -0.35 & -0.34 & -0.10 & -0.05 & {\color{ForestGreen}\textbf{0.51}} & -0.03 & 0.12 & -0.53 & -0.67 & {\textbf{0.15}} \\
      \rowcolor{RowAlt}
      Sharpe
      & -0.20 & {\textbf{0.58}} & -0.05 & 0.12 & 0.31 & 0.20 & 0.49 & 0.52 & 0.34 & {\color{ForestGreen}\textbf{0.60}} & 0.15 &  & -0.49 & -0.71 & -0.84 & -0.45 & -0.14 & {\color{ForestGreen}\textbf{1.40}} & -0.05 & 0.32 & -0.64 & -0.65 & {\textbf{0.36}} \\
      Sortino
      & -0.22 & {\color{ForestGreen}\textbf{4.62}} & -0.06 & 0.16 & 1.10 & 0.36 & 2.65 & 2.01 & 1.44 & {\textbf{4.15}} & 0.34 &  & -0.49 & -0.71 & -0.84 & -0.50 & -0.16 & {\color{ForestGreen}\textbf{9.97}} & -0.05 & 0.45 & -0.67 & -0.67 & {\textbf{0.51}} \\
      \rowcolor{RowAlt}
      Omega
      & 0.53 & {\textbf{6.29}} & 0.89 & 1.38 & 2.17 & 1.80 & 5.84 & 3.87 & 2.40 & {\color{ForestGreen}\textbf{7.12}} & 1.43 &  & 0.10 & 0.30 & 0.32 & 0.53 & 0.81 & {\color{ForestGreen}\textbf{11.95}} & 0.91 & 1.50 & 0.34 & 0.32 & {\textbf{1.73}} \\
      MaxDD
      & -25.56 & -0.67 & -10.61 & -8.10 & {\color{ForestGreen}\textbf{-0.49}} & -5.04 & -1.84 & -2.74 & {\textbf{-0.57}} & -2.46 & -6.23 &  & -3.36 & -7.67 & -4.80 & -3.09 & -4.02 & {\color{ForestGreen}\textbf{-0.08}} & -4.36 & {\textbf{-0.58}} & -12.64 & -16.40 & -3.17 \\
      \rowcolor{RowAlt}
      Edge
      & 0.07 & 0.07 & 0.01 & 0.00 & 0.07 & {\textbf{0.13}} & {\color{ForestGreen}\textbf{0.87}} & 0.11 & 0.05 & 0.12 & 0.03 &  & 0.13 & 0.00 & 0.00 & 0.00 & {\textbf{0.34}} & {\color{ForestGreen}\textbf{3.10}} & 0.13 & 0.09 & 0.00 & 0.12 & 0.00 \\

    \addlinespace[0.5em]
    \midrule
    \addlinespace[0.5em]
    \multicolumn{24}{l}{\textcolor{dpd2}{\textbf{Panel B: Absolute Error}}} \\
    \addlinespace[0.3em]

      Return
      & -0.13 & 0.17 & 0.01 & 0.03 & 0.08 & 0.13 & {\textbf{0.21}} & 0.13 & 0.05 & {\color{ForestGreen}\textbf{0.26}} & 0.03 &  & -0.42 & -0.07 & -0.11 & 0.03 & 0.04 & {\color{ForestGreen}\textbf{0.28}} & 0.13 & 0.13 & -0.05 & -0.04 & {\textbf{0.19}} \\
      \rowcolor{RowAlt}
      Sharpe
      & -0.25 & 0.68 & 0.03 & 0.11 & 0.26 & 0.44 & {\color{ForestGreen}\textbf{0.82}} & 0.51 & 0.19 & {\textbf{0.80}} & 0.10 &  & -0.47 & -0.30 & -0.45 & 0.19 & 0.26 & {\color{ForestGreen}\textbf{1.20}} & 0.53 & 0.49 & -0.15 & -0.09 & {\textbf{0.86}} \\
      Sortino
      & -0.28 & 1.97 & 0.04 & 0.15 & 0.59 & 1.24 & {\color{ForestGreen}\textbf{2.27}} & 0.90 & 0.37 & {\textbf{2.14}} & 0.15 &  & -0.49 & -0.36 & -0.54 & 0.29 & 0.38 & {\color{ForestGreen}\textbf{3.81}} & 0.89 & 0.77 & -0.20 & -0.13 & {\textbf{1.71}} \\
      \rowcolor{RowAlt}
      Omega
      & 0.61 & 2.83 & 1.05 & 1.22 & 1.48 & 2.25 & {\color{ForestGreen}\textbf{3.84}} & 2.02 & 1.30 & {\textbf{3.72}} & 1.15 &  & 0.30 & 0.69 & 0.58 & 1.24 & 1.39 & {\color{ForestGreen}\textbf{5.09}} & 2.02 & 1.78 & 0.82 & 0.88 & {\textbf{2.93}} \\
      MaxDD
      & -8.26 & -0.91 & -4.77 & -2.89 & {\textbf{-0.65}} & -1.71 & -1.82 & -3.15 & {\color{ForestGreen}\textbf{-0.64}} & -3.02 & -3.08 &  & -1.93 & -2.58 & -1.83 & -0.91 & -0.66 & {\color{ForestGreen}\textbf{-0.17}} & -0.65 & {\textbf{-0.37}} & -3.99 & -5.21 & -0.72 \\
      \rowcolor{RowAlt}
      Edge
      & 0.20 & 0.08 & 0.10 & 0.03 & 0.23 & 0.32 & {\color{ForestGreen}\textbf{0.39}} & 0.17 & 0.20 & {\textbf{0.35}} & 0.17 &  & 0.33 & 0.01 & 0.00 & 0.03 & 0.33 & {\color{ForestGreen}\textbf{2.26}} & 0.29 & 0.34 & 0.00 & {\textbf{0.64}} & 0.09 \\

    \addlinespace[0.5em]
    \midrule
    \addlinespace[0.5em]
    \multicolumn{24}{l}{\textcolor{dpd2}{\textbf{Panel C: Classical Forecast Accuracy}}} \\
    \addlinespace[0.3em]

      RMSE
      & 1.14 & 0.83 & 1.01 & 0.95 & 0.91 & 0.94 & {\textbf{0.80}} & 0.86 & 0.91 & {\color{ForestGreen}\textbf{0.76}} & 0.95 &  & 1.98 & 1.16 & 1.16 & 1.05 & 1.02 & {\color{ForestGreen}\textbf{0.70}} & 1.01 & 0.94 & 1.24 & 1.29 & {\textbf{0.92}} \\
      \rowcolor{RowAlt}
      MAE
      & 1.13 & 0.83 & 0.99 & 0.97 & 0.92 & 0.87 & {\textbf{0.79}} & 0.87 & 0.95 & {\color{ForestGreen}\textbf{0.74}} & 0.97 &  & 1.42 & 1.07 & 1.11 & 0.97 & 0.96 & {\color{ForestGreen}\textbf{0.72}} & 0.87 & 0.87 & 1.05 & 1.04 & {\textbf{0.81}} \\
      $\rho(1)$
      & 0.27 & 0.31 & 0.37 & 0.27 & 0.34 & 0.37 & {\textbf{0.25}} & 0.27 & 0.39 & {\color{ForestGreen}\textbf{0.24}} & 0.34 &  & {\color{ForestGreen}\textbf{0.22}} & 0.81 & 0.72 & 0.75 & 0.75 & {\textbf{0.45}} & 0.58 & 0.50 & 0.81 & 0.87 & 0.65 \\
      \rowcolor{RowAlt}
      DM $t$-stat
      & -0.67 & {\textbf{1.65}} & -0.16 & 0.51 & 1.11 & 0.65 & 1.62 & 1.54 & 1.02 & {\color{ForestGreen}\textbf{1.71}} & 0.59 &  & -1.01 & -1.04 & -1.72 & -0.94 & -0.24 & {\color{ForestGreen}\textbf{1.48}} & -0.11 & 0.58 & -0.85 & -0.99 & {\textbf{0.60}} \\

    \addlinespace[0.3em]
    \bottomrule
  \end{tabular}%
  }

  \vspace{0.25em}
  \parbox{\linewidth}{\scriptsize
    \textit{Notes}: Panels A--B report risk-adjusted metrics; Panel C reports classical forecast accuracy metrics. $\rho(1)$ = first-order autocorrelation of errors.
    Best: \textcolor{ForestGreen}{\textbf{bold green}}; second-best: \textbf{bold}.
  }
  }

\end{table}
\end{landscape}

%% file: 03_tables/infl__h4.tex
% ------------------------------------------------------------
%  EW VERSION - THREE-PANEL TABLE (SE / AE / Classical)
%  Template: dpd2 color + RowAlt alternating rows
% ------------------------------------------------------------

\begin{landscape}
\begin{table}[t!]
  \centering
  \caption{\normalsize Inflation ($h=4$)}
  \vspace*{-0.65em}
  \label{tab:infl_h4}

  {\fontfamily{phv}\selectfont
  \resizebox{\linewidth}{!}{%
  \scriptsize
  \setlength{\tabcolsep}{0.3em}
  \renewcommand{\arraystretch}{1.65}
  \begin{tabular}{l r r r r r r r r r r r c r r r r r r r r r r r}
    \toprule

    & \multicolumn{11}{c}{\textcolor{dpd2}{\textbf{2007Q2 -- 2019Q4}}}
    && \multicolumn{11}{c}{\textcolor{dpd2}{\textbf{2021Q1 -- 2025Q1}}}
    \\
    \cmidrule(lr){2-12}
    \cmidrule(lr){14-24}
    & FAAR & RF & LGB & LGB$^{\texttt{A}}$+ & LGB+ & KRR & NN & HNN & RR & SPF & TabPFN 
    && FAAR & RF & LGB & LGB$^{\texttt{A}}$+ & LGB+ & KRR & NN & HNN & RR & SPF & TabPFN \\

    \midrule
    \addlinespace[0.5em]
    \multicolumn{24}{l}{\textcolor{dpd2}{\textbf{Panel A: Squared Error}}} \\
    \addlinespace[0.3em]

      Return
      & -1.15 & 0.18 & 0.24 & 0.24 & 0.18 & 0.22 & 0.24 & {\textbf{0.29}} & 0.07 & {\color{ForestGreen}\textbf{0.36}} & -0.26 &  & -0.10 & -0.50 & -0.56 & -0.43 & -0.40 & {\color{ForestGreen}\textbf{0.42}} & -0.14 & -0.53 & -0.59 & {\textbf{0.00}} & -0.02 \\
      \rowcolor{RowAlt}
      Sharpe
      & -0.39 & 0.41 & 0.46 & 0.43 & 0.41 & 0.39 & {\textbf{0.48}} & 0.46 & 0.15 & {\color{ForestGreen}\textbf{0.54}} & -0.29 &  & -0.19 & -1.37 & -0.69 & -0.86 & -0.84 & {\color{ForestGreen}\textbf{0.94}} & -0.21 & -0.41 & -0.58 & {\textbf{0.01}} & -0.05 \\
      Sortino
      & -0.39 & 1.28 & {\textbf{1.91}} & 1.17 & 1.45 & 0.97 & 1.47 & 1.42 & 0.30 & {\color{ForestGreen}\textbf{2.85}} & -0.34 &  & -0.30 & -1.16 & -0.79 & -0.82 & -0.92 & {\color{ForestGreen}\textbf{8.69}} & -0.24 & -0.43 & -0.59 & {\textbf{0.01}} & -0.08 \\
      \rowcolor{RowAlt}
      Omega
      & 0.12 & 2.38 & 3.20 & 2.63 & 2.62 & 2.53 & {\textbf{3.21}} & 2.74 & 1.39 & {\color{ForestGreen}\textbf{4.74}} & 0.52 &  & 0.74 & 0.06 & 0.38 & 0.16 & 0.31 & {\color{ForestGreen}\textbf{8.36}} & 0.67 & 0.38 & 0.30 & {\textbf{1.01}} & 0.92 \\
      MaxDD
      & -60.05 & {\textbf{-0.42}} & -1.00 & -4.41 & {\color{ForestGreen}\textbf{-0.25}} & -0.78 & -0.56 & -3.57 & -0.58 & -2.19 & -21.50 &  & -4.55 & -7.55 & -13.52 & -7.71 & -8.16 & {\color{ForestGreen}\textbf{-0.09}} & -6.81 & -11.22 & -14.19 & -5.13 & {\textbf{-4.36}} \\
      \rowcolor{RowAlt}
      Edge
      & 0.07 & 0.00 & 0.03 & 0.07 & 0.11 & 0.07 & 0.00 & {\textbf{0.57}} & 0.00 & {\color{ForestGreen}\textbf{1.37}} & 0.10 &  & {\textbf{0.50}} & 0.00 & 0.04 & 0.00 & 0.00 & {\color{ForestGreen}\textbf{2.78}} & 0.03 & 0.09 & 0.02 & 0.13 & 0.01 \\

    \addlinespace[0.5em]
    \midrule
    \addlinespace[0.5em]
    \multicolumn{24}{l}{\textcolor{dpd2}{\textbf{Panel B: Absolute Error}}} \\
    \addlinespace[0.3em]

      Return
      & -0.33 & 0.06 & 0.09 & 0.12 & 0.06 & 0.11 & 0.12 & {\textbf{0.12}} & -0.06 & {\color{ForestGreen}\textbf{0.18}} & -0.22 &  & -0.14 & -0.31 & -0.36 & -0.18 & -0.29 & {\color{ForestGreen}\textbf{0.14}} & 0.08 & -0.06 & -0.04 & {\textbf{0.09}} & 0.06 \\
      \rowcolor{RowAlt}
      Sharpe
      & -0.44 & 0.17 & 0.26 & {\textbf{0.38}} & 0.18 & 0.30 & 0.37 & 0.29 & -0.21 & {\color{ForestGreen}\textbf{0.51}} & -0.49 &  & -0.52 & -1.68 & -0.87 & -0.85 & -1.18 & {\color{ForestGreen}\textbf{0.59}} & 0.23 & -0.14 & -0.09 & {\textbf{0.25}} & 0.19 \\
      Sortino
      & -0.45 & 0.28 & 0.48 & 0.64 & 0.35 & 0.45 & {\textbf{0.70}} & 0.52 & -0.27 & {\color{ForestGreen}\textbf{1.09}} & -0.52 &  & -0.73 & -1.31 & -1.00 & -0.87 & -1.19 & {\color{ForestGreen}\textbf{1.66}} & 0.38 & -0.18 & -0.12 & {\textbf{0.41}} & 0.36 \\
      \rowcolor{RowAlt}
      Omega
      & 0.34 & 1.28 & 1.47 & 1.68 & 1.33 & 1.58 & {\textbf{1.75}} & 1.46 & 0.72 & {\color{ForestGreen}\textbf{2.11}} & 0.49 &  & 0.49 & 0.06 & 0.35 & 0.28 & 0.24 & {\color{ForestGreen}\textbf{2.43}} & {\textbf{1.40}} & 0.83 & 0.88 & 1.38 & 1.28 \\
      MaxDD
      & -17.35 & {\textbf{-0.55}} & -4.05 & -4.01 & {\color{ForestGreen}\textbf{-0.54}} & -0.79 & -0.58 & -3.35 & -7.50 & -2.48 & -13.58 &  & -3.54 & -4.52 & -8.34 & -2.98 & -5.55 & {\color{ForestGreen}\textbf{-0.29}} & -2.38 & {\textbf{-2.21}} & -4.51 & -2.78 & -2.41 \\
      \rowcolor{RowAlt}
      Edge
      & 0.22 & 0.00 & 0.17 & {\textbf{0.25}} & 0.18 & 0.17 & 0.00 & {\color{ForestGreen}\textbf{0.75}} & 0.01 & 0.21 & 0.18 &  & 0.37 & 0.00 & 0.18 & 0.00 & 0.00 & {\color{ForestGreen}\textbf{0.99}} & 0.19 & 0.29 & 0.08 & {\textbf{0.69}} & 0.05 \\

    \addlinespace[0.5em]
    \midrule
    \addlinespace[0.5em]
    \multicolumn{24}{l}{\textcolor{dpd2}{\textbf{Panel C: Classical Forecast Accuracy}}} \\
    \addlinespace[0.3em]

      RMSE
      & 1.46 & 0.90 & 0.87 & 0.87 & 0.91 & 0.88 & 0.87 & {\textbf{0.84}} & 0.96 & {\color{ForestGreen}\textbf{0.80}} & 1.12 &  & 1.05 & 1.22 & 1.25 & 1.20 & 1.18 & {\color{ForestGreen}\textbf{0.76}} & 1.07 & 1.24 & 1.26 & {\textbf{1.00}} & 1.01 \\
      \rowcolor{RowAlt}
      MAE
      & 1.33 & 0.94 & 0.91 & 0.88 & 0.94 & 0.89 & 0.88 & {\textbf{0.88}} & 1.06 & {\color{ForestGreen}\textbf{0.82}} & 1.22 &  & 1.14 & 1.31 & 1.36 & 1.18 & 1.29 & {\color{ForestGreen}\textbf{0.86}} & 0.92 & 1.06 & 1.04 & {\textbf{0.91}} & 0.94 \\
      $\rho(1)$
      & 0.39 & 0.37 & 0.35 & 0.30 & 0.38 & 0.39 & 0.29 & {\color{ForestGreen}\textbf{0.22}} & 0.41 & {\textbf{0.24}} & 0.44 &  & {\textbf{0.57}} & 0.87 & 0.70 & 0.79 & 0.81 & 0.79 & 0.69 & {\color{ForestGreen}\textbf{0.42}} & 0.76 & 0.87 & 0.85 \\
      \rowcolor{RowAlt}
      DM $t$-stat
      & -1.12 & 1.52 & 1.63 & 1.44 & 1.49 & 1.49 & {\textbf{1.71}} & 1.66 & 0.49 & {\color{ForestGreen}\textbf{1.91}} & -0.96 &  & -0.47 & -2.84 & -1.43 & -1.71 & -1.49 & {\color{ForestGreen}\textbf{1.79}} & -0.45 & -0.88 & -0.95 & {\textbf{0.01}} & -0.07 \\

    \addlinespace[0.3em]
    \bottomrule
  \end{tabular}%
  }

  \vspace{0.25em}
  \parbox{\linewidth}{\scriptsize
    \textit{Notes}: Panels A--B report risk-adjusted metrics; Panel C reports classical forecast accuracy metrics. $\rho(1)$ = first-order autocorrelation of errors.
    Best: \textcolor{ForestGreen}{\textbf{bold green}}; second-best: \textbf{bold}.
  }
  }

\end{table}
\end{landscape}

%% file: 03_tables/houst__h1.tex
% ------------------------------------------------------------
%  EW VERSION - THREE-PANEL TABLE (SE / AE / Classical)
%  Template: dpd2 color + RowAlt alternating rows
% ------------------------------------------------------------

\begin{landscape}
\begin{table}[t!]
  \centering
  \caption{\normalsize Housing Starts ($h=1$)}
  \vspace*{-0.65em}
  \label{tab:houst_h1}

  {\fontfamily{phv}\selectfont
  \resizebox{\linewidth}{!}{%
  \scriptsize
  \setlength{\tabcolsep}{0.3em}
  \renewcommand{\arraystretch}{1.65}
  \begin{tabular}{l r r r r r r r r r r c r r r r r r r r r r}
    \toprule

    & \multicolumn{10}{c}{\textcolor{dpd2}{\textbf{2007Q2 -- 2019Q4}}}
    && \multicolumn{10}{c}{\textcolor{dpd2}{\textbf{2021Q1 -- 2025Q1}}}
    \\
    \cmidrule(lr){2-11}
    \cmidrule(lr){13-22}
    & FAAR & RF & LGB & LGB$^{\texttt{A}}$+ & LGB+ & KRR & NN & RR & SPF & TabPFN 
    && FAAR & RF & LGB & LGB$^{\texttt{A}}$+ & LGB+ & KRR & NN & RR & SPF & TabPFN \\

    \midrule
    \addlinespace[0.5em]
    \multicolumn{22}{l}{\textcolor{dpd2}{\textbf{Panel A: Squared Error}}} \\
    \addlinespace[0.3em]

      Return
      & -1.05 & -0.03 & -0.17 & {\textbf{0.00}} & {\color{ForestGreen}\textbf{0.06}} & -0.12 & -0.27 & -0.17 & -0.04 & -0.06 &  & -4.87 & 0.17 & 0.09 & {\color{ForestGreen}\textbf{0.31}} & 0.08 & 0.12 & -0.29 & 0.20 & {\textbf{0.30}} & -0.24 \\
      \rowcolor{RowAlt}
      Sharpe
      & -0.69 & -0.04 & -0.19 & {\textbf{0.00}} & {\color{ForestGreen}\textbf{0.09}} & -0.17 & -0.27 & -0.21 & -0.08 & -0.08 &  & -1.04 & 0.32 & 0.13 & {\textbf{0.48}} & 0.11 & 0.18 & -0.33 & 0.43 & {\color{ForestGreen}\textbf{0.83}} & -0.23 \\
      Sortino
      & -0.67 & -0.04 & -0.20 & {\textbf{0.00}} & {\color{ForestGreen}\textbf{0.10}} & -0.18 & -0.28 & -0.22 & -0.11 & -0.08 &  & -0.94 & 0.55 & 0.17 & {\textbf{0.95}} & 0.18 & 0.30 & -0.43 & 0.81 & {\color{ForestGreen}\textbf{2.13}} & -0.27 \\
      \rowcolor{RowAlt}
      Omega
      & 0.12 & 0.89 & 0.63 & {\textbf{1.01}} & {\color{ForestGreen}\textbf{1.23}} & 0.67 & 0.49 & 0.55 & 0.87 & 0.81 &  & 0.03 & 1.61 & 1.21 & {\textbf{2.02}} & 1.19 & 1.29 & 0.64 & 1.89 & {\color{ForestGreen}\textbf{4.02}} & 0.67 \\
      MaxDD
      & -53.41 & -12.61 & -15.76 & {\color{ForestGreen}\textbf{-7.19}} & -7.83 & -13.69 & -21.88 & -14.85 & {\textbf{-7.34}} & -13.39 &  & -67.71 & -0.67 & -0.78 & {\textbf{-0.53}} & -0.76 & -4.92 & -7.63 & -3.22 & {\color{ForestGreen}\textbf{-0.31}} & -8.59 \\
      \rowcolor{RowAlt}
      Edge
      & 0.03 & 0.00 & {\textbf{0.23}} & 0.06 & 0.03 & 0.01 & 0.07 & 0.04 & {\color{ForestGreen}\textbf{1.00}} & 0.07 &  & 0.01 & 0.00 & {\textbf{0.20}} & 0.01 & 0.07 & 0.01 & 0.19 & 0.03 & {\color{ForestGreen}\textbf{0.38}} & 0.00 \\

    \addlinespace[0.5em]
    \midrule
    \addlinespace[0.5em]
    \multicolumn{22}{l}{\textcolor{dpd2}{\textbf{Panel B: Absolute Error}}} \\
    \addlinespace[0.3em]

      Return
      & -0.40 & {\color{ForestGreen}\textbf{0.14}} & 0.06 & 0.07 & {\textbf{0.13}} & 0.01 & -0.03 & 0.01 & 0.05 & 0.08 &  & -1.15 & 0.13 & 0.12 & {\color{ForestGreen}\textbf{0.18}} & 0.08 & 0.05 & -0.09 & 0.08 & {\textbf{0.18}} & -0.11 \\
      \rowcolor{RowAlt}
      Sharpe
      & -0.98 & {\textbf{0.51}} & 0.21 & 0.33 & {\color{ForestGreen}\textbf{0.52}} & 0.03 & -0.09 & 0.05 & 0.17 & 0.30 &  & -1.32 & 0.43 & 0.32 & {\textbf{0.44}} & 0.19 & 0.14 & -0.18 & 0.30 & {\color{ForestGreen}\textbf{0.79}} & -0.23 \\
      Sortino
      & -0.93 & {\textbf{0.75}} & 0.28 & 0.52 & {\color{ForestGreen}\textbf{0.88}} & 0.04 & -0.11 & 0.06 & 0.26 & 0.41 &  & -1.13 & {\textbf{0.82}} & 0.50 & 0.78 & 0.33 & 0.22 & -0.25 & 0.50 & {\color{ForestGreen}\textbf{2.38}} & -0.29 \\
      \rowcolor{RowAlt}
      Omega
      & 0.22 & {\color{ForestGreen}\textbf{2.16}} & 1.35 & 1.61 & {\textbf{2.13}} & 1.05 & 0.89 & 1.07 & 1.24 & 1.56 &  & 0.09 & {\textbf{1.78}} & 1.48 & 1.78 & 1.32 & 1.18 & 0.80 & 1.50 & {\color{ForestGreen}\textbf{3.39}} & 0.72 \\
      MaxDD
      & -19.40 & -2.77 & -2.84 & {\textbf{-2.19}} & {\color{ForestGreen}\textbf{-0.85}} & -4.46 & -7.31 & -4.13 & -3.56 & -3.98 &  & -16.49 & -0.55 & {\textbf{-0.50}} & -0.51 & -0.61 & -3.21 & -3.67 & -2.38 & {\color{ForestGreen}\textbf{-0.30}} & -4.02 \\
      \rowcolor{RowAlt}
      Edge
      & 0.05 & 0.07 & {\textbf{0.43}} & 0.07 & 0.13 & 0.07 & 0.16 & 0.07 & {\color{ForestGreen}\textbf{1.59}} & 0.15 &  & 0.07 & 0.00 & 0.32 & 0.06 & 0.27 & 0.09 & {\color{ForestGreen}\textbf{0.49}} & 0.10 & {\textbf{0.38}} & 0.00 \\

    \addlinespace[0.5em]
    \midrule
    \addlinespace[0.5em]
    \multicolumn{22}{l}{\textcolor{dpd2}{\textbf{Panel C: Classical Forecast Accuracy}}} \\
    \addlinespace[0.3em]

      RMSE
      & 1.43 & 1.02 & 1.08 & {\textbf{1.00}} & {\color{ForestGreen}\textbf{0.97}} & 1.06 & 1.12 & 1.08 & 1.02 & 1.03 &  & 2.42 & 0.91 & 0.95 & {\color{ForestGreen}\textbf{0.83}} & 0.96 & 0.94 & 1.13 & 0.89 & {\textbf{0.84}} & 1.11 \\
      \rowcolor{RowAlt}
      MAE
      & 1.40 & {\color{ForestGreen}\textbf{0.86}} & 0.94 & 0.93 & {\textbf{0.87}} & 0.99 & 1.03 & 0.99 & 0.95 & 0.92 &  & 2.15 & 0.87 & 0.88 & {\color{ForestGreen}\textbf{0.82}} & 0.92 & 0.95 & 1.09 & 0.92 & {\textbf{0.82}} & 1.11 \\
      $\rho(1)$
      & 0.48 & 0.30 & 0.26 & 0.26 & 0.24 & 0.37 & {\textbf{0.23}} & 0.41 & 0.34 & {\color{ForestGreen}\textbf{0.23}} &  & 0.34 & -0.14 & -0.26 & {\color{ForestGreen}\textbf{-0.04}} & {\textbf{-0.09}} & 0.11 & 0.14 & 0.09 & 0.17 & 0.22 \\
      \rowcolor{RowAlt}
      DM $t$-stat
      & -2.46 & -0.14 & -0.67 & {\textbf{0.01}} & {\color{ForestGreen}\textbf{0.36}} & -0.60 & -0.98 & -0.69 & -0.31 & -0.28 &  & -1.88 & 0.61 & 0.26 & {\textbf{1.02}} & 0.21 & 0.33 & -0.66 & 0.79 & {\color{ForestGreen}\textbf{1.87}} & -0.46 \\

    \addlinespace[0.3em]
    \bottomrule
  \end{tabular}%
  }

  \vspace{0.25em}
  \parbox{\linewidth}{\scriptsize
    \textit{Notes}: Panels A--B report risk-adjusted metrics; Panel C reports classical forecast accuracy metrics. $\rho(1)$ = first-order autocorrelation of errors.
    Best: \textcolor{ForestGreen}{\textbf{bold green}}; second-best: \textbf{bold}.
  }
  }

\end{table}
\end{landscape}

%% file: 03_tables/houst__h2.tex
% ------------------------------------------------------------
%  EW VERSION - THREE-PANEL TABLE (SE / AE / Classical)
%  Template: dpd2 color + RowAlt alternating rows
% ------------------------------------------------------------

\begin{landscape}
\begin{table}[t!]
  \centering
  \caption{\normalsize Housing Starts ($h=2$)}
  \vspace*{-0.65em}
  \label{tab:houst_h2}

  {\fontfamily{phv}\selectfont
  \resizebox{\linewidth}{!}{%
  \scriptsize
  \setlength{\tabcolsep}{0.3em}
  \renewcommand{\arraystretch}{1.65}
  \begin{tabular}{l r r r r r r r r r r c r r r r r r r r r r}
    \toprule

    & \multicolumn{10}{c}{\textcolor{dpd2}{\textbf{2007Q2 -- 2019Q4}}}
    && \multicolumn{10}{c}{\textcolor{dpd2}{\textbf{2021Q1 -- 2025Q1}}}
    \\
    \cmidrule(lr){2-11}
    \cmidrule(lr){13-22}
    & FAAR & RF & LGB & LGB$^{\texttt{A}}$+ & LGB+ & KRR & NN & RR & SPF & TabPFN 
    && FAAR & RF & LGB & LGB$^{\texttt{A}}$+ & LGB+ & KRR & NN & RR & SPF & TabPFN \\

    \midrule
    \addlinespace[0.5em]
    \multicolumn{22}{l}{\textcolor{dpd2}{\textbf{Panel A: Squared Error}}} \\
    \addlinespace[0.3em]

      Return
      & -0.18 & -0.06 & -0.06 & -0.07 & -0.09 & {\color{ForestGreen}\textbf{-0.04}} & -0.23 & {\textbf{-0.05}} & -0.06 & -0.05 &  & -0.02 & {\textbf{0.09}} & -0.22 & -0.33 & -0.28 & 0.04 & -0.94 & {\color{ForestGreen}\textbf{0.13}} & 0.02 & 0.06 \\
      \rowcolor{RowAlt}
      Sharpe
      & -0.25 & -0.13 & -0.13 & -0.16 & -0.15 & {\color{ForestGreen}\textbf{-0.09}} & -0.36 & -0.12 & -0.12 & {\textbf{-0.11}} &  & -0.02 & {\textbf{0.18}} & -0.46 & -0.47 & -0.30 & 0.05 & -0.48 & {\color{ForestGreen}\textbf{0.23}} & 0.05 & 0.10 \\
      Sortino
      & -0.28 & -0.14 & -0.15 & -0.18 & -0.17 & {\color{ForestGreen}\textbf{-0.12}} & -0.39 & -0.14 & -0.16 & {\textbf{-0.13}} &  & -0.03 & {\textbf{0.27}} & -0.54 & -0.55 & -0.36 & 0.07 & -0.51 & {\color{ForestGreen}\textbf{0.34}} & 0.06 & 0.14 \\
      \rowcolor{RowAlt}
      Omega
      & 0.62 & 0.78 & 0.81 & 0.76 & 0.78 & {\color{ForestGreen}\textbf{0.87}} & 0.50 & 0.80 & {\textbf{0.83}} & 0.83 &  & 0.97 & {\textbf{1.27}} & 0.48 & 0.53 & 0.63 & 1.08 & 0.39 & {\color{ForestGreen}\textbf{1.43}} & 1.07 & 1.15 \\
      MaxDD
      & -17.66 & -10.26 & -9.81 & -10.78 & -16.55 & {\textbf{-8.82}} & -18.99 & {\color{ForestGreen}\textbf{-7.92}} & -9.64 & -9.77 &  & -7.28 & {\textbf{-0.81}} & -5.89 & -8.42 & -9.67 & -6.89 & -22.55 & -3.87 & {\color{ForestGreen}\textbf{-0.79}} & -4.16 \\
      \rowcolor{RowAlt}
      Edge
      & 0.29 & 0.03 & 0.02 & 0.01 & {\color{ForestGreen}\textbf{0.73}} & 0.07 & 0.00 & 0.00 & {\textbf{0.48}} & 0.00 &  & {\color{ForestGreen}\textbf{0.29}} & 0.00 & 0.00 & 0.07 & {\textbf{0.17}} & 0.00 & 0.12 & 0.00 & 0.01 & 0.04 \\

    \addlinespace[0.5em]
    \midrule
    \addlinespace[0.5em]
    \multicolumn{22}{l}{\textcolor{dpd2}{\textbf{Panel B: Absolute Error}}} \\
    \addlinespace[0.3em]

      Return
      & 0.02 & {\textbf{0.04}} & -0.03 & 0.03 & 0.00 & -0.01 & -0.06 & 0.02 & {\color{ForestGreen}\textbf{0.07}} & 0.01 &  & 0.03 & -0.02 & -0.10 & -0.21 & -0.12 & {\textbf{0.03}} & -0.22 & 0.03 & -0.01 & {\color{ForestGreen}\textbf{0.03}} \\
      \rowcolor{RowAlt}
      Sharpe
      & 0.06 & {\textbf{0.17}} & -0.11 & 0.10 & 0.01 & -0.05 & -0.23 & 0.08 & {\color{ForestGreen}\textbf{0.21}} & 0.05 &  & 0.05 & -0.07 & -0.33 & -0.50 & -0.25 & 0.07 & -0.29 & {\color{ForestGreen}\textbf{0.09}} & -0.05 & {\textbf{0.09}} \\
      Sortino
      & 0.08 & {\textbf{0.26}} & -0.14 & 0.14 & 0.02 & -0.07 & -0.29 & 0.13 & {\color{ForestGreen}\textbf{0.34}} & 0.08 &  & 0.07 & -0.11 & -0.45 & -0.60 & -0.32 & 0.10 & -0.34 & {\textbf{0.13}} & -0.08 & {\color{ForestGreen}\textbf{0.13}} \\
      \rowcolor{RowAlt}
      Omega
      & 1.08 & {\textbf{1.29}} & 0.86 & 1.15 & 1.02 & 0.94 & 0.73 & 1.12 & {\color{ForestGreen}\textbf{1.32}} & 1.07 &  & 1.08 & 0.92 & 0.62 & 0.53 & 0.71 & 1.09 & 0.64 & {\color{ForestGreen}\textbf{1.13}} & 0.93 & {\textbf{1.13}} \\
      MaxDD
      & -5.50 & -3.71 & -4.74 & -6.02 & -7.86 & -4.83 & -7.08 & {\color{ForestGreen}\textbf{-2.91}} & {\textbf{-3.44}} & -4.79 &  & -3.96 & {\color{ForestGreen}\textbf{-0.76}} & -3.44 & -5.02 & -5.23 & -4.13 & -7.83 & -2.61 & {\textbf{-0.81}} & -0.97 \\
      \rowcolor{RowAlt}
      Edge
      & 0.41 & 0.13 & 0.11 & 0.08 & {\textbf{0.70}} & 0.15 & 0.01 & 0.03 & {\color{ForestGreen}\textbf{0.99}} & 0.01 &  & {\color{ForestGreen}\textbf{0.45}} & 0.00 & 0.00 & 0.09 & {\textbf{0.40}} & 0.05 & 0.28 & 0.00 & 0.08 & 0.04 \\

    \addlinespace[0.5em]
    \midrule
    \addlinespace[0.5em]
    \multicolumn{22}{l}{\textcolor{dpd2}{\textbf{Panel C: Classical Forecast Accuracy}}} \\
    \addlinespace[0.3em]

      RMSE
      & 1.09 & 1.03 & 1.03 & 1.04 & 1.04 & {\color{ForestGreen}\textbf{1.02}} & 1.11 & {\textbf{1.02}} & 1.03 & 1.02 &  & 1.01 & {\textbf{0.96}} & 1.10 & 1.15 & 1.13 & 0.98 & 1.39 & {\color{ForestGreen}\textbf{0.94}} & 0.99 & 0.97 \\
      \rowcolor{RowAlt}
      MAE
      & 0.98 & {\textbf{0.96}} & 1.03 & 0.97 & 1.00 & 1.01 & 1.06 & 0.98 & {\color{ForestGreen}\textbf{0.93}} & 0.99 &  & 0.97 & 1.02 & 1.10 & 1.21 & 1.12 & {\textbf{0.97}} & 1.22 & 0.97 & 1.01 & {\color{ForestGreen}\textbf{0.97}} \\
      $\rho(1)$
      & 0.50 & 0.42 & 0.44 & 0.51 & 0.49 & {\color{ForestGreen}\textbf{0.41}} & 0.45 & 0.48 & 0.43 & {\textbf{0.42}} &  & 0.28 & 0.17 & 0.15 & 0.34 & 0.34 & 0.12 & {\color{ForestGreen}\textbf{0.06}} & {\textbf{0.07}} & 0.17 & 0.11 \\
      \rowcolor{RowAlt}
      DM $t$-stat
      & -0.69 & -0.43 & -0.46 & -0.42 & -0.46 & {\color{ForestGreen}\textbf{-0.30}} & -0.96 & -0.40 & -0.44 & {\textbf{-0.37}} &  & -0.04 & {\textbf{0.40}} & -1.00 & -0.97 & -0.54 & 0.09 & -1.05 & {\color{ForestGreen}\textbf{0.42}} & 0.10 & 0.18 \\

    \addlinespace[0.3em]
    \bottomrule
  \end{tabular}%
  }

  \vspace{0.25em}
  \parbox{\linewidth}{\scriptsize
    \textit{Notes}: Panels A--B report risk-adjusted metrics; Panel C reports classical forecast accuracy metrics. $\rho(1)$ = first-order autocorrelation of errors.
    Best: \textcolor{ForestGreen}{\textbf{bold green}}; second-best: \textbf{bold}.
  }
  }

\end{table}
\end{landscape}

%% file: 03_tables/houst__h4.tex
% ------------------------------------------------------------
%  EW VERSION - THREE-PANEL TABLE (SE / AE / Classical)
%  Template: dpd2 color + RowAlt alternating rows
% ------------------------------------------------------------

\begin{landscape}
\begin{table}[t!]
  \centering
  \caption{\normalsize Housing Starts ($h=4$)}
  \vspace*{-0.65em}
  \label{tab:houst_h4}

  {\fontfamily{phv}\selectfont
  \resizebox{\linewidth}{!}{%
  \scriptsize
  \setlength{\tabcolsep}{0.3em}
  \renewcommand{\arraystretch}{1.65}
  \begin{tabular}{l r r r r r r r r r r c r r r r r r r r r r}
    \toprule

    & \multicolumn{10}{c}{\textcolor{dpd2}{\textbf{2007Q2 -- 2019Q4}}}
    && \multicolumn{10}{c}{\textcolor{dpd2}{\textbf{2021Q1 -- 2025Q1}}}
    \\
    \cmidrule(lr){2-11}
    \cmidrule(lr){13-22}
    & FAAR & RF & LGB & LGB$^{\texttt{A}}$+ & LGB+ & KRR & NN & RR & SPF & TabPFN 
    && FAAR & RF & LGB & LGB$^{\texttt{A}}$+ & LGB+ & KRR & NN & RR & SPF & TabPFN \\

    \midrule
    \addlinespace[0.5em]
    \multicolumn{22}{l}{\textcolor{dpd2}{\textbf{Panel A: Squared Error}}} \\
    \addlinespace[0.3em]

      Return
      & -0.11 & 0.01 & -0.07 & -0.04 & -0.10 & {\color{ForestGreen}\textbf{0.04}} & -0.09 & -0.00 & -0.08 & {\textbf{0.04}} &  & -2.23 & -0.28 & -1.09 & -1.11 & -0.62 & {\color{ForestGreen}\textbf{0.01}} & -0.94 & {\textbf{-0.02}} & -0.31 & -0.21 \\
      \rowcolor{RowAlt}
      Sharpe
      & -0.30 & 0.03 & -0.17 & -0.10 & -0.23 & {\color{ForestGreen}\textbf{0.39}} & -0.24 & -0.02 & -0.17 & {\textbf{0.14}} &  & -0.66 & -0.43 & -0.91 & -0.89 & -0.80 & {\color{ForestGreen}\textbf{0.02}} & -1.09 & {\textbf{-0.05}} & -0.74 & -0.68 \\
      Sortino
      & -0.33 & 0.04 & -0.20 & -0.12 & -0.25 & {\color{ForestGreen}\textbf{0.83}} & -0.25 & -0.02 & -0.21 & {\textbf{0.18}} &  & -0.64 & -0.50 & -0.88 & -0.86 & -0.81 & {\color{ForestGreen}\textbf{0.03}} & -1.00 & {\textbf{-0.06}} & -0.75 & -0.81 \\
      \rowcolor{RowAlt}
      Omega
      & 0.58 & 1.05 & 0.76 & 0.84 & 0.68 & {\color{ForestGreen}\textbf{1.80}} & 0.62 & 0.97 & 0.76 & {\textbf{1.27}} &  & 0.08 & 0.51 & 0.21 & 0.22 & 0.27 & {\color{ForestGreen}\textbf{1.03}} & 0.17 & {\textbf{0.94}} & 0.32 & 0.41 \\
      MaxDD
      & -8.19 & -3.85 & -7.38 & -7.81 & -10.87 & {\color{ForestGreen}\textbf{-0.45}} & -8.73 & {\textbf{-3.29}} & -10.08 & -4.50 &  & -38.87 & -5.80 & -15.56 & -21.27 & -10.77 & {\color{ForestGreen}\textbf{-0.90}} & -18.58 & {\textbf{-2.36}} & -5.72 & -4.26 \\
      \rowcolor{RowAlt}
      Edge
      & 0.17 & 0.20 & {\textbf{0.46}} & 0.01 & 0.17 & 0.00 & 0.10 & 0.00 & {\color{ForestGreen}\textbf{1.14}} & 0.10 &  & 0.00 & 0.00 & 0.04 & {\textbf{0.38}} & 0.04 & 0.00 & 0.08 & {\color{ForestGreen}\textbf{0.63}} & 0.37 & 0.00 \\

    \addlinespace[0.5em]
    \midrule
    \addlinespace[0.5em]
    \multicolumn{22}{l}{\textcolor{dpd2}{\textbf{Panel B: Absolute Error}}} \\
    \addlinespace[0.3em]

      Return
      & 0.00 & {\textbf{0.04}} & -0.01 & {\color{ForestGreen}\textbf{0.05}} & -0.00 & 0.03 & -0.03 & 0.00 & 0.04 & 0.03 &  & -0.67 & -0.17 & -0.49 & -0.33 & -0.29 & {\color{ForestGreen}\textbf{0.02}} & -0.39 & {\textbf{-0.01}} & -0.14 & -0.15 \\
      \rowcolor{RowAlt}
      Sharpe
      & 0.01 & {\textbf{0.24}} & -0.03 & 0.23 & -0.02 & {\color{ForestGreen}\textbf{0.29}} & -0.14 & 0.01 & 0.12 & 0.17 &  & -0.78 & -0.41 & -0.88 & -0.67 & -0.75 & {\color{ForestGreen}\textbf{0.13}} & -0.97 & {\textbf{-0.05}} & -0.57 & -0.70 \\
      Sortino
      & 0.02 & {\textbf{0.41}} & -0.04 & 0.33 & -0.02 & {\color{ForestGreen}\textbf{0.49}} & -0.16 & 0.02 & 0.17 & 0.22 &  & -0.75 & -0.55 & -0.93 & -0.74 & -0.83 & {\color{ForestGreen}\textbf{0.20}} & -0.96 & {\textbf{-0.06}} & -0.66 & -0.85 \\
      \rowcolor{RowAlt}
      Omega
      & 1.02 & {\textbf{1.37}} & 0.96 & 1.36 & 0.97 & {\color{ForestGreen}\textbf{1.49}} & 0.80 & 1.02 & 1.17 & 1.29 &  & 0.15 & 0.58 & 0.32 & 0.43 & 0.37 & {\color{ForestGreen}\textbf{1.20}} & 0.30 & {\textbf{0.93}} & 0.49 & 0.43 \\
      MaxDD
      & -3.11 & {\color{ForestGreen}\textbf{-0.34}} & -0.89 & -0.95 & -4.38 & {\textbf{-0.42}} & -4.73 & -2.40 & -4.54 & -0.86 &  & -11.15 & -3.02 & -6.33 & -7.69 & -4.96 & {\color{ForestGreen}\textbf{-0.62}} & -7.89 & {\textbf{-1.54}} & -2.52 & -2.67 \\
      \rowcolor{RowAlt}
      Edge
      & 0.34 & 0.19 & {\textbf{0.57}} & 0.07 & 0.36 & 0.02 & 0.05 & 0.00 & {\color{ForestGreen}\textbf{1.84}} & 0.06 &  & 0.00 & 0.06 & 0.15 & {\color{ForestGreen}\textbf{0.93}} & 0.13 & 0.07 & 0.19 & 0.26 & {\textbf{0.45}} & 0.00 \\

    \addlinespace[0.5em]
    \midrule
    \addlinespace[0.5em]
    \multicolumn{22}{l}{\textcolor{dpd2}{\textbf{Panel C: Classical Forecast Accuracy}}} \\
    \addlinespace[0.3em]

      RMSE
      & 1.05 & 1.00 & 1.03 & 1.02 & 1.05 & {\color{ForestGreen}\textbf{0.98}} & 1.04 & 1.00 & 1.04 & {\textbf{0.98}} &  & 1.80 & 1.13 & 1.44 & 1.45 & 1.27 & {\color{ForestGreen}\textbf{1.00}} & 1.39 & {\textbf{1.01}} & 1.15 & 1.10 \\
      \rowcolor{RowAlt}
      MAE
      & 1.00 & {\textbf{0.96}} & 1.01 & {\color{ForestGreen}\textbf{0.95}} & 1.00 & 0.97 & 1.03 & 1.00 & 0.96 & 0.97 &  & 1.67 & 1.17 & 1.49 & 1.33 & 1.29 & {\color{ForestGreen}\textbf{0.98}} & 1.39 & {\textbf{1.01}} & 1.14 & 1.15 \\
      $\rho(1)$
      & 0.43 & 0.42 & 0.38 & {\textbf{0.36}} & 0.44 & 0.40 & 0.44 & 0.42 & 0.46 & {\color{ForestGreen}\textbf{0.28}} &  & {\color{ForestGreen}\textbf{0.06}} & 0.25 & 0.19 & 0.42 & 0.29 & 0.15 & 0.22 & {\textbf{0.09}} & 0.24 & 0.24 \\
      \rowcolor{RowAlt}
      DM $t$-stat
      & -1.19 & 0.09 & -0.54 & -0.37 & -0.79 & {\color{ForestGreen}\textbf{1.46}} & -0.80 & -0.07 & -0.59 & {\textbf{0.35}} &  & -1.15 & -1.06 & -1.94 & -2.06 & -2.95 & {\color{ForestGreen}\textbf{0.06}} & -2.22 & {\textbf{-0.10}} & -1.73 & -2.16 \\

    \addlinespace[0.3em]
    \bottomrule
  \end{tabular}%
  }

  \vspace{0.25em}
  \parbox{\linewidth}{\scriptsize
    \textit{Notes}: Panels A--B report risk-adjusted metrics; Panel C reports classical forecast accuracy metrics. $\rho(1)$ = first-order autocorrelation of errors.
    Best: \textcolor{ForestGreen}{\textbf{bold green}}; second-best: \textbf{bold}.
  }
  }

\end{table}
\end{landscape}

%% file: 03_tables/indpro__h1.tex
% ------------------------------------------------------------
%  EW VERSION - THREE-PANEL TABLE (SE / AE / Classical)
%  Template: dpd2 color + RowAlt alternating rows
% ------------------------------------------------------------

\begin{landscape}
\begin{table}[t!]
  \centering
  \caption{\normalsize Industrial Production ($h=1$)}
  \vspace*{-0.65em}
  \label{tab:indpro_h1}

  {\fontfamily{phv}\selectfont
  \resizebox{\linewidth}{!}{%
  \scriptsize
  \setlength{\tabcolsep}{0.3em}
  \renewcommand{\arraystretch}{1.65}
  \begin{tabular}{l r r r r r r r r r r c r r r r r r r r r r}
    \toprule

    & \multicolumn{10}{c}{\textcolor{dpd2}{\textbf{2007Q2 -- 2019Q4}}}
    && \multicolumn{10}{c}{\textcolor{dpd2}{\textbf{2022Q1 -- 2025Q1}}}
    \\
    \cmidrule(lr){2-11}
    \cmidrule(lr){13-22}
    & FAAR & RF & LGB & LGB$^{\texttt{A}}$+ & LGB+ & KRR & NN & RR & SPF & TabPFN 
    && FAAR & RF & LGB & LGB$^{\texttt{A}}$+ & LGB+ & KRR & NN & RR & SPF & TabPFN \\

    \midrule
    \addlinespace[0.5em]
    \multicolumn{22}{l}{\textcolor{dpd2}{\textbf{Panel A: Squared Error}}} \\
    \addlinespace[0.3em]

      Return
      & 0.02 & -0.02 & -0.03 & {\color{ForestGreen}\textbf{0.23}} & -0.01 & -0.12 & {\textbf{0.03}} & -0.25 & -0.46 & -0.25 &  & -5.53 & -0.12 & -0.64 & -1.85 & 0.06 & -0.85 & -1.60 & {\textbf{0.08}} & {\color{ForestGreen}\textbf{0.21}} & -0.87 \\
      \rowcolor{RowAlt}
      Sharpe
      & 0.02 & -0.07 & -0.04 & {\color{ForestGreen}\textbf{0.42}} & -0.04 & -0.25 & {\textbf{0.04}} & -0.43 & -0.49 & -0.38 &  & -0.98 & -0.12 & -0.88 & -0.93 & 0.06 & -0.66 & -0.62 & {\textbf{0.14}} & {\color{ForestGreen}\textbf{0.44}} & -0.41 \\
      Sortino
      & 0.04 & -0.10 & -0.05 & {\color{ForestGreen}\textbf{1.57}} & -0.05 & -0.29 & {\textbf{0.06}} & -0.44 & -0.51 & -0.39 &  & -0.91 & -0.15 & -0.86 & -0.88 & 0.07 & -0.67 & -0.62 & {\textbf{0.20}} & {\color{ForestGreen}\textbf{1.05}} & -0.43 \\
      \rowcolor{RowAlt}
      Omega
      & 1.05 & 0.89 & 0.92 & {\color{ForestGreen}\textbf{2.77}} & 0.95 & 0.63 & {\textbf{1.08}} & 0.37 & 0.29 & 0.40 &  & 0.07 & 0.84 & 0.26 & 0.16 & 1.10 & 0.29 & 0.21 & {\textbf{1.20}} & {\color{ForestGreen}\textbf{1.99}} & 0.37 \\
      MaxDD
      & -16.59 & -10.16 & {\textbf{-9.18}} & {\color{ForestGreen}\textbf{-0.30}} & -11.58 & -13.46 & -14.62 & -17.61 & -26.70 & -19.54 &  & -73.70 & -7.67 & -10.06 & -26.17 & -6.82 & -14.93 & -25.45 & {\textbf{-2.85}} & {\color{ForestGreen}\textbf{-1.90}} & -16.91 \\
      \rowcolor{RowAlt}
      Edge
      & {\textbf{0.70}} & 0.00 & 0.01 & 0.36 & 0.07 & 0.06 & {\color{ForestGreen}\textbf{2.12}} & 0.00 & 0.04 & 0.21 &  & 0.06 & {\color{ForestGreen}\textbf{0.18}} & {\textbf{0.13}} & 0.00 & 0.02 & 0.00 & 0.00 & 0.00 & 0.00 & 0.12 \\

    \addlinespace[0.5em]
    \midrule
    \addlinespace[0.5em]
    \multicolumn{22}{l}{\textcolor{dpd2}{\textbf{Panel B: Absolute Error}}} \\
    \addlinespace[0.3em]

      Return
      & -0.15 & -0.08 & -0.08 & {\color{ForestGreen}\textbf{0.03}} & -0.17 & -0.06 & {\textbf{-0.01}} & -0.14 & -0.17 & -0.10 &  & -1.13 & -0.01 & -0.30 & -0.66 & {\textbf{0.07}} & -0.35 & -0.36 & -0.03 & {\color{ForestGreen}\textbf{0.11}} & -0.23 \\
      \rowcolor{RowAlt}
      Sharpe
      & -0.39 & -0.29 & -0.25 & {\color{ForestGreen}\textbf{0.12}} & -0.55 & -0.18 & {\textbf{-0.03}} & -0.47 & -0.46 & -0.30 &  & -1.02 & -0.02 & -0.77 & -0.98 & {\textbf{0.13}} & -0.64 & -0.48 & -0.07 & {\color{ForestGreen}\textbf{0.37}} & -0.33 \\
      Sortino
      & -0.51 & -0.38 & -0.31 & {\color{ForestGreen}\textbf{0.19}} & -0.64 & -0.24 & {\textbf{-0.04}} & -0.55 & -0.53 & -0.35 &  & -0.95 & -0.03 & -0.79 & -0.95 & {\textbf{0.17}} & -0.68 & -0.50 & -0.09 & {\color{ForestGreen}\textbf{0.65}} & -0.37 \\
      \rowcolor{RowAlt}
      Omega
      & 0.59 & 0.66 & 0.70 & {\color{ForestGreen}\textbf{1.17}} & 0.50 & 0.78 & {\textbf{0.97}} & 0.53 & 0.51 & 0.66 &  & 0.19 & 0.97 & 0.36 & 0.23 & {\textbf{1.21}} & 0.38 & 0.41 & 0.91 & {\color{ForestGreen}\textbf{1.66}} & 0.56 \\
      MaxDD
      & -14.60 & -9.98 & -9.27 & {\color{ForestGreen}\textbf{-7.49}} & -14.76 & {\textbf{-8.96}} & -10.23 & -10.72 & -10.75 & -11.43 &  & -16.05 & -4.53 & -5.65 & -9.84 & -3.95 & -6.72 & -7.31 & {\textbf{-2.46}} & {\color{ForestGreen}\textbf{-0.70}} & -6.31 \\
      \rowcolor{RowAlt}
      Edge
      & {\textbf{0.41}} & 0.00 & 0.05 & 0.29 & 0.10 & 0.23 & {\color{ForestGreen}\textbf{1.48}} & 0.00 & 0.23 & 0.34 &  & {\color{ForestGreen}\textbf{0.28}} & {\textbf{0.26}} & 0.25 & 0.00 & 0.05 & 0.00 & 0.00 & 0.00 & 0.02 & 0.25 \\

    \addlinespace[0.5em]
    \midrule
    \addlinespace[0.5em]
    \multicolumn{22}{l}{\textcolor{dpd2}{\textbf{Panel C: Classical Forecast Accuracy}}} \\
    \addlinespace[0.3em]

      RMSE
      & 0.99 & 1.01 & 1.01 & {\color{ForestGreen}\textbf{0.88}} & 1.01 & 1.06 & {\textbf{0.99}} & 1.12 & 1.21 & 1.12 &  & 2.56 & 1.06 & 1.28 & 1.69 & 0.97 & 1.36 & 1.61 & {\textbf{0.96}} & {\color{ForestGreen}\textbf{0.89}} & 1.37 \\
      \rowcolor{RowAlt}
      MAE
      & 1.15 & 1.08 & 1.08 & {\color{ForestGreen}\textbf{0.97}} & 1.17 & 1.06 & {\textbf{1.01}} & 1.14 & 1.17 & 1.10 &  & 2.13 & 1.01 & 1.30 & 1.66 & {\textbf{0.93}} & 1.35 & 1.36 & 1.03 & {\color{ForestGreen}\textbf{0.89}} & 1.23 \\
      $\rho(1)$
      & 0.46 & 0.53 & 0.48 & 0.48 & 0.59 & 0.52 & {\color{ForestGreen}\textbf{0.31}} & 0.62 & 0.71 & {\textbf{0.36}} &  & 0.58 & 0.45 & 0.53 & 0.58 & 0.40 & 0.63 & 0.54 & {\textbf{0.33}} & {\color{ForestGreen}\textbf{0.08}} & 0.44 \\
      \rowcolor{RowAlt}
      DM $t$-stat
      & 0.08 & -0.19 & -0.32 & {\color{ForestGreen}\textbf{1.61}} & -0.08 & -0.84 & {\textbf{0.10}} & -1.21 & -1.23 & -0.82 &  & -1.40 & -0.25 & -1.39 & -1.71 & 0.10 & -0.99 & -0.93 & {\textbf{0.32}} & {\color{ForestGreen}\textbf{0.99}} & -0.74 \\

    \addlinespace[0.3em]
    \bottomrule
  \end{tabular}%
  }

  \vspace{0.25em}
  \parbox{\linewidth}{\scriptsize
    \textit{Notes}: Panels A--B report risk-adjusted metrics; Panel C reports classical forecast accuracy metrics. $\rho(1)$ = first-order autocorrelation of errors.
    Best: \textcolor{ForestGreen}{\textbf{bold green}}; second-best: \textbf{bold}.
  }
  }

\end{table}
\end{landscape}

%% file: 03_tables/indpro__h2.tex
% ------------------------------------------------------------
%  EW VERSION - THREE-PANEL TABLE (SE / AE / Classical)
%  Template: dpd2 color + RowAlt alternating rows
% ------------------------------------------------------------

\begin{landscape}
\begin{table}[t!]
  \centering
  \caption{\normalsize Industrial Production ($h=2$)}
  \vspace*{-0.65em}
  \label{tab:indpro_h2}

  {\fontfamily{phv}\selectfont
  \resizebox{\linewidth}{!}{%
  \scriptsize
  \setlength{\tabcolsep}{0.3em}
  \renewcommand{\arraystretch}{1.65}
  \begin{tabular}{l r r r r r r r r r r c r r r r r r r r r r}
    \toprule

    & \multicolumn{10}{c}{\textcolor{dpd2}{\textbf{2007Q2 -- 2019Q4}}}
    && \multicolumn{10}{c}{\textcolor{dpd2}{\textbf{2022Q1 -- 2025Q1}}}
    \\
    \cmidrule(lr){2-11}
    \cmidrule(lr){13-22}
    & FAAR & RF & LGB & LGB$^{\texttt{A}}$+ & LGB+ & KRR & NN & RR & SPF & TabPFN 
    && FAAR & RF & LGB & LGB$^{\texttt{A}}$+ & LGB+ & KRR & NN & RR & SPF & TabPFN \\

    \midrule
    \addlinespace[0.5em]
    \multicolumn{22}{l}{\textcolor{dpd2}{\textbf{Panel A: Squared Error}}} \\
    \addlinespace[0.3em]

      Return
      & -0.28 & 0.02 & {\textbf{0.11}} & -0.05 & -0.10 & -0.05 & {\color{ForestGreen}\textbf{0.18}} & 0.06 & -0.10 & -0.07 &  & -2.98 & 0.09 & -3.00 & -1.61 & 0.12 & -0.48 & -0.99 & {\textbf{0.27}} & {\color{ForestGreen}\textbf{0.39}} & -2.44 \\
      \rowcolor{RowAlt}
      Sharpe
      & -0.44 & 0.06 & {\color{ForestGreen}\textbf{0.24}} & -0.13 & -0.32 & -0.12 & {\textbf{0.23}} & 0.21 & -0.38 & -0.22 &  & -1.86 & 0.09 & -0.60 & -0.97 & 0.13 & -0.46 & -0.65 & {\textbf{0.48}} & {\color{ForestGreen}\textbf{0.73}} & -1.17 \\
      Sortino
      & -0.55 & 0.11 & {\textbf{0.54}} & -0.15 & -0.35 & -0.17 & {\color{ForestGreen}\textbf{0.72}} & 0.39 & -0.48 & -0.26 &  & -1.41 & 0.15 & -0.59 & -0.96 & 0.20 & -0.55 & -0.66 & {\textbf{1.55}} & {\color{ForestGreen}\textbf{1.81}} & -1.06 \\
      \rowcolor{RowAlt}
      Omega
      & 0.49 & 1.12 & {\textbf{1.67}} & 0.79 & 0.57 & 0.82 & {\color{ForestGreen}\textbf{1.69}} & 1.48 & 0.45 & 0.69 &  & 0.07 & 1.13 & 0.12 & 0.30 & 1.19 & 0.56 & 0.32 & {\textbf{2.30}} & {\color{ForestGreen}\textbf{3.17}} & 0.18 \\
      MaxDD
      & -25.26 & -0.87 & {\color{ForestGreen}\textbf{-0.48}} & -7.54 & -8.23 & -7.29 & {\textbf{-0.54}} & -0.54 & -5.61 & -6.90 &  & -39.13 & -5.08 & -40.11 & -21.20 & -6.38 & -9.94 & -16.57 & {\textbf{-1.22}} & {\color{ForestGreen}\textbf{-0.39}} & -33.69 \\
      \rowcolor{RowAlt}
      Edge
      & 0.22 & 0.00 & 0.00 & 0.03 & 0.00 & 0.14 & {\color{ForestGreen}\textbf{2.78}} & 0.00 & 0.05 & {\textbf{0.29}} &  & 0.09 & {\textbf{0.13}} & 0.00 & 0.04 & 0.03 & 0.09 & 0.00 & 0.01 & {\color{ForestGreen}\textbf{0.24}} & 0.00 \\

    \addlinespace[0.5em]
    \midrule
    \addlinespace[0.5em]
    \multicolumn{22}{l}{\textcolor{dpd2}{\textbf{Panel B: Absolute Error}}} \\
    \addlinespace[0.3em]

      Return
      & -0.29 & -0.07 & {\textbf{-0.03}} & -0.07 & -0.12 & -0.07 & -0.04 & {\color{ForestGreen}\textbf{-0.02}} & -0.08 & -0.07 &  & -1.22 & {\textbf{0.08}} & -0.56 & -0.71 & 0.03 & -0.36 & -0.40 & 0.04 & {\color{ForestGreen}\textbf{0.21}} & -0.95 \\
      \rowcolor{RowAlt}
      Sharpe
      & -0.67 & -0.36 & -0.12 & -0.24 & -0.55 & -0.23 & {\textbf{-0.11}} & {\color{ForestGreen}\textbf{-0.11}} & -0.40 & -0.21 &  & -1.97 & {\textbf{0.13}} & -0.65 & -0.88 & 0.06 & -0.62 & -0.67 & 0.09 & {\color{ForestGreen}\textbf{0.63}} & -1.19 \\
      Sortino
      & -0.75 & -0.45 & {\textbf{-0.17}} & -0.31 & -0.64 & -0.32 & -0.18 & {\color{ForestGreen}\textbf{-0.15}} & -0.52 & -0.28 &  & -1.46 & {\textbf{0.20}} & -0.67 & -0.93 & 0.09 & -0.71 & -0.73 & 0.19 & {\color{ForestGreen}\textbf{1.40}} & -1.09 \\
      \rowcolor{RowAlt}
      Omega
      & 0.44 & 0.64 & 0.86 & 0.75 & 0.51 & 0.75 & {\textbf{0.86}} & {\color{ForestGreen}\textbf{0.87}} & 0.58 & 0.74 &  & 0.10 & {\textbf{1.18}} & 0.30 & 0.37 & 1.08 & 0.49 & 0.42 & 1.16 & {\color{ForestGreen}\textbf{2.35}} & 0.22 \\
      MaxDD
      & -23.01 & -8.12 & -6.60 & -9.49 & -10.97 & -9.60 & -11.56 & {\textbf{-5.79}} & {\color{ForestGreen}\textbf{-4.97}} & -8.99 &  & -16.57 & {\textbf{-0.94}} & -8.19 & -9.92 & -4.18 & -6.54 & -7.43 & -1.49 & {\color{ForestGreen}\textbf{-0.34}} & -12.84 \\
      \rowcolor{RowAlt}
      Edge
      & 0.30 & 0.00 & 0.04 & 0.09 & 0.05 & 0.16 & {\color{ForestGreen}\textbf{0.86}} & 0.01 & 0.26 & {\textbf{0.48}} &  & 0.29 & {\textbf{0.39}} & 0.00 & 0.16 & 0.17 & 0.16 & 0.00 & 0.07 & {\color{ForestGreen}\textbf{0.50}} & 0.00 \\

    \addlinespace[0.5em]
    \midrule
    \addlinespace[0.5em]
    \multicolumn{22}{l}{\textcolor{dpd2}{\textbf{Panel C: Classical Forecast Accuracy}}} \\
    \addlinespace[0.3em]

      RMSE
      & 1.13 & 0.99 & {\textbf{0.94}} & 1.02 & 1.05 & 1.02 & {\color{ForestGreen}\textbf{0.91}} & 0.97 & 1.05 & 1.03 &  & 1.99 & 0.95 & 2.00 & 1.61 & 0.94 & 1.22 & 1.41 & {\textbf{0.85}} & {\color{ForestGreen}\textbf{0.78}} & 1.85 \\
      \rowcolor{RowAlt}
      MAE
      & 1.29 & 1.07 & {\textbf{1.03}} & 1.07 & 1.12 & 1.07 & 1.04 & {\color{ForestGreen}\textbf{1.02}} & 1.08 & 1.07 &  & 2.22 & {\textbf{0.92}} & 1.56 & 1.71 & 0.97 & 1.36 & 1.40 & 0.96 & {\color{ForestGreen}\textbf{0.79}} & 1.95 \\
      $\rho(1)$
      & {\textbf{0.54}} & 0.66 & 0.64 & 0.67 & 0.70 & 0.71 & {\color{ForestGreen}\textbf{0.49}} & 0.71 & 0.74 & 0.68 &  & 0.52 & 0.38 & 0.31 & 0.68 & {\textbf{0.22}} & 0.64 & 0.64 & 0.29 & {\color{ForestGreen}\textbf{0.07}} & 0.86 \\
      \rowcolor{RowAlt}
      DM $t$-stat
      & -1.44 & 0.20 & {\color{ForestGreen}\textbf{0.71}} & -0.47 & -1.06 & -0.41 & 0.58 & {\textbf{0.68}} & -1.36 & -0.87 &  & -3.37 & 0.17 & -1.08 & -1.79 & 0.22 & -0.82 & -0.94 & {\textbf{0.87}} & {\color{ForestGreen}\textbf{1.32}} & -2.04 \\

    \addlinespace[0.3em]
    \bottomrule
  \end{tabular}%
  }

  \vspace{0.25em}
  \parbox{\linewidth}{\scriptsize
    \textit{Notes}: Panels A--B report risk-adjusted metrics; Panel C reports classical forecast accuracy metrics. $\rho(1)$ = first-order autocorrelation of errors.
    Best: \textcolor{ForestGreen}{\textbf{bold green}}; second-best: \textbf{bold}.
  }
  }

\end{table}
\end{landscape}

%% file: 03_tables/indpro__h4.tex
% ------------------------------------------------------------
%  EW VERSION - THREE-PANEL TABLE (SE / AE / Classical)
%  Template: dpd2 color + RowAlt alternating rows
% ------------------------------------------------------------

\begin{landscape}
\begin{table}[t!]
  \centering
  \caption{\normalsize Industrial Production ($h=4$)}
  \vspace*{-0.65em}
  \label{tab:indpro_h4}

  {\fontfamily{phv}\selectfont
  \resizebox{\linewidth}{!}{%
  \scriptsize
  \setlength{\tabcolsep}{0.3em}
  \renewcommand{\arraystretch}{1.65}
  \begin{tabular}{l r r r r r r r r r r c r r r r r r r r r r}
    \toprule

    & \multicolumn{10}{c}{\textcolor{dpd2}{\textbf{2007Q2 -- 2019Q4}}}
    && \multicolumn{10}{c}{\textcolor{dpd2}{\textbf{2022Q1 -- 2025Q1}}}
    \\
    \cmidrule(lr){2-11}
    \cmidrule(lr){13-22}
    & FAAR & RF & LGB & LGB$^{\texttt{A}}$+ & LGB+ & KRR & NN & RR & SPF & TabPFN 
    && FAAR & RF & LGB & LGB$^{\texttt{A}}$+ & LGB+ & KRR & NN & RR & SPF & TabPFN \\

    \midrule
    \addlinespace[0.5em]
    \multicolumn{22}{l}{\textcolor{dpd2}{\textbf{Panel A: Squared Error}}} \\
    \addlinespace[0.3em]

      Return
      & -0.48 & 0.01 & -0.09 & -0.14 & -0.14 & {\color{ForestGreen}\textbf{0.12}} & 0.01 & {\textbf{0.10}} & -0.03 & -0.02 &  & 0.36 & -0.25 & 0.15 & -1.53 & 0.05 & {\color{ForestGreen}\textbf{0.67}} & -0.28 & {\textbf{0.60}} & 0.54 & -0.74 \\
      \rowcolor{RowAlt}
      Sharpe
      & -0.96 & 0.05 & -0.20 & -0.49 & -0.65 & {\color{ForestGreen}\textbf{0.63}} & 0.03 & {\textbf{0.60}} & -0.28 & -0.04 &  & 0.74 & -0.27 & 0.20 & -1.12 & 0.07 & {\color{ForestGreen}\textbf{1.38}} & -0.28 & {\textbf{1.32}} & 1.22 & -0.74 \\
      Sortino
      & -0.88 & 0.07 & -0.28 & -0.55 & -0.69 & {\color{ForestGreen}\textbf{1.80}} & 0.04 & {\textbf{1.53}} & -0.33 & -0.04 &  & 3.30 & -0.32 & 0.27 & -1.04 & 0.10 & {\textbf{5.20}} & -0.32 & {\color{ForestGreen}\textbf{8.09}} & 4.93 & -0.78 \\
      \rowcolor{RowAlt}
      Omega
      & 0.15 & 1.08 & 0.65 & 0.50 & 0.38 & {\color{ForestGreen}\textbf{3.05}} & 1.05 & {\textbf{2.89}} & 0.62 & 0.92 &  & 3.87 & 0.67 & 1.36 & 0.19 & 1.10 & {\textbf{7.42}} & 0.68 & {\color{ForestGreen}\textbf{9.09}} & 7.02 & 0.38 \\
      MaxDD
      & -26.20 & -3.28 & -7.47 & -9.23 & -9.20 & {\textbf{-0.15}} & -7.99 & {\color{ForestGreen}\textbf{-0.13}} & -3.07 & -10.27 &  & -0.17 & -5.23 & -0.57 & -19.42 & -0.68 & {\textbf{-0.08}} & -8.58 & {\color{ForestGreen}\textbf{-0.05}} & -0.09 & -8.54 \\
      \rowcolor{RowAlt}
      Edge
      & 0.00 & 0.03 & 0.88 & 0.24 & 0.02 & 0.01 & {\textbf{0.89}} & 0.01 & 0.05 & {\color{ForestGreen}\textbf{1.83}} &  & {\textbf{0.15}} & 0.00 & 0.05 & 0.11 & 0.00 & 0.04 & 0.01 & 0.00 & {\color{ForestGreen}\textbf{0.81}} & 0.02 \\

    \addlinespace[0.5em]
    \midrule
    \addlinespace[0.5em]
    \multicolumn{22}{l}{\textcolor{dpd2}{\textbf{Panel B: Absolute Error}}} \\
    \addlinespace[0.3em]

      Return
      & -0.37 & 0.02 & -0.12 & -0.16 & -0.10 & {\textbf{0.07}} & -0.06 & {\color{ForestGreen}\textbf{0.07}} & -0.01 & -0.02 &  & 0.24 & -0.12 & 0.11 & -0.54 & 0.01 & {\color{ForestGreen}\textbf{0.37}} & -0.13 & 0.30 & {\textbf{0.34}} & -0.37 \\
      \rowcolor{RowAlt}
      Sharpe
      & -1.03 & 0.06 & -0.45 & -0.49 & -0.39 & {\textbf{0.39}} & -0.20 & {\color{ForestGreen}\textbf{0.45}} & -0.11 & -0.05 &  & 0.73 & -0.23 & 0.26 & -0.78 & 0.02 & {\color{ForestGreen}\textbf{1.12}} & -0.25 & 1.00 & {\textbf{1.07}} & -0.65 \\
      Sortino
      & -0.96 & 0.09 & -0.50 & -0.60 & -0.54 & {\textbf{0.63}} & -0.27 & {\color{ForestGreen}\textbf{0.85}} & -0.16 & -0.06 &  & 2.71 & -0.30 & 0.42 & -0.84 & 0.02 & {\textbf{3.00}} & -0.30 & 2.98 & {\color{ForestGreen}\textbf{3.04}} & -0.72 \\
      \rowcolor{RowAlt}
      Omega
      & 0.22 & 1.09 & 0.51 & 0.55 & 0.63 & {\textbf{1.61}} & 0.77 & {\color{ForestGreen}\textbf{1.81}} & 0.86 & 0.94 &  & 3.44 & 0.73 & 1.44 & 0.38 & 1.02 & {\textbf{3.84}} & 0.73 & 3.67 & {\color{ForestGreen}\textbf{4.07}} & 0.45 \\
      MaxDD
      & -19.70 & -4.26 & -8.16 & -11.20 & -9.07 & {\textbf{-0.47}} & -8.46 & {\color{ForestGreen}\textbf{-0.33}} & -2.34 & -7.04 &  & {\textbf{-0.15}} & -2.67 & -0.51 & -7.71 & -0.73 & {\color{ForestGreen}\textbf{-0.12}} & -4.66 & -0.18 & -0.15 & -4.87 \\
      \rowcolor{RowAlt}
      Edge
      & 0.02 & 0.19 & 0.23 & {\textbf{0.36}} & 0.11 & 0.05 & 0.27 & 0.08 & 0.29 & {\color{ForestGreen}\textbf{1.49}} &  & 0.50 & 0.00 & 0.23 & {\textbf{0.54}} & 0.00 & 0.05 & 0.03 & 0.00 & {\color{ForestGreen}\textbf{1.04}} & 0.13 \\

    \addlinespace[0.5em]
    \midrule
    \addlinespace[0.5em]
    \multicolumn{22}{l}{\textcolor{dpd2}{\textbf{Panel C: Classical Forecast Accuracy}}} \\
    \addlinespace[0.3em]

      RMSE
      & 1.22 & 1.00 & 1.04 & 1.07 & 1.07 & {\color{ForestGreen}\textbf{0.94}} & 1.00 & {\textbf{0.95}} & 1.02 & 1.01 &  & 0.80 & 1.12 & 0.92 & 1.59 & 0.97 & {\color{ForestGreen}\textbf{0.58}} & 1.13 & {\textbf{0.63}} & 0.68 & 1.32 \\
      \rowcolor{RowAlt}
      MAE
      & 1.37 & 0.98 & 1.12 & 1.16 & 1.10 & {\textbf{0.93}} & 1.06 & {\color{ForestGreen}\textbf{0.93}} & 1.01 & 1.02 &  & 0.76 & 1.12 & 0.89 & 1.54 & 0.99 & {\color{ForestGreen}\textbf{0.63}} & 1.13 & 0.70 & {\textbf{0.66}} & 1.37 \\
      $\rho(1)$
      & 0.64 & 0.74 & {\textbf{0.63}} & 0.74 & 0.79 & 0.73 & 0.70 & 0.73 & 0.75 & {\color{ForestGreen}\textbf{0.55}} &  & 0.23 & 0.67 & 0.32 & 0.81 & 0.47 & {\color{ForestGreen}\textbf{-0.00}} & 0.62 & {\textbf{0.16}} & 0.38 & 0.63 \\
      \rowcolor{RowAlt}
      DM $t$-stat
      & -2.58 & 0.19 & -0.92 & -1.77 & -1.73 & {\textbf{1.54}} & 0.09 & {\color{ForestGreen}\textbf{1.85}} & -1.20 & -0.14 &  & 1.33 & -0.54 & 0.42 & -1.87 & 0.16 & {\color{ForestGreen}\textbf{2.57}} & -0.51 & {\textbf{2.50}} & 2.05 & -1.69 \\

    \addlinespace[0.3em]
    \bottomrule
  \end{tabular}%
  }

  \vspace{0.25em}
  \parbox{\linewidth}{\scriptsize
    \textit{Notes}: Panels A--B report risk-adjusted metrics; Panel C reports classical forecast accuracy metrics. $\rho(1)$ = first-order autocorrelation of errors.
    Best: \textcolor{ForestGreen}{\textbf{bold green}}; second-best: \textbf{bold}.
  }
  }

\end{table}
\end{landscape}

%% file: 03_tables/dSpread__h1.tex
% ------------------------------------------------------------
%  EW VERSION - THREE-PANEL TABLE (SE / AE / Classical)
%  Template: dpd2 color + RowAlt alternating rows
% ------------------------------------------------------------

\begin{landscape}
\begin{table}[t!]
  \centering
  \caption{\normalsize $\Delta$Spread ($h=1$)}
  \vspace*{-0.65em}
  \label{tab:dSpread_h1}

  {\fontfamily{phv}\selectfont
  \resizebox{\linewidth}{!}{%
  \scriptsize
  \setlength{\tabcolsep}{0.3em}
  \renewcommand{\arraystretch}{1.65}
  \begin{tabular}{l r r r r r r r r r r c r r r r r r r r r r}
    \toprule

    & \multicolumn{10}{c}{\textcolor{dpd2}{\textbf{2007Q2 -- 2019Q4}}}
    && \multicolumn{10}{c}{\textcolor{dpd2}{\textbf{2021Q1 -- 2025Q1}}}
    \\
    \cmidrule(lr){2-11}
    \cmidrule(lr){13-22}
    & FAAR & RF & LGB & LGB$^{\texttt{A}}$+ & LGB+ & KRR & NN & RR & SPF & TabPFN 
    && FAAR & RF & LGB & LGB$^{\texttt{A}}$+ & LGB+ & KRR & NN & RR & SPF & TabPFN \\

    \midrule
    \addlinespace[0.5em]
    \multicolumn{22}{l}{\textcolor{dpd2}{\textbf{Panel A: Squared Error}}} \\
    \addlinespace[0.3em]

      Return
      & -0.52 & {\color{ForestGreen}\textbf{0.08}} & -0.12 & {\textbf{-0.01}} & -0.12 & -0.14 & -0.44 & -0.04 & {---} & -0.10 &  & -1.98 & 0.03 & -0.45 & 0.04 & {\color{ForestGreen}\textbf{0.36}} & 0.18 & 0.32 & 0.06 & {---} & {\textbf{0.33}} \\
      \rowcolor{RowAlt}
      Sharpe
      & -0.53 & {\color{ForestGreen}\textbf{0.11}} & -0.16 & {\textbf{-0.01}} & -0.13 & -0.15 & -0.26 & -0.06 & {---} & -0.14 &  & -0.54 & 0.05 & -0.25 & 0.06 & {\color{ForestGreen}\textbf{0.49}} & 0.35 & 0.39 & 0.15 & {---} & {\textbf{0.42}} \\
      Sortino
      & -0.55 & {\color{ForestGreen}\textbf{0.16}} & -0.19 & {\textbf{-0.01}} & -0.16 & -0.17 & -0.28 & -0.07 & {---} & -0.16 &  & -0.54 & 0.09 & -0.27 & 0.13 & {\color{ForestGreen}\textbf{1.98}} & 1.20 & {\textbf{1.35}} & 0.34 & {---} & 1.09 \\
      \rowcolor{RowAlt}
      Omega
      & 0.34 & {\color{ForestGreen}\textbf{1.29}} & 0.72 & {\textbf{0.99}} & 0.76 & 0.67 & 0.42 & 0.86 & {---} & 0.76 &  & 0.06 & 1.09 & 0.52 & 1.11 & {\color{ForestGreen}\textbf{2.56}} & 1.86 & {\textbf{2.32}} & 1.27 & {---} & 2.10 \\
      MaxDD
      & -34.75 & {\color{ForestGreen}\textbf{-0.90}} & -13.76 & {\textbf{-7.90}} & -15.41 & -10.76 & -24.34 & -8.48 & {---} & -13.71 &  & -3.88 & -4.22 & -15.28 & -2.64 & -1.94 & -1.57 & -2.42 & {\textbf{-1.36}} & {---} & {\color{ForestGreen}\textbf{-0.44}} \\
      \rowcolor{RowAlt}
      Edge
      & {\textbf{0.35}} & 0.30 & 0.02 & 0.03 & 0.13 & 0.02 & 0.07 & 0.03 & {---} & {\color{ForestGreen}\textbf{0.51}} &  & 0.03 & 0.00 & 0.00 & 0.11 & 0.14 & 0.00 & {\color{ForestGreen}\textbf{1.72}} & 0.00 & {---} & {\textbf{0.60}} \\

    \addlinespace[0.5em]
    \midrule
    \addlinespace[0.5em]
    \multicolumn{22}{l}{\textcolor{dpd2}{\textbf{Panel B: Absolute Error}}} \\
    \addlinespace[0.3em]

      Return
      & -0.19 & {\color{ForestGreen}\textbf{0.08}} & -0.02 & {\textbf{0.06}} & 0.04 & 0.02 & -0.05 & 0.03 & {---} & 0.02 &  & -0.52 & -0.03 & -0.05 & -0.07 & {\textbf{0.07}} & -0.06 & 0.05 & -0.11 & {---} & {\color{ForestGreen}\textbf{0.10}} \\
      \rowcolor{RowAlt}
      Sharpe
      & -0.47 & {\color{ForestGreen}\textbf{0.28}} & -0.07 & {\textbf{0.18}} & 0.11 & 0.05 & -0.12 & 0.11 & {---} & 0.05 &  & -0.65 & -0.11 & -0.09 & -0.17 & {\textbf{0.20}} & -0.25 & 0.14 & -0.54 & {---} & {\color{ForestGreen}\textbf{0.23}} \\
      Sortino
      & -0.55 & {\color{ForestGreen}\textbf{0.48}} & -0.10 & {\textbf{0.31}} & 0.17 & 0.07 & -0.15 & 0.18 & {---} & 0.07 &  & -0.64 & -0.17 & -0.11 & -0.25 & {\textbf{0.39}} & -0.37 & 0.25 & -0.65 & {---} & {\color{ForestGreen}\textbf{0.40}} \\
      \rowcolor{RowAlt}
      Omega
      & 0.50 & {\color{ForestGreen}\textbf{1.53}} & 0.90 & {\textbf{1.32}} & 1.18 & 1.08 & 0.80 & 1.22 & {---} & 1.07 &  & 0.19 & 0.87 & 0.86 & 0.79 & {\textbf{1.30}} & 0.74 & 1.23 & 0.52 & {---} & {\color{ForestGreen}\textbf{1.39}} \\
      MaxDD
      & -14.34 & {\color{ForestGreen}\textbf{-0.45}} & -6.00 & {\textbf{-0.52}} & -0.88 & -2.68 & -4.68 & -0.96 & {---} & -5.74 &  & -2.37 & -1.98 & -4.02 & -2.69 & -2.01 & {\textbf{-1.85}} & -2.26 & -1.93 & {---} & {\color{ForestGreen}\textbf{-0.51}} \\
      \rowcolor{RowAlt}
      Edge
      & {\textbf{0.60}} & 0.28 & 0.08 & 0.08 & 0.52 & 0.13 & 0.17 & 0.14 & {---} & {\color{ForestGreen}\textbf{0.60}} &  & 0.22 & 0.00 & 0.00 & 0.25 & 0.07 & 0.00 & {\textbf{1.09}} & 0.00 & {---} & {\color{ForestGreen}\textbf{1.14}} \\

    \addlinespace[0.5em]
    \midrule
    \addlinespace[0.5em]
    \multicolumn{22}{l}{\textcolor{dpd2}{\textbf{Panel C: Classical Forecast Accuracy}}} \\
    \addlinespace[0.3em]

      RMSE
      & 1.23 & {\color{ForestGreen}\textbf{0.96}} & 1.06 & {\textbf{1.00}} & 1.06 & 1.07 & 1.20 & 1.02 & {---} & 1.05 &  & 1.73 & 0.98 & 1.21 & 0.98 & {\color{ForestGreen}\textbf{0.80}} & 0.90 & 0.82 & 0.97 & {---} & {\textbf{0.82}} \\
      \rowcolor{RowAlt}
      MAE
      & 1.19 & {\color{ForestGreen}\textbf{0.92}} & 1.02 & {\textbf{0.94}} & 0.96 & 0.98 & 1.05 & 0.97 & {---} & 0.98 &  & 1.52 & 1.03 & 1.05 & 1.07 & {\textbf{0.93}} & 1.06 & 0.95 & 1.11 & {---} & {\color{ForestGreen}\textbf{0.90}} \\
      $\rho(1)$
      & {\textbf{0.15}} & 0.23 & 0.27 & 0.21 & 0.31 & 0.22 & {\color{ForestGreen}\textbf{0.10}} & 0.21 & {---} & 0.18 &  & {\textbf{0.11}} & 0.37 & 0.15 & 0.30 & 0.12 & 0.21 & {\color{ForestGreen}\textbf{0.08}} & 0.27 & {---} & 0.16 \\
      \rowcolor{RowAlt}
      DM $t$-stat
      & -1.98 & {\color{ForestGreen}\textbf{0.76}} & -0.89 & {\textbf{-0.05}} & -0.56 & -1.11 & -1.35 & -0.48 & {---} & -0.60 &  & -1.11 & 0.10 & -0.51 & 0.14 & {\color{ForestGreen}\textbf{1.20}} & 0.84 & 0.90 & 0.40 & {---} & {\textbf{1.08}} \\

    \addlinespace[0.3em]
    \bottomrule
  \end{tabular}%
  }

  \vspace{0.25em}
  \parbox{\linewidth}{\scriptsize
    \textit{Notes}: Panels A--B report risk-adjusted metrics; Panel C reports classical forecast accuracy metrics. $\rho(1)$ = first-order autocorrelation of errors.
    Best: \textcolor{ForestGreen}{\textbf{bold green}}; second-best: \textbf{bold}.
  }
  }

\end{table}
\end{landscape}

%% file: 03_tables/dSpread__h2.tex
% ------------------------------------------------------------
%  EW VERSION - THREE-PANEL TABLE (SE / AE / Classical)
%  Template: dpd2 color + RowAlt alternating rows
% ------------------------------------------------------------

\begin{landscape}
\begin{table}[t!]
  \centering
  \caption{\normalsize $\Delta$Spread ($h=2$)}
  \vspace*{-0.65em}
  \label{tab:dSpread_h2}

  {\fontfamily{phv}\selectfont
  \resizebox{\linewidth}{!}{%
  \scriptsize
  \setlength{\tabcolsep}{0.3em}
  \renewcommand{\arraystretch}{1.65}
  \begin{tabular}{l r r r r r r r r r r c r r r r r r r r r r}
    \toprule

    & \multicolumn{10}{c}{\textcolor{dpd2}{\textbf{2007Q2 -- 2019Q4}}}
    && \multicolumn{10}{c}{\textcolor{dpd2}{\textbf{2021Q1 -- 2025Q1}}}
    \\
    \cmidrule(lr){2-11}
    \cmidrule(lr){13-22}
    & FAAR & RF & LGB & LGB$^{\texttt{A}}$+ & LGB+ & KRR & NN & RR & SPF & TabPFN 
    && FAAR & RF & LGB & LGB$^{\texttt{A}}$+ & LGB+ & KRR & NN & RR & SPF & TabPFN \\

    \midrule
    \addlinespace[0.5em]
    \multicolumn{22}{l}{\textcolor{dpd2}{\textbf{Panel A: Squared Error}}} \\
    \addlinespace[0.3em]

      Return
      & -1.45 & 0.14 & -0.35 & {\color{ForestGreen}\textbf{0.22}} & -0.10 & 0.17 & 0.09 & 0.13 & {---} & {\textbf{0.21}} &  & -4.03 & 0.01 & -0.08 & 0.08 & {\color{ForestGreen}\textbf{0.21}} & 0.19 & 0.05 & 0.11 & {---} & {\textbf{0.20}} \\
      \rowcolor{RowAlt}
      Sharpe
      & -0.66 & {\textbf{0.40}} & -0.33 & {\color{ForestGreen}\textbf{0.55}} & -0.12 & 0.30 & 0.12 & 0.32 & {---} & 0.37 &  & -0.51 & 0.02 & -0.11 & 0.19 & 0.27 & {\color{ForestGreen}\textbf{0.51}} & 0.08 & {\textbf{0.42}} & {---} & 0.29 \\
      Sortino
      & -0.64 & {\textbf{0.69}} & -0.35 & {\color{ForestGreen}\textbf{1.24}} & -0.14 & 0.47 & 0.15 & 0.51 & {---} & 0.59 &  & -0.51 & 0.02 & -0.14 & 0.36 & 0.59 & {\color{ForestGreen}\textbf{1.65}} & 0.15 & {\textbf{1.15}} & {---} & 0.52 \\
      \rowcolor{RowAlt}
      Omega
      & 0.10 & 1.90 & 0.46 & {\color{ForestGreen}\textbf{2.51}} & 0.79 & 1.69 & 1.26 & {\textbf{1.92}} & {---} & 1.91 &  & 0.02 & 1.03 & 0.84 & 1.33 & 1.57 & {\color{ForestGreen}\textbf{2.25}} & 1.13 & {\textbf{1.92}} & {---} & 1.53 \\
      MaxDD
      & -78.01 & -0.64 & -28.59 & {\color{ForestGreen}\textbf{-0.20}} & -16.96 & {\textbf{-0.37}} & -0.62 & -0.43 & {---} & -0.60 &  & -4.91 & -2.36 & -6.07 & -2.58 & -3.82 & -1.32 & -4.57 & {\textbf{-1.16}} & {---} & {\color{ForestGreen}\textbf{-0.82}} \\
      \rowcolor{RowAlt}
      Edge
      & 0.01 & 0.08 & 0.11 & 0.24 & 0.09 & 0.08 & {\textbf{0.64}} & 0.00 & {---} & {\color{ForestGreen}\textbf{0.73}} &  & 0.00 & 0.10 & 0.16 & 0.00 & {\textbf{0.57}} & 0.00 & 0.06 & 0.00 & {---} & {\color{ForestGreen}\textbf{1.11}} \\

    \addlinespace[0.5em]
    \midrule
    \addlinespace[0.5em]
    \multicolumn{22}{l}{\textcolor{dpd2}{\textbf{Panel B: Absolute Error}}} \\
    \addlinespace[0.3em]

      Return
      & -0.43 & 0.11 & -0.12 & 0.11 & 0.02 & {\textbf{0.12}} & 0.11 & 0.09 & {---} & {\color{ForestGreen}\textbf{0.17}} &  & -0.67 & {\color{ForestGreen}\textbf{0.04}} & -0.00 & {\textbf{0.02}} & -0.01 & 0.02 & -0.06 & -0.04 & {---} & 0.02 \\
      \rowcolor{RowAlt}
      Sharpe
      & -0.73 & {\textbf{0.54}} & -0.28 & 0.45 & 0.05 & 0.43 & 0.33 & 0.42 & {---} & {\color{ForestGreen}\textbf{0.60}} &  & -0.59 & {\color{ForestGreen}\textbf{0.15}} & -0.01 & 0.08 & -0.02 & {\textbf{0.09}} & -0.16 & -0.23 & {---} & 0.04 \\
      Sortino
      & -0.70 & {\textbf{0.90}} & -0.31 & 0.73 & 0.07 & 0.73 & 0.50 & 0.74 & {---} & {\color{ForestGreen}\textbf{1.02}} &  & -0.58 & {\color{ForestGreen}\textbf{0.21}} & -0.02 & 0.11 & -0.03 & {\textbf{0.14}} & -0.21 & -0.31 & {---} & 0.07 \\
      \rowcolor{RowAlt}
      Omega
      & 0.20 & {\textbf{2.01}} & 0.63 & 1.85 & 1.08 & 1.80 & 1.60 & 1.97 & {---} & {\color{ForestGreen}\textbf{2.24}} &  & 0.12 & {\color{ForestGreen}\textbf{1.18}} & 0.99 & 1.10 & 0.98 & {\textbf{1.11}} & 0.82 & 0.76 & {---} & 1.06 \\
      MaxDD
      & -24.04 & -0.72 & -13.61 & -0.39 & -7.63 & {\color{ForestGreen}\textbf{-0.26}} & -0.39 & {\textbf{-0.38}} & {---} & -0.61 &  & -3.24 & {\color{ForestGreen}\textbf{-0.92}} & -2.24 & -2.38 & -2.80 & -1.63 & -3.88 & {\textbf{-1.41}} & {---} & -3.12 \\
      \rowcolor{RowAlt}
      Edge
      & 0.04 & 0.16 & 0.27 & 0.26 & 0.16 & 0.07 & {\textbf{0.62}} & 0.00 & {---} & {\color{ForestGreen}\textbf{1.01}} &  & 0.01 & 0.29 & {\textbf{0.75}} & 0.06 & 0.23 & 0.04 & 0.08 & 0.00 & {---} & {\color{ForestGreen}\textbf{0.93}} \\

    \addlinespace[0.5em]
    \midrule
    \addlinespace[0.5em]
    \multicolumn{22}{l}{\textcolor{dpd2}{\textbf{Panel C: Classical Forecast Accuracy}}} \\
    \addlinespace[0.3em]

      RMSE
      & 1.56 & 0.93 & 1.16 & {\color{ForestGreen}\textbf{0.89}} & 1.05 & 0.91 & 0.95 & 0.93 & {---} & {\textbf{0.89}} &  & 2.24 & 1.00 & 1.04 & 0.96 & {\color{ForestGreen}\textbf{0.89}} & 0.90 & 0.98 & 0.94 & {---} & {\textbf{0.90}} \\
      \rowcolor{RowAlt}
      MAE
      & 1.43 & 0.89 & 1.12 & 0.89 & 0.98 & {\textbf{0.88}} & 0.89 & 0.91 & {---} & {\color{ForestGreen}\textbf{0.83}} &  & 1.67 & {\color{ForestGreen}\textbf{0.96}} & 1.00 & {\textbf{0.98}} & 1.01 & 0.98 & 1.06 & 1.04 & {---} & 0.98 \\
      $\rho(1)$
      & {\color{ForestGreen}\textbf{0.04}} & 0.19 & 0.27 & 0.14 & 0.21 & 0.18 & 0.29 & 0.25 & {---} & {\textbf{0.05}} &  & {\color{ForestGreen}\textbf{0.06}} & 0.34 & {\textbf{0.13}} & 0.29 & 0.41 & 0.19 & 0.18 & 0.22 & {---} & 0.16 \\
      \rowcolor{RowAlt}
      DM $t$-stat
      & -1.75 & 1.23 & -1.27 & {\color{ForestGreen}\textbf{1.96}} & -0.40 & 1.12 & 0.48 & {\textbf{1.41}} & {---} & 1.04 &  & -1.06 & 0.04 & -0.22 & 0.40 & 0.58 & {\color{ForestGreen}\textbf{1.08}} & 0.15 & {\textbf{0.97}} & {---} & 0.66 \\

    \addlinespace[0.3em]
    \bottomrule
  \end{tabular}%
  }

  \vspace{0.25em}
  \parbox{\linewidth}{\scriptsize
    \textit{Notes}: Panels A--B report risk-adjusted metrics; Panel C reports classical forecast accuracy metrics. $\rho(1)$ = first-order autocorrelation of errors.
    Best: \textcolor{ForestGreen}{\textbf{bold green}}; second-best: \textbf{bold}.
  }
  }

\end{table}
\end{landscape}

%% file: 03_tables/dSpread__h4.tex
% ------------------------------------------------------------
%  EW VERSION - THREE-PANEL TABLE (SE / AE / Classical)
%  Template: dpd2 color + RowAlt alternating rows
% ------------------------------------------------------------

\begin{landscape}
\begin{table}[t!]
  \centering
  \caption{\normalsize $\Delta$Spread ($h=4$)}
  \vspace*{-0.65em}
  \label{tab:dSpread_h4}

  {\fontfamily{phv}\selectfont
  \resizebox{\linewidth}{!}{%
  \scriptsize
  \setlength{\tabcolsep}{0.3em}
  \renewcommand{\arraystretch}{1.65}
  \begin{tabular}{l r r r r r r r r r r c r r r r r r r r r r}
    \toprule

    & \multicolumn{10}{c}{\textcolor{dpd2}{\textbf{2007Q2 -- 2019Q4}}}
    && \multicolumn{10}{c}{\textcolor{dpd2}{\textbf{2021Q1 -- 2025Q1}}}
    \\
    \cmidrule(lr){2-11}
    \cmidrule(lr){13-22}
    & FAAR & RF & LGB & LGB$^{\texttt{A}}$+ & LGB+ & KRR & NN & RR & SPF & TabPFN 
    && FAAR & RF & LGB & LGB$^{\texttt{A}}$+ & LGB+ & KRR & NN & RR & SPF & TabPFN \\

    \midrule
    \addlinespace[0.5em]
    \multicolumn{22}{l}{\textcolor{dpd2}{\textbf{Panel A: Squared Error}}} \\
    \addlinespace[0.3em]

      Return
      & -0.51 & 0.09 & -0.04 & -0.69 & -0.45 & {\textbf{0.10}} & -0.16 & {\color{ForestGreen}\textbf{0.11}} & {---} & 0.01 &  & -8.16 & -0.41 & -0.43 & -0.25 & {\color{ForestGreen}\textbf{0.10}} & -0.03 & {\textbf{0.07}} & -0.05 & {---} & -0.31 \\
      \rowcolor{RowAlt}
      Sharpe
      & -0.47 & 0.24 & -0.07 & -0.46 & -0.42 & {\textbf{0.25}} & -0.19 & {\color{ForestGreen}\textbf{0.32}} & {---} & 0.04 &  & -0.66 & -0.76 & -0.38 & -0.41 & {\color{ForestGreen}\textbf{0.19}} & -0.24 & {\textbf{0.16}} & -0.27 & {---} & -0.60 \\
      Sortino
      & -0.49 & 0.34 & -0.09 & -0.47 & -0.44 & {\textbf{0.39}} & -0.22 & {\color{ForestGreen}\textbf{0.46}} & {---} & 0.06 &  & -0.65 & -0.74 & -0.40 & -0.50 & {\color{ForestGreen}\textbf{0.35}} & -0.29 & {\textbf{0.31}} & -0.34 & {---} & -0.61 \\
      \rowcolor{RowAlt}
      Omega
      & 0.38 & 1.49 & 0.90 & 0.30 & 0.38 & {\textbf{1.57}} & 0.69 & {\color{ForestGreen}\textbf{1.71}} & {---} & 1.07 &  & 0.02 & 0.21 & 0.43 & 0.55 & {\textbf{1.32}} & 0.72 & {\color{ForestGreen}\textbf{1.35}} & 0.70 & {---} & 0.33 \\
      MaxDD
      & -31.20 & -0.89 & -8.82 & -41.78 & -26.72 & {\color{ForestGreen}\textbf{-0.66}} & -16.69 & {\textbf{-0.79}} & {---} & -0.81 &  & -138.84 & -7.67 & -12.13 & -5.84 & -2.52 & {\color{ForestGreen}\textbf{-0.81}} & -2.18 & {\textbf{-0.88}} & {---} & -6.34 \\
      \rowcolor{RowAlt}
      Edge
      & 0.24 & 0.03 & 0.26 & 0.23 & {\color{ForestGreen}\textbf{0.29}} & 0.01 & {\textbf{0.27}} & 0.05 & {---} & 0.02 &  & 0.00 & 0.00 & {\textbf{0.16}} & 0.01 & {\color{ForestGreen}\textbf{0.87}} & 0.02 & 0.09 & 0.04 & {---} & 0.08 \\

    \addlinespace[0.5em]
    \midrule
    \addlinespace[0.5em]
    \multicolumn{22}{l}{\textcolor{dpd2}{\textbf{Panel B: Absolute Error}}} \\
    \addlinespace[0.3em]

      Return
      & -0.14 & {\color{ForestGreen}\textbf{0.10}} & 0.02 & -0.16 & -0.10 & 0.07 & 0.00 & {\textbf{0.08}} & {---} & 0.05 &  & -1.16 & -0.21 & -0.12 & -0.16 & {\color{ForestGreen}\textbf{0.02}} & {\textbf{0.01}} & 0.01 & -0.04 & {---} & -0.19 \\
      \rowcolor{RowAlt}
      Sharpe
      & -0.31 & {\color{ForestGreen}\textbf{0.44}} & 0.05 & -0.35 & -0.26 & 0.35 & 0.01 & {\textbf{0.39}} & {---} & 0.24 &  & -0.74 & -0.65 & -0.25 & -0.46 & {\textbf{0.05}} & {\color{ForestGreen}\textbf{0.12}} & 0.05 & -0.27 & {---} & -0.62 \\
      Sortino
      & -0.35 & {\color{ForestGreen}\textbf{0.69}} & 0.07 & -0.38 & -0.31 & 0.51 & 0.02 & {\textbf{0.54}} & {---} & 0.36 &  & -0.71 & -0.67 & -0.28 & -0.52 & {\textbf{0.08}} & {\color{ForestGreen}\textbf{0.17}} & 0.07 & -0.35 & {---} & -0.65 \\
      \rowcolor{RowAlt}
      Omega
      & 0.62 & {\color{ForestGreen}\textbf{1.80}} & 1.08 & 0.55 & 0.67 & 1.66 & 1.02 & {\textbf{1.74}} & {---} & 1.37 &  & 0.07 & 0.36 & 0.66 & 0.57 & 1.06 & {\color{ForestGreen}\textbf{1.16}} & {\textbf{1.08}} & 0.71 & {---} & 0.36 \\
      MaxDD
      & -11.38 & -1.96 & -4.75 & -12.17 & -7.56 & {\textbf{-0.62}} & -5.47 & -0.78 & {---} & {\color{ForestGreen}\textbf{-0.58}} &  & -19.79 & -4.47 & -5.23 & -4.33 & -2.05 & {\color{ForestGreen}\textbf{-0.35}} & -0.95 & {\textbf{-0.80}} & {---} & -4.00 \\
      \rowcolor{RowAlt}
      Edge
      & {\textbf{0.48}} & 0.18 & 0.40 & 0.39 & 0.44 & 0.06 & {\color{ForestGreen}\textbf{0.56}} & 0.09 & {---} & 0.07 &  & 0.00 & 0.00 & {\textbf{0.73}} & 0.09 & {\color{ForestGreen}\textbf{0.88}} & 0.06 & 0.15 & 0.06 & {---} & 0.06 \\

    \addlinespace[0.5em]
    \midrule
    \addlinespace[0.5em]
    \multicolumn{22}{l}{\textcolor{dpd2}{\textbf{Panel C: Classical Forecast Accuracy}}} \\
    \addlinespace[0.3em]

      RMSE
      & 1.23 & 0.95 & 1.02 & 1.30 & 1.21 & {\textbf{0.95}} & 1.08 & {\color{ForestGreen}\textbf{0.94}} & {---} & 0.99 &  & 3.03 & 1.19 & 1.20 & 1.12 & {\color{ForestGreen}\textbf{0.95}} & 1.02 & {\textbf{0.96}} & 1.02 & {---} & 1.14 \\
      \rowcolor{RowAlt}
      MAE
      & 1.14 & {\color{ForestGreen}\textbf{0.90}} & 0.98 & 1.16 & 1.10 & 0.93 & 1.00 & {\textbf{0.92}} & {---} & 0.95 &  & 2.16 & 1.21 & 1.12 & 1.16 & {\color{ForestGreen}\textbf{0.98}} & {\textbf{0.99}} & 0.99 & 1.04 & {---} & 1.19 \\
      $\rho(1)$
      & 0.35 & 0.25 & 0.22 & {\textbf{0.10}} & 0.18 & 0.27 & {\color{ForestGreen}\textbf{0.02}} & 0.22 & {---} & 0.14 &  & -0.45 & {\textbf{0.19}} & 0.23 & {\color{ForestGreen}\textbf{0.17}} & 0.27 & 0.28 & 0.25 & 0.26 & {---} & 0.22 \\
      \rowcolor{RowAlt}
      DM $t$-stat
      & -1.67 & 0.96 & -0.31 & -1.08 & -1.65 & {\textbf{1.07}} & -0.62 & {\color{ForestGreen}\textbf{1.57}} & {---} & 0.14 &  & -1.16 & -1.41 & -0.71 & -0.78 & {\color{ForestGreen}\textbf{0.41}} & -0.50 & {\textbf{0.32}} & -0.69 & {---} & -1.09 \\

    \addlinespace[0.3em]
    \bottomrule
  \end{tabular}%
  }

  \vspace{0.25em}
  \parbox{\linewidth}{\scriptsize
    \textit{Notes}: Panels A--B report risk-adjusted metrics; Panel C reports classical forecast accuracy metrics. $\rho(1)$ = first-order autocorrelation of errors.
    Best: \textcolor{ForestGreen}{\textbf{bold green}}; second-best: \textbf{bold}.
  }
  }

\end{table}
\end{landscape}